\documentclass[12pt,a4paper]{report}
\usepackage{epsfig,subfigure,multirow}
\usepackage{fancyheadings,calc}
\usepackage{amsmath,amssymb,amsfonts}
\usepackage{pstricks,pst-node,pst-tree}
\usepackage{tipa}
\usepackage{algorithm}
\usepackage{algpseudocode}
\usepackage{tikz}
\usepackage{bbding}
\usepackage{pifont}
\usepackage{CJKutf8}
\usepackage{array}
\usepackage{pdflscape}

\normalsize

\newpsobject{showgrid}{psgrid}{subgriddiv=1,griddots=10,gridlabels=6pt}

\begin{document}

\pagenumbering{roman}

\begin{center}

\LARGE{\textbf{ \\
\vspace{1cm}
Unsupervised Spoken Term Discovery on Untranscribed Speech}} \\
\vspace{3cm}
\Large{SUNG, Man Ling} \\
\vspace{3cm}
\large{A Thesis Submitted in Partial Fulfillment \\
of the Requirements for the Degree of \\
Master of Philosophy \\
in \\ Electronic Engineering \\ 
\vspace{4cm}
The Chinese University of Hong Kong \\
September 2019 } \\
\end{center}


\pagestyle{plain}
\setcounter{page}{1}



\begin{center}\textbf{\large{Abstract}}\end{center}

%
%
%
%

Speech technology is becoming mature recent years mostly contributed to the development of deep neural network (DNN). 
However, under the situation when 1) there are not enough training data, and 2) phonetic information is absent, traditional acoustic modelling technique is no longer applicable. This is a common problem when processing low/zero-resource languages data, broadcasts, lectures, meetings, which are mostly untranscribed data. 

In this thesis, we investigate the use of unsupervised spoken term discovery in tackling this problem. 
Unsupervised spoken term discovery aims to discover topic-related terminologies in a speech without knowing the phonetic properties of the language and content. It can be further divided into two parts: Acoustic segment modelling (ASM) and unsupervised pattern discovery. ASM learns the phonetic structures of zero-resource language audio with no phonetic knowledge available, generating self-derived ``phonemes''.
The audio are labelled with these ``phonemes'' to obtain ``phoneme'' sequences. Unsupervised pattern discovery searches for repetitive patterns in the ``phoneme'' sequences. The discovered patterns can be grouped to determine the keywords of the audio. 

Multilingual neural network with bottleneck layer is used for feature extraction. Experiments show that bottleneck features facilitate the training of ASM compared to conventional features such as MFCC.

The unsupervised spoken term discovery system is experimented with online lectures covering different topics by different speakers. It is shown that the system learns the phonetic information of the language and can discover frequent spoken terms that align with text  transcription. By using information retrieval technology such as word embedding and TFIDF, it is shown that the discovered keywords can be further used for topic comparison.
\begin{CJK*}{UTF8}{bsmi}
\begin{center}\textbf{\large{摘要}}\end{center}

關鍵詞發現（spoken term discovery）是在大量語音數據庫中，發現在其中重複出現的相關詞語的技術。關鍵詞發現近年在學術研究及技術應用領域都有高度的關注及興趣。
傳統的聲學模型多適用於資源豐富的語言，但難以處理資源缺乏的語言或過於大量而未標示的語音。
本文關注處理如何在目標語言沒有足夠訓練資源下，以關鍵詞發現對語音及其內容作出整理。

本文先採用聲學片段模型（ASM－Acoustic Segment Model）框架來無監督訓練語音識別器，生成子詞標註。
再以序列比對（sequence alignment）選出重複子詞序列，並以聚類發現在語音中多次出現的關鍵詞。

我們提出以多語言神經網絡提取瓶頸特徵進行語音片段聚類，並測試不同的聚類方法。實驗證明多語言神經網絡生成的瓶頸特徵比傳統MFCC及FBANK特徵更適合用來訓練ASM。


本文使用網上的課堂錄音進行測試。實驗証明，此模型能有效處理無標籤的課堂錄音，並發現和文本相應的多次出現關鍵詞。發現的關鍵詞，在使用資訊檢索的技術，可以用作課堂內容分類及比較。

\end{CJK*}

\clearpage

\begin{centering}
\large{\textbf{Acknowledgements}}\\
\vspace{\baselineskip}
\end{centering}

I would like to express my sincere gratitude to my supervisor, Professor Tan Lee, for his guidance, support and patience during my Mphil. study. He provided a lot of insights and placed a lot of efforts on guiding students to think independently. 
My special thanks to Dr. Man-Hung Siu for the opportunities and support during the internship with Raytheon BBN technology.
I am also thankful to Professor Wing-Kin Ma, Professor Pak-Chung Ching for the inspiration on research field through seminars and conversations.

I would also like to thank all the colleages in DSP-STL lab. It is my pleasure to work with Siyuan, ying Qin, Matthew, Herman, Shuiyang, Jiarui, Yuzhong, Yuanyuan, Dehua,
Ryan, Mingjie, Keung, Lawrence, and more. Thanks to Arthur for all the technical support and maintenance. Thanks to 
David, Hoi-To Wai, Shing Yu and Gary for the time to enjoy coffee and lunch.

Finally I would like to thank my family for their love and care. 
Thanks for the support and prayer from friends, brothers and sisters in church and CCC.
Thank you Michael for the understanding and encouragement, as well as knowledge in working on Linux platform.

\clearpage


\tableofcontents

\addcontentsline{toc}{chapter}{List of Tables}
\listoftables\newpage
\addcontentsline{toc}{chapter}{List of Figures}
\listoffigures\newpage




\clearpage
\pagestyle{fancy}
\pagenumbering{arabic}

\chapter{Introduction}

\section{Background}
For a number of decades, great efforts have been put toward developing computing systems that are able to recognize and understand human speech. The relevant technology is known as automatic speech recognition (ASR).
State-of-the-art ASR systems are well developed for most of the major languages in the world. It can be arguably said that they are close to human performance in terms of recognition accuracy \cite{saon2017english, kim2017joint}.

A typical speech recognition system has two key components, namely, acoustic model and language model. The acoustic model (AM) maps input speech signals to phonemes or other linguistic units. The language model (LM) governs how to derive a word sequence from a phoneme sequence. Both the AM and LM are in the form of statistical models or neural network models that are learned from data with properly represented contents.
For an ASR system to achieve state-of-the-art performance, a large amount of training data are indispensable \cite{dahl2012context}. To accomplish effective modeling of a given specific language, the training data must be well defined 
-- with the lexicon information and word-by-word transcriptions being accurately provided. When such kinds of data or knowledge resources are not available, which is commonly known as the ``low-resource'' or ``zero-resource'' scenario, training a high-performance model remains a great challenge.

Among the 7,000 languages in the world \cite{ethnologue}, the top 23 major languages are spoken by more than half of the global population. However, half of these 23 languages are still considered as low-resource languages as there are no well developed recognition systems at the moment due to limited data, hence the community has to look into alternative speech technologies which require less data \cite{babelwebsite, strassel2016lorelei}. 

Building up the data resource for a new language is not feasible in terms of the time, manpower, and linguistic expertise required. In the latest collection of Linguistic Data Consortium (LDC), only 102 languages are covered \cite{ldc2019database}. Therefore, there has been increasing research interest in non-traditional speech modelling techniques. The Zero Resource Speech Challenges have been organized regularly since 2015 to encourage bench-marking and research exchange on spoken language technology for low-resource languages \cite{zerospeech}.
The Low Resource Languages for Emergent Incidents Program (LORELEI) of DARPA in 2015 aimed at language-universal technology that does not rely on huge, manually translated, transcribed or annotated corpora, and is able to efficiently handle practical incidences in low-resource scenarios \cite{dapraLORELEI, Christianson2018}.







%
%
Research on low-resource languages can be categorized according to the following three assumed scenarios:
\begin{enumerate}
\item A large amount of un-transcribed data is available with only limited transcribed data are available;
\item Phonetic knowledge about the language is provided, but the available speech data are too little to training a statistical model;
\item Phonetic knowledge about the language is not available, which is referred to as the zero-resource case.
\end{enumerate}


In the first scenario, a common approach is to locate a small subset of speech data that is informative and representative, e.g., containing typical content, good coverage of phonetic variations, and/or few confusing words. These data are then manually transcribed to facilitate so-called \textbf{active learning} \cite{hakkani2002active}. Another approach is \textbf{semi-supervised learning}, in which a seed model is first trained with a small set of transcribed data to learn the hypothesis of the language, it will then decode the transcription of all unlabelled data \cite{vesely2013semi}. 

In the case of limited transcribed data, \textbf{transfer learning} methods can be applied. The idea is to transfer linguistic knowledge from a high-resource language in processing the target low-resource language. Transfer learning models are trained with transcribed data from one or more high-resource language(s) and refined with the data from the low-resource language. When a deep neural network model (DNN) is adopted, the hidden layers are shared, and the softmax outputs represent phonemes of the languages separately. 
By joint training with high-resource data, the ASR system could achieve a better performance than training with limited low-resource data \cite{huang2013cross}.

In the extreme case that both phonetic knowledge and transcriptions are absent, \textbf{unsupervised learning} is needed to learn the constituting elements and structure of the language completely from audio recordings. Specifically, the elements to be learned could be subword units \cite{liu2018completely}, word-like units \cite{lee2015unsupervised}, or phrase-level units \cite{park2005towards}. Automatic discovery of multi-word phrases has been receiving most interest over the past years \cite{babelwebsite,trmal2017kaldi}. For example, the MIT CSAIL group investigated methods of unsupervised pattern discovery on classroom lectures \cite{park2005towards}. 



Without requiring any prior linguistic knowledge, unsupervised learning methods can be applied to any low-resource language. They are also useful in a wider range of real-world applications that may involve popular spoken languages. These applications may involve multi-lingual, code-mixing, and/or accented speech that contain many colloquial terms and non-speech sounds, with unknown and complicated acoustic conditions. It is generally impractical and unnecessary to make effort on obtaining formal and accurate transcriptions for such kinds of speech data. Numerous studies have been done in this area \cite{park2005towards, hermann2018multilingual}. The key technical problem is known as unsupervised acoustic modelling.

Applications of unsupervised learning can also be applied to non-speech data. There are works on audio event detection that search for occurrence of pattern of automatically learnt acoustic units \cite{kumar2012audio}, music pattern analysis that learn the structure of music (e.g. ABAB) through pattern of music notes and can be applied to different genres such as jazz, classical, etc \cite{dannenberg2003pattern}.
Real world recordings are complex, with environments, acoustic elements and pattern durations that are changing and unpredictable from each of them. Traditional modelling methods have not considered enough variation of all the elements, and we do not have complete information yet, unsupervised learning can be considered to be a feasible approach.

Moreover, considering rapid increasing of information on the Internet, nowadays it is easy to get access to several million terabytes of data \cite{internetdata}. It is however impossible to apply traditional learning methods on these data as they are mostly unlabelled. Exploring the potential of unsupervised learning on these data has a great deal of implications as 
online data can be fully utilized, which is a better alternative than producing more labelled data. Through discovery of repeated data pattern, we can also avoid spending too much time in reading through every single bit of data. Summerization on the patterns can provide us useful information that can be represented in much less bytes.




\subsection{Thesis objective}

One main interest of this thesis is to explore the use of unsupervised learning on audio that are extracted from Internet, as it is easier to collect data for analysis that are recorded in real world scenario. 

A system is built such that when raw audio recordings without any phonetic and transcribed information are provided as input. The system automatically learns the phonetic units of the language, then performs pattern discovery on the phonetic units to obtain repeated word phrases of the recordings. The word phrases are then compared across recordings for topic comparison.

\begin{figure}
    \centering
    \hspace{-1cm}
  \includegraphics[width=15cm, height=4cm]{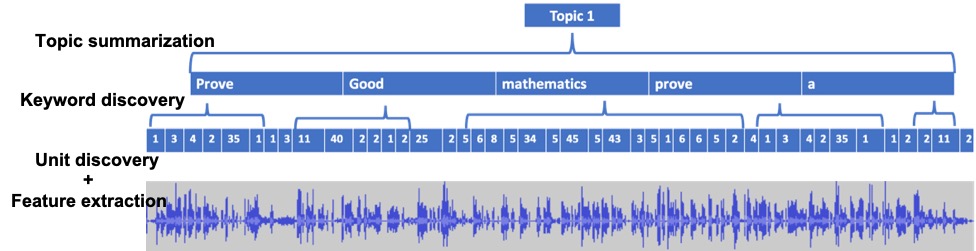}
   \caption{Bottom up approach of the fully unsupervised system.}
	\label{fig:system}
\end{figure}

In the system, a bottom-up approach that contains different levels of unsupervised learning are researched and used as shown in Figure \ref{fig:system}. The hierarchical structure is described as follow:

\begin{enumerate}
\item \textbf{Unsupervised feature extraction} that learns to extract representative linguistic representations from zero-resource language.
\item \textbf{Unsupervised units discovery} and \textbf{segmentation} that learn the linguistic units and the units boundary information.
\item \textbf{Unsupervised pattern discovery} that discovers keyword phrases of each recording through searching for repeated unit sequences (patterns).
\item \textbf{Unsupervised topic comparison} on each recording to determine how similar or different the recordings are.
\end{enumerate}

\section{Organization of Thesis}

The following chapters of the thesis will be presented as follow:

Chapter 2 reviews on recent projects related to unsupervised acoustic modelling in the research field and their applications. Different approaches related to traditional acoustic speech recognition systems and the use of deep neural networks are also introduced.

Chapter 3 introduces the first part of the system, unsupervised acoustic modelling, which discovers phonetic unit information of the recording. Approaches in extracting language independent features and different clustering methods are discussed.

Chapter 4 introduces the second part of the system, pattern discovery of the unit sequences generated from the unsupervised acoustic model. The proposed metric, algorithm and their effectiveness in discovering patterns are discussed.

Chapter 5 compares patterns discovered from opencourse lectures. Relationship of the patterns and lecture topics is also evaluated. Suggestions on potential applications in topic comparison are also given. 

Chapter 6 concludes the whole work and discusses the areas of improvement and future work.
\chapter{Background}

This chapter provides the general background of research on spoken pattern discovery and reviews related previous studies. We will start by describing conventional automatic speech recognition (ASR) system design, and then focus on acoustic modeling in unsupervised scenario. Representative works on spoken term detection are also discussed.

\section{Fundamentals of ASR}

Automatic speech recognition (ASR) is a technology that enables computers to analyze and convert sound waves of speech into text. A typical ASR system is trained to have the ability to map an audio input to a sequence of phonemes or words.

\subsection{Probabilistic framework}

Let $O$ be an observed audio signal and $W$ be a word sequence or phoneme sequence. $P(W|O)$ denotes the conditional probability of $W$ given $O$, indicating how likely $W$ is the cause of $O$. The goal of ASR is to determine the most likely word sequence $W^*$ when the observation $O$ is given, i.e.,
\begin{equation}
W^* = {arg\,max}_W P(W|O)
\end{equation}

Following the Bayes' Theorem, we have

\begin{equation}
{arg\,max}_W P(W|O) = {arg\,max}_W  \underbrace{P(O|W)}_{\makebox[0pt]{\tiny{acoustic model}}} \overbrace{P(W)}^{\makebox[0pt]{\tiny{language model}}},
\end{equation}
in which the maximization is applied on two parts of models:
\begin{enumerate}
\item \textbf{Acoustic model (AM)}, which describes the mapping between acoustic observation $O$ to linguistic representation $W$. $P(O|W)$ measures the probability of $O$ being obtained when $W$ is spoken.
\item \textbf{Language model (LM)}, which represents the language rules/properties governing $W$. $P(W)$ basically measures how likely $W^*$ is valid in the language.
\end{enumerate}

\begin{figure}[h]
    \centering
    \hspace{-1cm}
  \includegraphics[width=15cm, height=3cm]{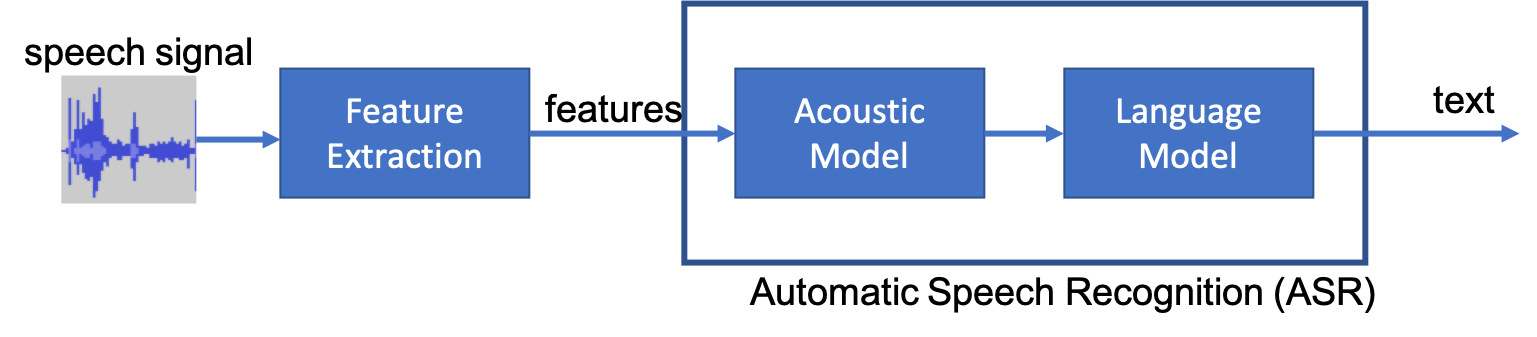}
   \caption{The process of ASR.}
	\label{fig:asr_system}
\end{figure}

The whole process of ASR is illustrated as in Figure \ref{fig:asr_system}. The acoustic observation $O$ is typically obtained from the raw audio via a feature extraction process. The extracted features $O$ are used as the input of the acoustic model to evaluate $P(O|W)$. Combining with the language model information, the most likely word sequence $W^*$ is determined.

\subsection{Feature extraction}
Feature extraction is the first step of processing raw input speech, aiming to obtain a meaningful representation for subsequent modeling and evaluation.
A general goal of feature extraction for ASR is to derive a low-dimension feature vector from each short-time frame of speech. The features are generally expected to be insensitive against changes of speaker and recording environment and be discriminative to phonemes \cite{gold2011speech}. Typically, features at frame level are computed every 10 ms with an analysis window of around 25 ms \cite{gales2008application}.

\begin{figure}[h]
    \centering
    \hspace{-1cm}
  \includegraphics[width=9cm, height=4.5cm]{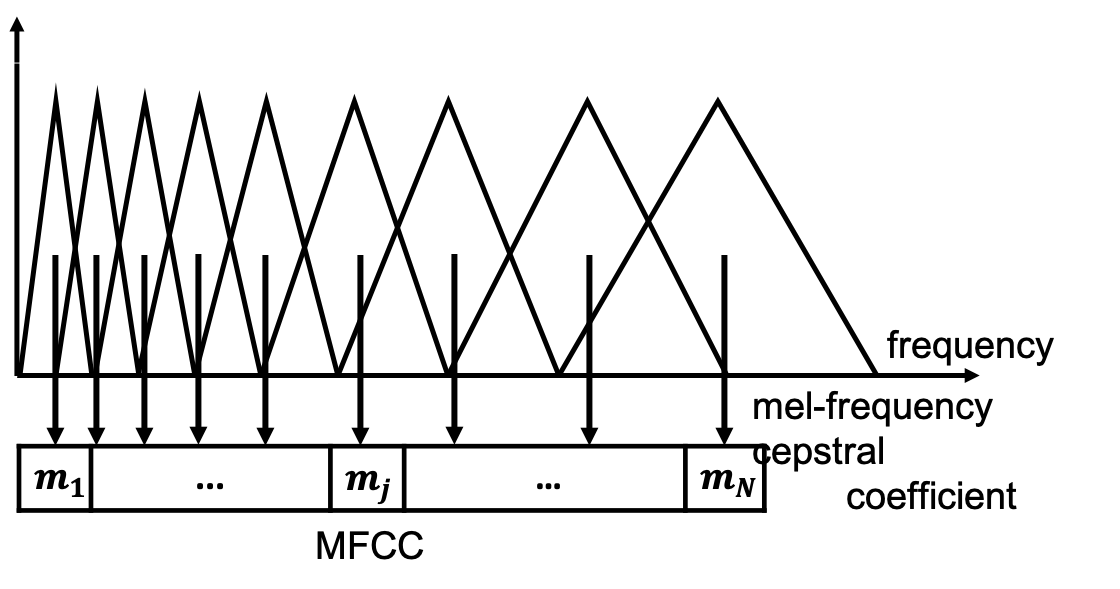}
   \caption{Triangle filters that generate MFCC features.}
	\label{fig:mfcc}
\end{figure}

\textbf{Mel-frequency cepstral coefficients (MFCCs)} are by far the best known and most commonly used feature for acoustic modeling of ASR \cite{davis1980comparison}. Spectral analysis of input speech is applied on the Mel scale, which was inspired by human auditory perception \cite{stevens1937scale}. The Mel-scaled filter-bank comprises a number of triangular filters as shown in Figure \ref{fig:mfcc}. MFCC features are computed by taking Discrete Cosine Transform on the log power of these filters' output.

Another commonly used feature is known as the \textbf{perceptual linear predictive (PLP)} coefficients \cite{hermansky1990perceptual}, which gives an estimate of auditory spectrum based on the concept of hearing psychophysics. The Fourier spectrum of speech signal passes through critical-band integration and re-sampling, and the result is multiplied with an equal-loudness curve and compressed with the power law in hearing. The inverse Fourier transform is applied to obtain the PLP coefficients.

MFCC and PLP both provide a representation of smoothed short-term spectrum that is compressed and normalized in the same way as human auditory perception. Previous research showed that PLP is more robust to noise than MFCC \cite{dave2013feature}.

\subsection{Acoustic model} \label{Acoustic Model}

Acoustic model represents the relationship between the audio signal and its corresponding phonemes. It learns the relationship with statistical representations of sounds $O$ that make up the word $W$. Given a word $W$, its pronunciation is formed by sequence of phones $Q = q_1q_2q_3...q_n$. The probability of observed sounds $O$ given $W$ is:


\begin{equation}
P(O|W) = \sum_Q P(O|Q) P(Q|W)
\end{equation}

The input audio is not limited to a single word, but can also be a sentence or paragraph. We denote the written form of corresponding words as transcription.

To model the relationship of input audio and transcription, two problems of probability computation are needed: 1) transition probability from phone $q_i$ to $q_{i+1}$, 2) the output observation probability from $O_t$ to $q_i$ at specific time $t$. In the next section we will explain how the statistical relationships are learnt by a conventional acoustic model that can determine the possible transcription of future incoming audio signal. 

\subsubsection{Phone-based GMM-HMM}

\begin{figure}[h]
    \centering
    \hspace{-1cm}
  \includegraphics[width=8cm, height=4.5cm]{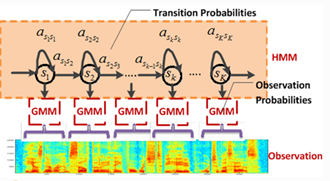}
   \caption{Illustration of the GMM-HMM model.}
	\label{fig:gmmhmm}
\end{figure}


As shown in Figure \ref{fig:gmmhmm}), each phone $q_i$ is represented by a continuous density HMM with states $s_1,s_2...,s_N$. At each time step $t$, the HMM may make a transition from its current state $i$ to the next connected state $j$. The transition probability from state $s_i$ to $s_j$ is denoted as $a_{ij}$. At the same time, the observation probability $b_j$ is generated at state $j$ with a specific statistical distribution associated with the state. The state output distribution can be modelled with a mixture of Gaussians, providing a highly flexible distribution to model speaker, accent and gender difference.

For the model to work properly, training is needed to adjust model parameters $\lambda$ of GMM-HMM such that the input audio signal can align with transcription. Therefore, exact $O$ and $W$ have to be known for training the model. This is called supervised training.

Expectation-maximisation (EM) can be used to search for suitable model parameters. It iterately calculates the likelihood of input-to-output-alignment given certain model parameters, and re-estimates the model parameters accordingly, until the parameters that give the maximum likelihood is reached, i.e. ${arg\,max}_\lambda P(O|W)$. 
The trained acoustic model is then capable to process new audio data from the same domain as the training data and recognize the corresponding phones and words.






\section{DNN based ASR} \label{Supervised DNN Acoustic Modelling}







The first attempt of using DNN in ASR was the phone recognition system reported in \cite{mohamed2009deep, dahl2010phone}. They demonstrated clearly better phone accuracy than well-tuned traditional GMM-HMM models. Since \cite{hinton2012deep}, 
DNN has been taking over GMM for high-performance acoustic modeling in ASR. Many new model structures have been developed, leading to significant and continuous performance improvement.


\subsection{Basics of DNN}

The basic form of DNN is a multi-layer feed-forward network built with simple computation units called neurons. Each neuron performs computation of a simple function $y_j = f(x_j)$, where $y_j$ and $x_j$ are the output and input of the neuron, respectively. In a feed-forward network, the input to a neuron at an intermediate layer is given as
\begin{equation}
    x_j = b_j + \sum_i y_iw_{ij},
\end{equation}
where $w_{ij}$ denotes the connection weight from neuron $i$ to $j$, and $b_j$ is known as the bias of neuron $j$.

Training of a DNN refers to the process of determining the values of weights and biases of all neurons in the network. This is typically done by minimizing a loss function $L(.)$ that quantifies the discrepancy between the actual output of the DNN and the desired output. The minimization can be done by stochastic gradient descent, which is an iterative optimization algorithm similar to EM, the derivative of the loss function of each training example is back-propagated, and the values of $w$ are updated and fine-tuned until local minimum of the loss function is met.




Essentially, the goal of DNN is to approximate non-linear function $f(.)$ that can produce $y$ from $x$ in $y = f(x)$, by learning from the training examples. It is commonly used when the function is too hard to be formulated and understood.

\subsection{DNN-HMM for ASR}

When the variations in the audio are too large and it is too hard to understand the relationship of the observed audio and states using GMM, DNN can be used to approximate the distribution instead. 
There is a variety of DNN models that have been applied to acoustic modeling. A brief review of these models is given below.

\subsubsection{Model structure and training}
The first attempt of using DNN in ASR is deep belief network (DBN) in phone recognition \cite{mohamed2009deep, dahl2010phone}, which is a method in training DNN by stacking pre-trained narrow networks (Restricted Boltzmann Machine) together to make it ``deep''.

It replaces traditional GMM from GMM-HMM to DNN-HMM \cite{dahl2012context} and achieves phone accuracies that are higher than well-tuned traditional models. It is now therefore the basic and standard model used in DNN acoustic modelling.

Compare with GMM, DNN does not require uncorrelated features such as MFCC, therefore other features such as filter bank (fbank) can be used in training DNN-HMM that give better representation to the speech data \cite{mohamed2012understanding}.


\subsubsection{Convolutional neural network (CNN)}
CNN is widely used in image processing \cite{lecun1995convolutional}. Different from normal DNN, activation functions are applied to the nodes that are fully connected. CNN also consists of convolution layers at the beginning of the network that convolute the input 2D image to the next layers. It also has pooling layers at the latter part that extract the maximum neighbour values to reduce layer resolution.

In speech recognition, CNN replaces DNN to form a CNN-HMM, with input feature being the 2D spectrogram of the audio signal with frequency and time information. The special structure of weight sharing, pooling and local connectivity of CNN enables invariability to slight changes in speech features, making it better in dealing with speaker and environment variations \cite{abdel2014convolutional}. 

\subsection{Temporal DNN}

In speech recognition, it is not only important to consider local region information, longer dependencies such as context, referencing from previously appeared words and language structures can also benefit the ASR training. This is especially important in time series applications such as speech, audio and video, compare to static data such as image processing. Several temporal acoustic models are therefore developed and become more widely used in speech recognition.

\subsubsection{Time delay neural network (TDNN)}
TDNN is first introduced in the application of phone recognition \cite{waibel1995phoneme}.
Concept of delay is introduced, extra weights representing the input delay are multiplied before
computing the total weighted sum to the unit.
With this design, the network is exposed to sequence of patterns and is able to relate and compare the current input with its past inputs, resulting in more powerful time series data processing ability.

However, as the structure gets more complicated, the network becomes more complex as well. A small TDNN network can consists of several millions parameters and large amount of training data is needed \cite{peddinti2015time}.


\subsubsection{Recurrent neural network (RNN)}

RNN is a DNN with self-connected hidden layers, allowing it to has ``memory'' on its previous states in processing the input sequence. The self-connecting edges have extra weights that determine the importance of the previous unit's states.  
However, when processing long sequential data especially in speech recognition, it faces the 
vanishing and the exploding gradient problems \cite{hochreiter2001gradient} and therefore is not widely used until the extension to LSTM 
 \cite{graves2013speech}.





\subsubsection{Long short-term memory (LSTM)}

To solve the vanishing gradient problem, modification to RNN is made by introducing regulating cells that control the flowing of data and error \cite{hochreiter1997long}. Input gate, output gate and forget gate are added to control whether the value should go into the unit, pass to the next unit or  reset in the unit respectively.

In speech processing applications, bidirectional LSTM (BILSTM) is used more often to consider both past and future events into account during sequence training \cite{graves2005framewise, sak2014long}. Despite higher ability in relating events among the sequence and gives better performance in learning the phoneme sequences, it takes much longer time to train the network.





\subsection{End-to-end speech recognition (E2E)}

End-to-end speech recognition trains the whole ASR with one neural network system. Training and optimizing acoustic model 
and language model 
separately will result in sub-optimal solution of the combined ASR. By training the whole ASR as one single system
, a better decoding result can be achieved.

Currently, E2E technique includes connectionist temporal classification (CTC), attention-based encoder decoder and hybrid of the two -- attention-based CTC \cite{kim2017joint}.

\subsubsection{Connectionist temporal classification (CTC)}

Temporal DNN acoustic model only determines the most possible phone of each utterance or frame, which is called framewise classification. However, CTC learns the probability of observing the corresponding labels at particular time. By multiplying the probability of each label at different time, possible sequence paths with their probabilities corresponding to the observed audio are obtained.
The best representative phone sequence can be obtained by choosing the sequence path with the highest path probability \cite{graves2006connectionist}.




In practice, RNN and LSTM are used in constructing CTC due to their sequence considering property \cite{kim2017joint, ochiai2017multichannel}. It has the same training objective as HMM but outperforms HMM \cite{graves2006connectionist}. CTC represents both acoustic model and language model as one model, and directly searches for model parameters that give the maximum likelihood of the input-output mapping. It can be used as end-to-end model that learns the whole ASR to map audio to text without learning intermediate phonemes \cite{graves2014towards}.
However, since it is a complete end-to-end model, it is hard to interpret intermediate information such as phones.

\subsubsection{Attention-based encoder decoder}

Attention is commonly used in sequence-to-sequence processing such as machine translation and natural language processing, it tells specifically which elements in the sequence should the model places more or less attentions when making decisions \cite{luong2015effective}.

Encoder-decoder model is used to tackle input and output with variable lengths \cite{sutskever2014sequence}. The encoder takes in input speech features and generates intermediate representations, and encoder takes in the representations to output the desired text sequence. Content-and-location-awareness is added into the attention mechanism to allow the model to output text with correct word order as the input speech \cite{chorowski2015attention, bahdanau2016end}. 
While being so flexible, it is difficult to predict proper alignment due to the lack of left-to-right constraints.




\section{DNN for feature extraction}

Besides acoustic modelling, DNN can be widely applied to other modelling techniques such as feature extraction and language modelling. It can also be used in domain adaptation with some modifications to the training process. 

When we do not have enough speech data for a specific language, known as low-resource language, it is hard for traditional DNN acoustic model to achieve satisfying performance when solely trained the low resource language compare to trained with rich resource languages. But recently, increasing effort has been put in developing methods to tackle the issues yielded by low resource language as the computation power improves significantly and more advanced technologies are discovered.

\subsection{Knowledge transfer}
One method to deal with limited data is knowledge transfer, in which resource-rich data from another domain is used to assist the modelling of resource-limited data. In this session, some methods for knowledge transfer are introduced.

\subsubsection{Knowledge distillation}
Another approach to limited data is knowledge distillation, also named as teacher-student model. After training classification model (teacher), a new model (student) is trained base on the posterior probability distribution output from the teacher. The student is expected to learn the decision boundary information from the teacher and therefore achieves performance as compatible as the teacher with much less model parameters \cite{hinton2015distilling}.

This training technique can be applied to model compression (from large model to small model with equal performance), model conversion (from statistical model to neural network model \cite{bbn2018nnconvert}) and domain adaptation (from one high-resource domain to another low-resource domain \cite{asami2017domain}).

\subsubsection{Multilingual DNN}

\begin{figure}[h]
    \centering
    \hspace{-1cm}
  \includegraphics[width=8cm, height=6cm]{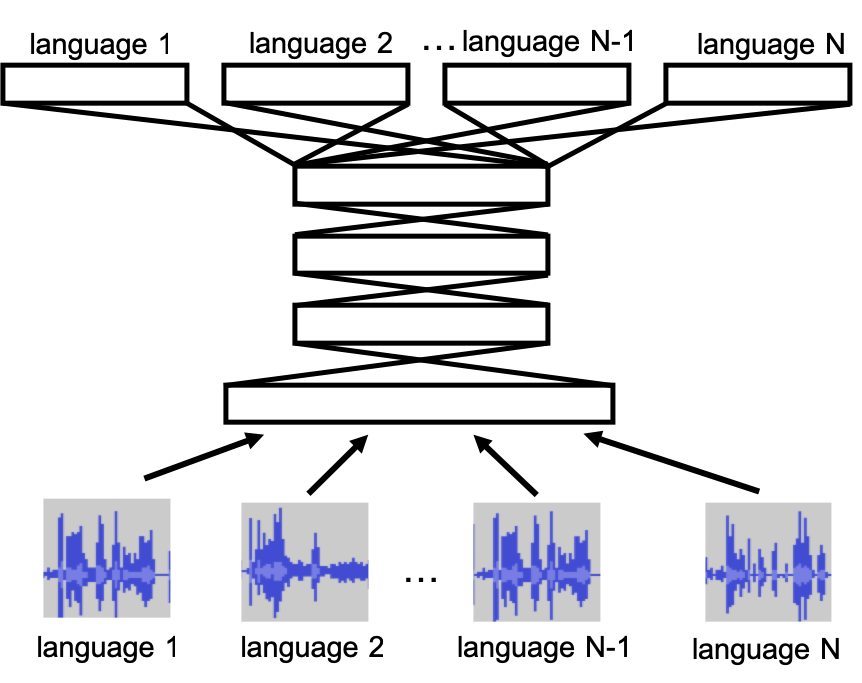}
   \caption{Multilingual DNN}
	\label{fig:Multilang_DNN}
\end{figure}

Besides training two separate models, another popular method in recognizing low resource language is transfer learning. A model is first trained with language with sufficient data, and fine-tune with target language with limited data. Transcription is expected to be available. 

It is experimented that training with more languages can significantly improve the result, therefore multilingual training is more preferable nowadays \cite{heigold2013multilingual}. In general, this process is also called multi-task learning. 
When training with more than one language(s), the hidden layers of acoustic models of different languages are shared, only the input features and output softmax layers are language specific (Figure \ref{fig:Multilang_DNN}).
It is expected that the rich resource language can help to provide language information for understanding the low resource language better. Pre-training with more languages can help the model to learn more generalized language properties and avoid overfitting, and is more suitable to apply to a new language.

Multi-task learning is a training technique which does not limited to DNN, either statistical models \cite{lu2014cross} or DNN \cite{nguyen2017transfer} models including CNN, LSTM, etc. can be used. 

\subsection{DNN as feature extractor}

DNN can also serve as feature extractor. The training process is similar to a phone recognizer. After training a phone classification network, the speech is feedforward to the network and vectors are extracted from the layer before the softmax, which are the posterior probabilities of the input speech frames corresponding to the phonemes. Then they are used as features to train on a GMM-HMM acoustic model. It is shown that compared with traditional MFCC, the features generated learn information than can benefit phone classification \cite{hermansky2000tandem}.

\subsubsection{Bottleneck layer and feature extraction}

Besides learning the posterior probability vectors, when one of the hidden layer is narrowed and with linear activation function, it compresses the data that passes through the layer \cite{kramer1991nonlinear}. Thus the DNN learns to forward the most important information to the next layers at the same time when solving the classification task without much result degradation \cite{grezl2008optimizing}. Features can then be extracted from the linear bottleneck layer by forwarding speech data to the model. The features generated are more data driven than in \cite{hermansky2000tandem}.
 
Bottleneck layer can be placed at 1) center layer, 2) layer before the softmax output. It is better to place at (2) to learn representations that are most sensitive to phoneme variations \cite{zhang2014extracting, you2015investigation}.

\subsection{Multilingual bottleneck network}

Multilingual network with bottleneck layer can extract language independent features for new input audio, preferably when each language is trained on separate softmax output layer \cite{vesely2012language}. 

\begin{figure}[h]
    \centering
    \hspace{-1cm}
  \includegraphics[width=8cm, height=6cm]{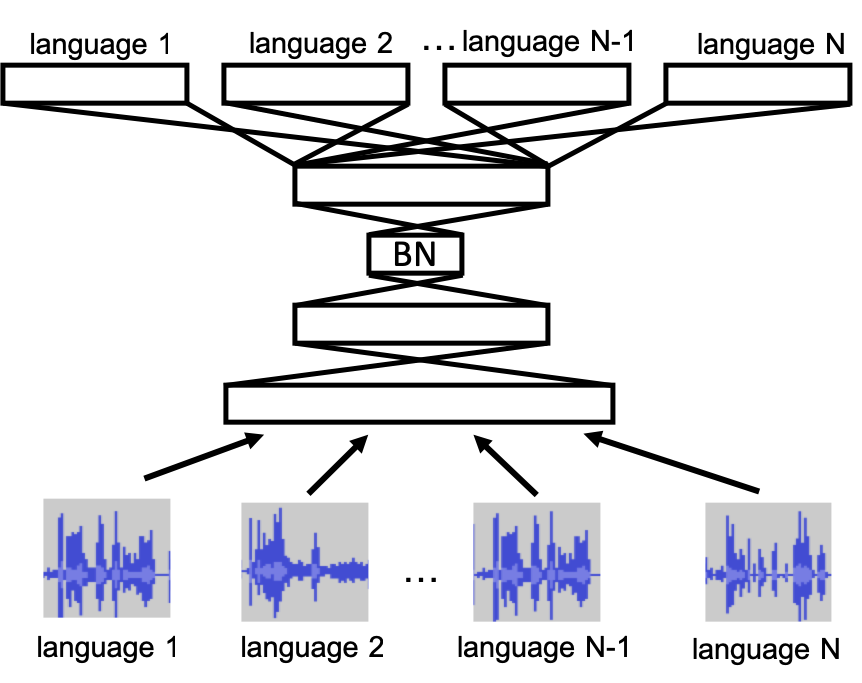}
   \caption{Multilingual DNN with bottleneck layer}
	\label{fig:Multilang_DNN_BN}
\end{figure}

Experiment \cite{vu2014investigating} has been done on investigating the learning effect of bottleneck feature from multilingual neural network, showing that even though the output phoneme labels of different languages are different, phonemes with same pronunciation are still projected to the same IPA symbol\footnote{IPA (International Phonetic Alphabet) is a standardized representation of the sounds of spoken language, it is independent to any language} area when represented as bottleneck feature. This concludes that multilingual bottleneck feature is able to learn phonetic information that are generalized and can represent the phonemes of a new untranscribed language.

Even though the process of training a multilingual bottleneck network is supervised and transcription is required in training, when training with many languages, the network with the bottleneck layer inserted learns the general phonetic properties that can be applied to any language. Multilingual bottleneck DNN can then be used as a feature extractor to extract language independent features from unlabelled new language data \cite{vu2012multilingual}.

\section{Acoustic segment modelling (ASM)}

Now we further look into the use of DNN in unsupervised acoustic modelling. When there is no linguistic knowledge nor transcription available, supervised acoustic modelling techniques are no longer applicable. One of the unsupervised acoustic modelling approach to tackle this scenario is 
\textbf{acoustic segment modelling (ASM)}, it discovers possible phonetic units and build an acoustic model accordingly.

ASM is first proposed in \cite{lee1988segment}, aiming to learn a self-derived acoustic model for isolated word recognition. The trained ASM on word recognition is shown compatible to supervised acoustic model. This workflow then becomes standardized for all ASM architecture.

\begin{figure}[h]
    \centering
    \hspace{-1cm}
  \includegraphics[width=14cm, height=1.7cm]{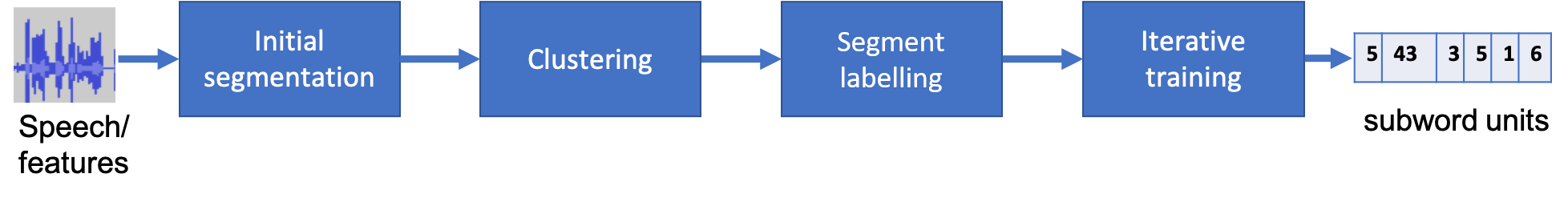}
   \caption{Components of Acoustic Segment Modelling (ASM)}
	\label{fig:ASM}
\end{figure}

Typically, ASM consists of three stages  (Figure \ref{fig:ASM}): 

\begin{enumerate}
\item \textbf{Initial segmentation} that identifies the potential phonetic units and their segmentation information from the input speech. Since the phonetic units discovered are not guaranteed to be the real phones of the language, they are called \textbf{subword units} instead.
\item \textbf{Segment clustering and labelling} that groups segments into clusters of subword units and labels the speech with corresponding subword units.
\item \textbf{Iterative training} of acoustic model with the discovered subword units.
\end{enumerate}

\subsection{Segmentation}

The first step for discovering phonemes from the speech is to identify all the potential phonetic unit boundaries, which is called segmentation. This process is relatively easier but is also very important. After segmentation, segment information such as durations, locations and segment features can be obtained for clustering into subword groups.

\textbf{Maximum likelihood segmentation}:
In \cite{qiao2008unsupervised}, number of segments can be determined based on the spectral distortion. The segment boundaries can be obtained by minimizing the overall likelihood distortion  using dynamic programming based Maximum likelihood (ML) segmentation.

\textbf{Dynamic programming algorithm (DP)} \cite{cohen1981segmenting}:
Statistical models for the speech data are built to model speaker, channel and speech information similar to an acoustic model. Then dynamic programming approach is used to identify the most probable segmentation.

\textbf{Maximum-margin segmentation} \cite{estevan2007finding}: 
Given the frame level feature vectors, maximum-margin clustering searches for the boundaries such that the margins between segments are maximized when grouping the frames into segment clusters.

\textbf{Bottom-up hierarchical clustering}: 
In \cite{qiao2008unsupervised}, time-constrained agglomerative clustering algorithm is used to find the optimal segmentation. It begins by treating each frame as an initial cluster and merges these clusters into larger segment clusters until the terminating criteria is met.


\textbf{Maximum spectral transition} \cite{dusan2006relation}: 
Phoneme can be analyzed in spectral space, maximum spectral transition can be used to determine the phoneme boundaries with the elimination of those with too short intervals.

\textbf{Graph-based observation space}:
The model in \cite{glass2003probabilistic} takes in features represented by observation space, such as graph or network, instead of temporal sequence space such as MFCC. This provides better information on the change of phonetic properties for segmentation. Boundaries are represented as arcs on the graph.




\textbf{Nonparametric Bayesian model}: 
Also name as Hidden Markov Model with Dirichlet process priors. A Dirichlet process (DP) is a discrete distribution of weighted sum of impulse functions. It is often used in Bayesian inference.
\cite{lee2012nonparametric} uses nonparametric Bayesian model with Dirichlet process priors to segment the utterances, and uses Gibbs sampler to estimate the segment boundaries.
In \cite{torbati2013speech}, hierarchical Dirichlet processes (HDP) is used, which is HMM with unbounded number of states to segment the utterances.



\textbf{Recognizers from other languages}: 
In \cite{feng2016exploiting}, the untranscribed speech is decoded with language mismatched phoneme recognizers which are trained with high resource languages. The segment boundaries produced by different recognizers are merged in form a single set of boundaries. The frame-level features in the same segments are averaged to form segment-level features.



\subsection{Clustering/Quantization} \label{clustering}
With segment-level boundaries available, segmental features can be obtained by combining the frame-level features within the boundaries. Segmental features are then grouped into subword clusters that are acoustically similar. The subword clusters are then labelled to the speech to generate initial segment sequence for subsequence segment modelling stage.

\textbf{Lloyd algorithm} \cite{lee1988segment}: 
Lloyd algorithm is often used in vector quantization. Cluster centroids are computed using a segment codebook. The goal of the segment codebook is to generate a set of vectors such that the accumulated segment distortion is minimized. 
Segments are assigned to their nearest cookbook entry to form groups of segments, then the distortion is minimized by updating centroids of the segment groups. The process is iterated until converge.

\textbf{Gaussian component clustering} \cite{wang2012acoustic}:
A GMM is trained on frame-level features, with the number of Gaussian components set to be the desire subword units. Clustering is then performed on the Gaussian components. Each cluster is then a small GMM of a subword unit, and the clusters can be used to score the speech segments. Segments are labelled with clusters of highest scores to generate the label sequence.

\textbf{Segmental Gaussian Mixture Model (SGMM)} \cite{siu2014unsupervised}:
Different from GMM, each term in an SGMM is a Gaussian whose mean is a vector trajectory in the cepstral feature space that varies over time to represent time varying characteristics of a sound.

Each segment is fitted with polynomial trajectory model. The pairwise distances between segments are calculated for clustering by binary centroid splitting algorithm. The clusters are then used as the basis for generating SGMM. SGMM is trained with EM algorithm and then the raw audio is labelled into initial label sequences for the next step.

\textbf{Spectral Clustering}:
With the class-by-segment posterior probabilities generated by recognizers or GMM models, spectral clustering can be used to cluster the speech segments, e.g. k-means clustering \cite{wang2015acoustic}.

If there are more than one set of segment posterior representations available, multiview spectral embedding can be used to embed the multiple representations into single posterior representation \cite{xia2010multiview}, and perform spectral clustering on the embedded vectors.


\subsection{Iterative modelling}
After obtaining all the initial segment information and segment clusters. ASM is trained iteratively to learn the finalized segment boundaries and clusters. Although different models can be used, the training process is consistence, where the speech audio is labelled with initial segment clusters, and trained with the ASM. The ASM is then decoded with the same set of audio. The decoding result is used as input labels again. The training process repeats until the training criteria is met.

First work in ASM \cite{lee1988segment} used HMM for iterative modelling, the concept of iterative training is the same as training an acoustic model with the initial segmentation and labels. Therefore any statistical model in Section \ref{Acoustic Model} and DNN architectures in Section \ref{Supervised DNN Acoustic Modelling} can be used for iterative modelling.






\section{Unsupervised word discovery}

Spoken term detection indexes speech based on the content efficiently. It aims at locating spoken terms that appeared in speech, especially if the speech is related to specific topics such as meetings, lectures and conversations. 

Examples of traditional spoken term detection are: spoken term detection using Large Vocabulary Continuous Speech Recognition (LVCSR), acoustic based keyword spotting, and query by example (QBE). They require the model to be trained in supervised manner and the target spoken terms are known \cite{mandal2014recent}.

However, in the problem of zero-resource language, there is zero understanding to the language, not to mention knowing the spoken terms we are looking for.\textbf{Spoken term discovery (STD)}, is a completely unsupervised method that exploits repeating patterns in the speech signal.

There are two main approaches in spoken term discovery, one combines the work of developing ASM follow by STD, another one is an integrated STD system. In an ASM-STD model, there are 2 main approaches as well: 1) query by example using template matching to discover spoken terms and 2) direct clustering of subword sequences into spoken terms.


\subsection{Query by example using template matching}
When both transcription and target keywords of the speech are unavailable, templates are learnt in unsupervised manner for QBE. The system first discovers all the possible spoken terms from the speech and saves them as templates. The templates are then compared with the speech to discover repeating segment sequences, which are the spoken terms in the recordings.

\subsubsection{Segmental DTW for template matching}

Segmental Dynamic Time Wrapping (DTW) is widely used in template matching, it is similar with frame-level DTW, despite the keywords are compared in segment level to provide more efficient computation \cite{chan2010unsupervised}. It scores the similarity of the two sequences.

One example of using segmental-DTW in QBE with template matching is BBN's work \cite{siu2011unsupervised}. They named the subword units discovered using HMM iterative model as self-organized units (SOU) and divided the pattern discovery process into 3 stages: SOU template discovery, template organization and audio segment clustering.

With the subword units generated by HMM model, SOU sequence can be generated by searching for the 1-best path decoded from the HMM. SOU templates can be located by searching for common SOU n-grams from lattices. They can be organized by merging similar templates into same groups. With the templates, audio segment clustering is done by comparing SOU lattices that match the templates.


Besides the lattices of SOU, segmental DTW can be applied on different segment representation sequences, such as segments represented by spectrogram \cite{chan2010unsupervised}, posteriorgrams generated from ASM or Gaussian mixture model \cite{wang2012acoustic}, spectrograms image that capture temporal, frequency and energy information \cite{barnwal2012spectrographic}.

\subsubsection{Other approaches}

A sliding window with similar length is used to compute segment features from the sequence. The features are then trained with positive and negative examples using SVM \cite{barnwal2012spectrographic}. However, each example requires a SVM classifier and performance decreases with increasing number of keywords.

\subsection{Direct clustering of subword sequences}

Instead of discovering the templates for segmental DTW, another approach is to directly cluster all the discovered subword sequences into sequence clusters by grouping similar sequences. Each cluster is expected to correspond to a specific spoken term.
\subsubsection{Local alignment with graph clustering}
Alex and James first proposed unsupervised pattern discovery, which applies local alignment follow by graph clustering on subword sequences of recording to discover acoustic patterns \cite{park2008unsupervised}.

Local alignment is a modification of segmental DTW, introducing shape constrain and different starting points for comparison. Different from segmental DTW which aligns the two complete sequences, local alignment tries to locate matching subsequences within two segment sequences.

After obtaining the subword subsequences, graph clustering is used to cluster the sequences into clusters. 
The segment positions and their similarities are formulated as graph, the nodes represent the segment locations in time and the edges represent the similarities between the nodes, edges with values larger than the threshold are removed to form clusters of segments. The spoken terms are obtained from the finalized clusters obtained by Newman algorithm.

Clustering techniques are not limited to graph clustering mentioned above. Once the subword subsequences are obtained, other clustering techniques such as those introduced in Section \ref{clustering} can be applied.

\subsection{Integrated STD}

Besides combining different subsystems to discover spoken terms, there are also models that directly search for the common spoken terms on word-segment-based instead of subword-unit-based. Arbitrary-length word segments are embedded to fixed length vectors that facilitates the clustering and topic classification process.

\textbf{Bayesian GMM model} \cite{kamper2017segmental}:
A single Bayesian GMM model is used to learn the best segmentation and discover spoken terms through iterative modelling. Syllable boundary detection is used to determine all the likely word boundaries. Then segmental features are extracted using correspondence autoencoder. The features are clustered and trained by the Bayesian GMM model. Re-segmentation, feature extraction and re-clustering of segments are done based on the performance of the currently trained model until the configuration that gives the optimal performance is reached.

\textbf{K Nearest Neighbour (KNN) clustering} \cite{kawakami2018unsupervised}:  Speech is pre-segmented into possible terms and fix-length term-embedding is applied to produce fix-length vectors of the segments. KNN instead of DTW is used to search for common segments. Clustering is then applied to group discovered segments into spoken terms.

\textbf{Embedded segmental k-means model} \cite{kamper2017embedded}:
There is also work on embedded segmental k-means model, which is very similar with k-mean clustering in learning the vector representations that group acoustic similar segments together.

\section{Applications of unsupervised spoken term discovery}

Spoken term discovery has been applied to different types of speech, besides corpora that are formally recorded, it can also be applied to real word recordings such as lectures available online \cite{park2008unsupervised}, child speech \cite{goldwater2009bayesian} and broadcast news \cite{chung2013unsupervised}.

In terms of applications, it can be used for topic comparison \cite{siu2014unsupervised}, language structure understanding \cite{lee2015unsupervised} and topic discovery \cite{schwartz2001unsupervised}.

\section{Summary}

In this chapter, we discussed the fundamental of acoustic recognition system, how a conventional acoustic model is trained in supervised manner and how deep neural network technology can be applied. 

However, in the scenario of zero resource language, without the transcript and pre-defined phonetic units for training. It is impossible to train an acoustic model with supervised methods. The more fundamental challenge to this problem becomes: how can we understand and process an untranscibed speech without any manual interpretation? This can be further divided into 2 main sub-problems.

\begin{enumerate}
\item How can phonetic units be discovered from the speech, such that the process of training an acoustic model based on the discovered units can well represent the untranscribed speech.
\item With the speech labelled based on the phonetic units discovered in (1), how can the content of the speech be interpreted? How can the topics be discovered and categorized? 
\end{enumerate}

These are what acoustic segment modelling is trying to tackle, it focuses on clustering speech segments into potential phonetic units (subword units), and transcribes the speech according to the phonetic information discovered in unsupervised manner. Spoken term discovery then discovers content-related terms from the labelled unit sequences for analyzing.

In the following chapters, we will discuss how subword units can be discovered using various clustering methods. A suitable feature that can well-represent the speech frames for clustering is also investigated. After obtaining descent transcription formed by the discovered units, spoken term discovery method is investigated.




\chapter{Acoustic Segment Modelling}


Acoustic segment model (ASM) is one of the main approaches to unsupervised acoustic modeling in the absence of speech transcription. It involves three sequential steps: initial segmentation, segment clustering and labelling, and iterative model training. Since the approach is totally data-driven without requiring any prior knowledge about input speech, effective feature representation plays a vital role in determining the system performance. In the present study, a multilingual DNN is trained to serve two purposes. On one hand, it is used to perform phone recognition from which an initial segmentation of input utterance can be obtained. On the other hand, bottleneck feature (BNF) representations extracted from the DNN are used for segment clustering. 




\section{Multilingual DNN} \label{multilingual_BNF}

As discussed in Chapter 2, a DNN can learn linguistic knowledge from one language and apply to another language. This approach could be exploited to achieve knowledge transfer from one or more resource-rich languages to a low-resource target language. In this section, two different structures of multilingual DNN are investigated. They are namely multilingual DNN with bottleneck layer (\textbf{Multilingual DNN-BN}) and multilingual DNN with stacked bottleneck layer (\textbf{Multilingual DNN-SBN}). The DNNs are trained by multi-task learning strategy with a number of existing speech corpora.

\subsection{Multilingual DNN with bottleneck layer} \label{multilingual-dnn-bn}

\begin{figure}[h]
\begin{center}
\includegraphics[width=8cm]{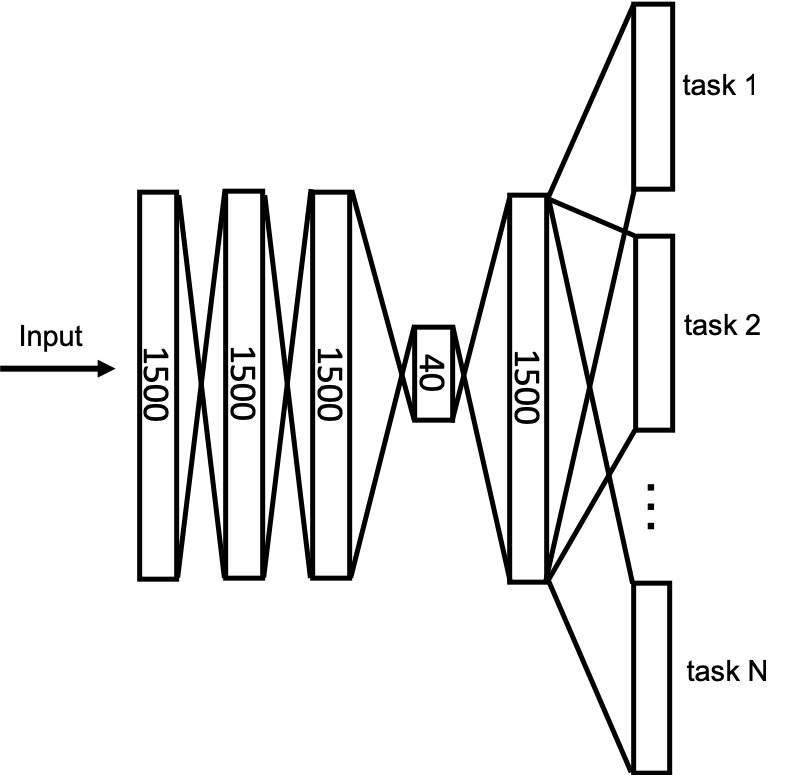}
\end{center}
\caption{Architecture of the Multilingual DNN-BN with $5$ learning tasks}
\label{fig:bnnet}
\end{figure}

As shown in Figure \ref{fig:bnnet}, the Multilingual DNN-BN consists of $4$ hidden layers. One of the hidden layers has a small dimension of $40$, which is a linear transformation layer named the bottleneck layer. The other hidden layers all are of dimension $1500$. The output layer contains $5$ blocks, each corresponding to one learning task. In this study, $5$ existing speech corpora from $4$ languages are applied to formulate the $5$ learning tasks. Details of the speech corpora are given as in Table \ref{table:corpora}. These corpora have similar characteristics, in terms of speaking style, speaker population, and channel condition (e.g., sampling rate of $16$ kHz). The four languages are selected with considerations on providing a good phonetic coverage and diversity, such that the DNN can learn better features for acoustic modeling of a new language.

 \begin{table}[h] 
 \centering
 \begin{tabular}{|c|c|}
 \hline
Corpus & Language\\\hline
TIMIT \cite{timit} & English \\\hline

WSJ \cite{paul1992design} & English \\\hline
 CUSENT \cite{lee2002spoken} & Cantonese \\\hline
 863 \cite{qian2004introduction} & Mandarin \\\hline
distant-speech database  \cite{radeck2015open}& German \\\hline
 \end{tabular} 
 \caption{Information of the corpora used}
 \label{table:corpora}
 \end{table}

For each of the $5$ learning tasks, a context-dependent GMM-HMM acoustic model is trained in a supervised manner with the respective speech corpora. Training of the CD-GMM-HMM follows standard Kaldi recipe, i.e., 
(1) $13$-dimension MFCC features extraction; 
(2) training a monophone model; 
(3) training a triphone model with delta features, followed by delta-delta features; 
(4) triphone HMM model trained with MFCC features transformed with Linear Discriminant Analysis (LDA) and Maximum Likelihood Linear Transform (MLLT);
(5) speaker adapted training (SAT), which the model is trained on feature-space Maximum Likelihood Linear Regression (fMLLR) adopted features.


Supervised training of Multilingual DNN-BN is carried out with the training data from all of the $5$ speech corpora and the state-level time alignment produced by the language-specific CD-GMM-HMM. Input features to the DNN cover a contextual window of $11$ frames and each frame is represented by $23$ Mel-scale filter-bank coefficients. 






\subsection{Multilingual DNN with stacked bottleneck layer} 
\label{multilingual-dnn-sbn}

Previous research shown that bottleneck features extracted from a DNN can be used as a compact and informative speech representation. In order to better capture temporal dependency in speech, bottleneck features extracted from a multilingual DNN can be stacked across a certain number of time frames and applied as input features to another DNN \cite{grezl2009investigation}. 

\begin{figure}[h]
\includegraphics[width=14cm]{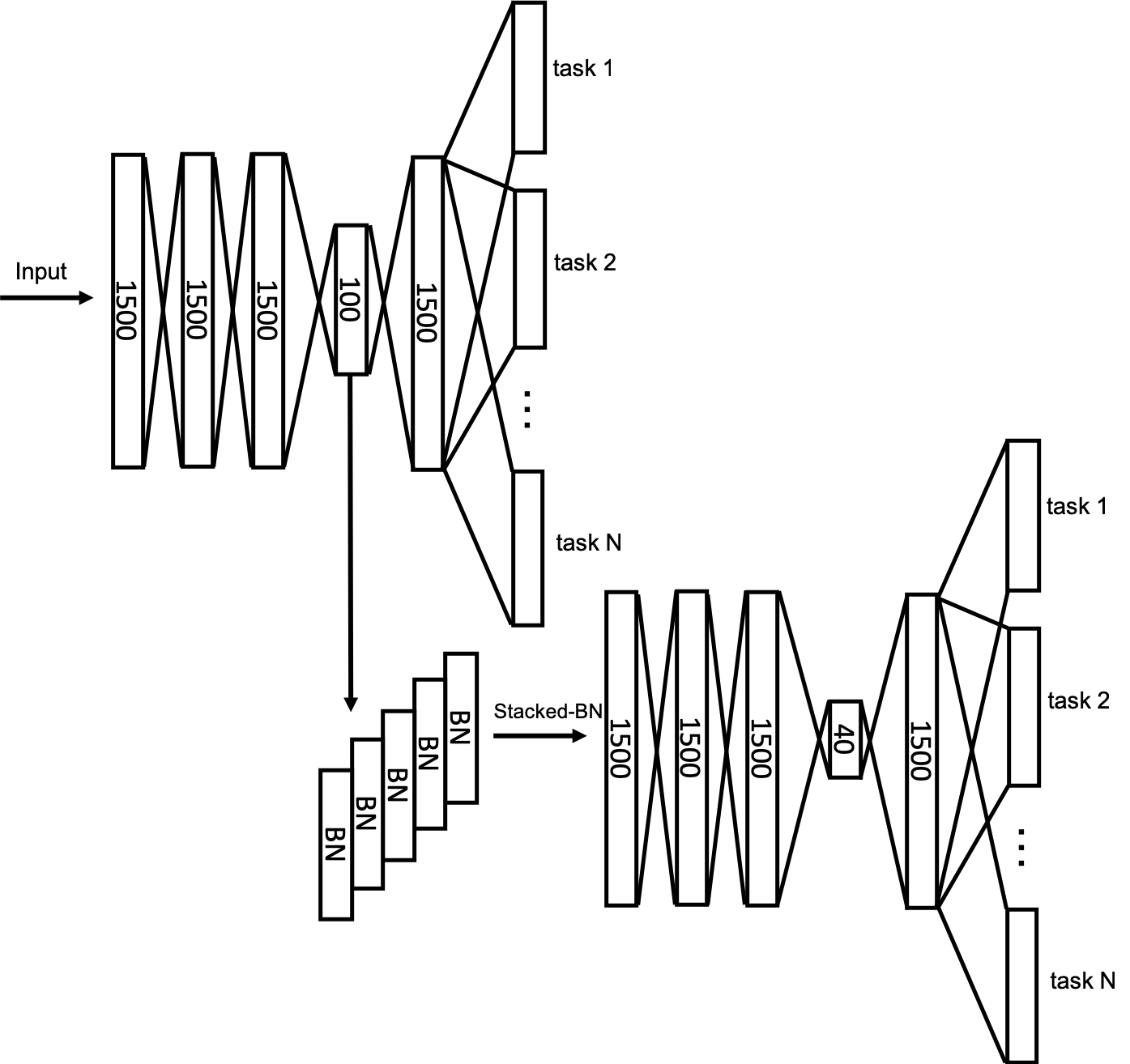}

\caption{Multilingual DNN with stacked bottleneck layer.}
\label{fig:sbnnet}
\end{figure}

This architecture, named Multilingual DNN-SBN, is illustrated as in Figure \ref{fig:sbnnet}. The first DNN adopts similar structure and training strategy to Multilingual DNN-BN as described in the last section, except that the dimension of bottleneck layer is changed to $100$ to include more information for training the second DNN. The input to the second DNN is obtained by stacking the $100$-dimension bottleneck layer output of the first DNN over a contextual window of $\pm 6$ frames. The second DNN uses a narrower bottleneck layer of $40$ dimensions, and the training procedure is similar to the first DNN. It takes in all the $5$ corpora for training. The $40$-dimension bottleneck features from the second DNN are named SBNF (stacked bottleneck features).

\subsection{Phone recognition with the multilingual DNN}


In this section, the two multilingual DNN models are evaluated in the task of phone recognition. They are compared with conventional monolingual acoustic models tested on $2$ of the mentioned speech corpora. All acoustic models are trained in a supervised manner. All $5$ corpora are used for training the multilingual DNNs, while only WSJ and CUSENT are used for training monolingual acoustic models. The test sets for both monolingual and multilingual DNNs are from WSJ and CUSENT.

For the monolingual acoustic models, the following model configurations were implemented and evaluated:
\begin{enumerate}
  \item CD-GMM-HMM
    \begin{itemize}
    \item The setting is the same as the CD-GMM-HMM described in section \ref{multilingual-dnn-bn}.
    \end{itemize}
  \item Subspace Gaussian Mixture Models (SGMMs)-HMM \cite{povey2011subspace}
    \begin{itemize}
    \item SGMMs represent the parameters of each state GMM as a vector mapping the stacked vectors to a subspace.
    \item Traing procedure: (1) fMLLR adopted features are used; (2) a trained GMM-HMM is available; (3) the Gaussians are clustered to initialize Universal Background Model (UBM); (4) SGMM is initialized with states’ pdfs equivalent to UBM and phone alignments obtained from GMM-HMM; (5) the SGMM-HMM is trained using EM algorithm. 
    
    \end{itemize}
  \item DNN/DBN-HMM 
    \begin{itemize}
    \item The DNN is trained using pre-trained DBN layers.
    \item The DNN has 6 hidden layers, each layer with dimension of 1024. The DNN is trained by first pre-training 6 individual DBN layers. Then the DBN layers are stacked to form an initial DNN structure. The DNN is then fine-tuned iteratively.
    \end{itemize}
\end{enumerate}

Each of the monolingual acoustic model is trained and tested with one same language only. For each model, two ASRs are trained and experimented with two corpora separately, one with WSJ and another one with CUSENT. The corpora are split into training sets and test sets, training sets are used to train the ASRs, while test sets are used for evaluation. 

Training procedure of Multilingual DNN-BN and Multilingual DNN-SBN is the same as in Section \ref{multilingual-dnn-bn} and \ref{multilingual-dnn-sbn}. The same $5$ corpora are used for training. 
The same test sets from CUSENT and WSJ used in evaluating monolingual ASRs are used to evaluate the multilingual ASRs.





The performances of ASRs are measured in terms of phone accuracy in decoding the test set.
The phone error rates attained with the monolingual models and the two multilingual DNNs are compared as in Table \ref{ASM_comparision}. It is noted that multilingual training really does help improve the performance of acoustic models. The monolingual tasks of both WSJ and CUSENT are considered to represent rich-resource training, and the achieved recognition performances are at a fairly high level. Nevertheless, by leveraging speech data from other languages, the multilingual DNNs were able to further reduce the phone error rate for WSJ. For CUSENT, multilingual DNNs do not perform as good as monolingual DNN, though they are better than all the GMM-HMM based models. This may probably due to the loss of certain language specific information, which are learnt in monolingual ASRs trained with resource-rich data. It is also seen that the Multilingual DNN-SBN performs better than multilingual DNN-BN, at the cost of significantly longer training time.

  \begin{table}[h] 
 \centering
 
 \begin{tabular}{|c|c|c|} 
 \hline
 \multicolumn{1}{|c|}{ } & \multicolumn{2}{|c|}{Phone error rate}\\\hline
Model &  WSJ & CUSENT \\\hline
CD-GMM-HMM &  6.98\% & 9.39\% \\
SGMMs-HMM & 6.95\% &8.78\%  \\
DNN/DBN-HMM & 6.26\% &7.02\% \\\hline 
Multilingual DNN-BN & 5.37\% & 8.32\% \\
Multilingual DNN-SBN &  5.33\% & 7.96\% \\\hline 
  
  \end{tabular}
  \caption{Performances on WSJ and CUSENT corpora using different models}
  \label{ASM_comparision}
 \end{table}

\subsection{Evaluation of multilingual bottleneck features by visualization} 
\label{sec:BNF_visual}

Given a trained multilingual DNN, frame-level bottleneck features could be computed from input speech of any language. In this section, we examine the effectiveness of multilingual bottleneck feature in unsupervised segment modelling of a new language and make comparison with 
conventional MFCC features.


We trained several DNNs and tested them with a different language to evaluate if the BNF extracted from DNNs are effective in unsupervised phone catagorization. Different types of BNFs are extracted from multilingual and monolingual DNNs. The DNNs are first trained in supervised manner, BNFs are then obtained by decoding a corpus with language different from training corpora. The DNNs experimented and their training procedures are described as follow:
\begin{enumerate}
    \item A single language DNN-BN
    \begin{itemize}
        \item With the same architecture as in section \ref{multilingual-dnn-bn}, expect it is trained on one corpus only: CUSENT (Cantonese).
    \end{itemize}
    \item Multilingual DNN-BN
    \begin{itemize}
        \item With the same architecture and training procedure as in Section \ref{multilingual-dnn-bn}. It is trained with all $5$ corpora.
    \end{itemize}
    \item Multilingual DNN-SBN
    \begin{itemize}
        \item With the same architecture and training procedure as in Section \ref{multilingual-dnn-sbn}. It is trained with all $5$ corpora.
    \end{itemize}
\end{enumerate}

After training the DNNs, the DNNs take in MFCC features of the new language. Here, telephone conversation corpus named Callhome Spanish \cite{callhome} is used. The corpus has complete new language that the DNNs haven't seen before, it also contains speaking tone, recording channel and sound events (e.g. laughter) much different from the training corpora. It is expected to evaluate the models' ability in processing new data recorded in different scenario. The layers beyond the bottleneck layer are removed, and the bottleneck layer output is extracted as BNFs. $40$ dimension BNFs extracted are compared with $13$ dimension MFCC features of the same corpus. The features are evaluated in frame-level.

\begin{figure}[H] 
\centering     
\subfigure[MFCC]{\label{fig:2a}\includegraphics[width=60mm]{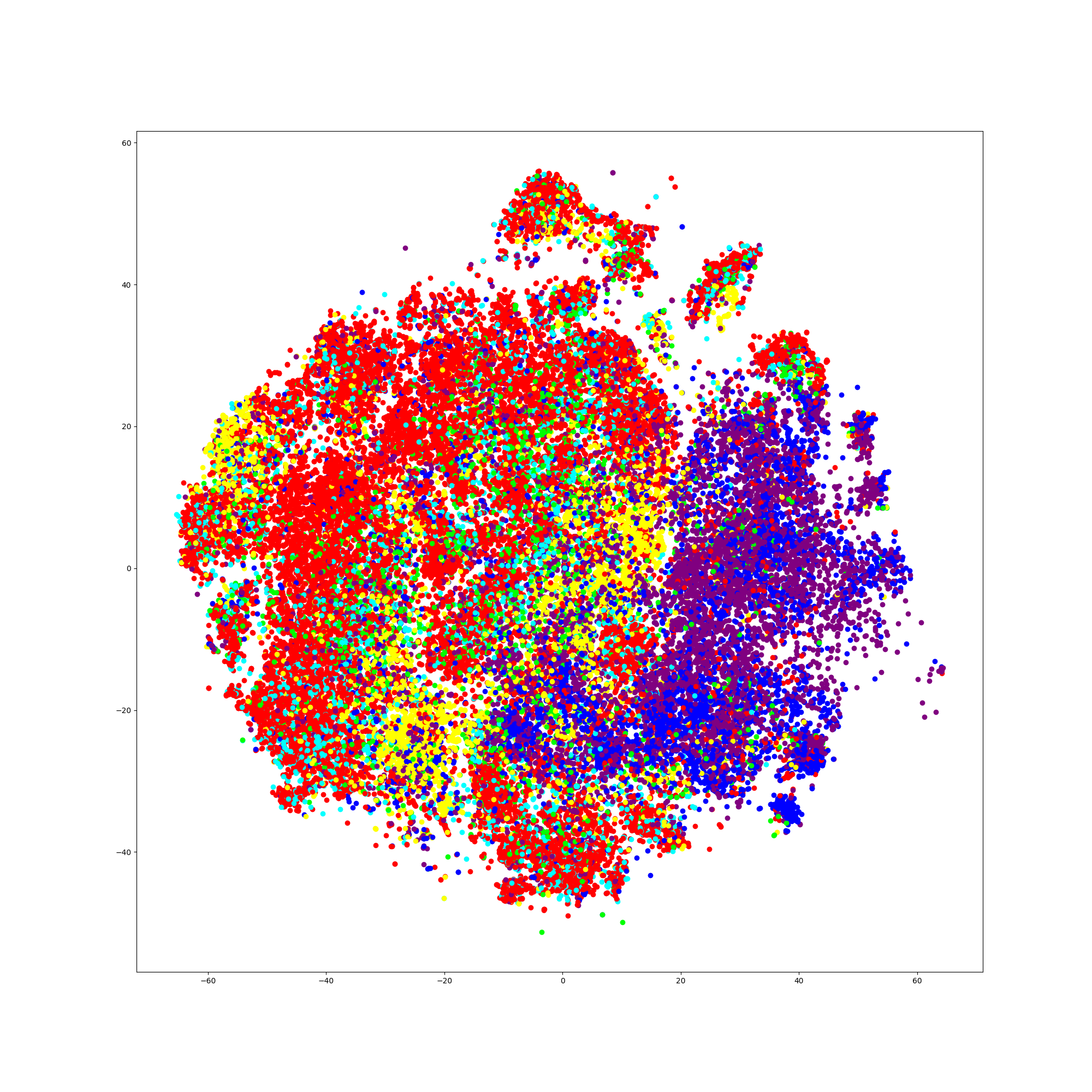}}
\subfigure[BNF from Single language DNN]{\label{fig:2b}\includegraphics[width=60mm]{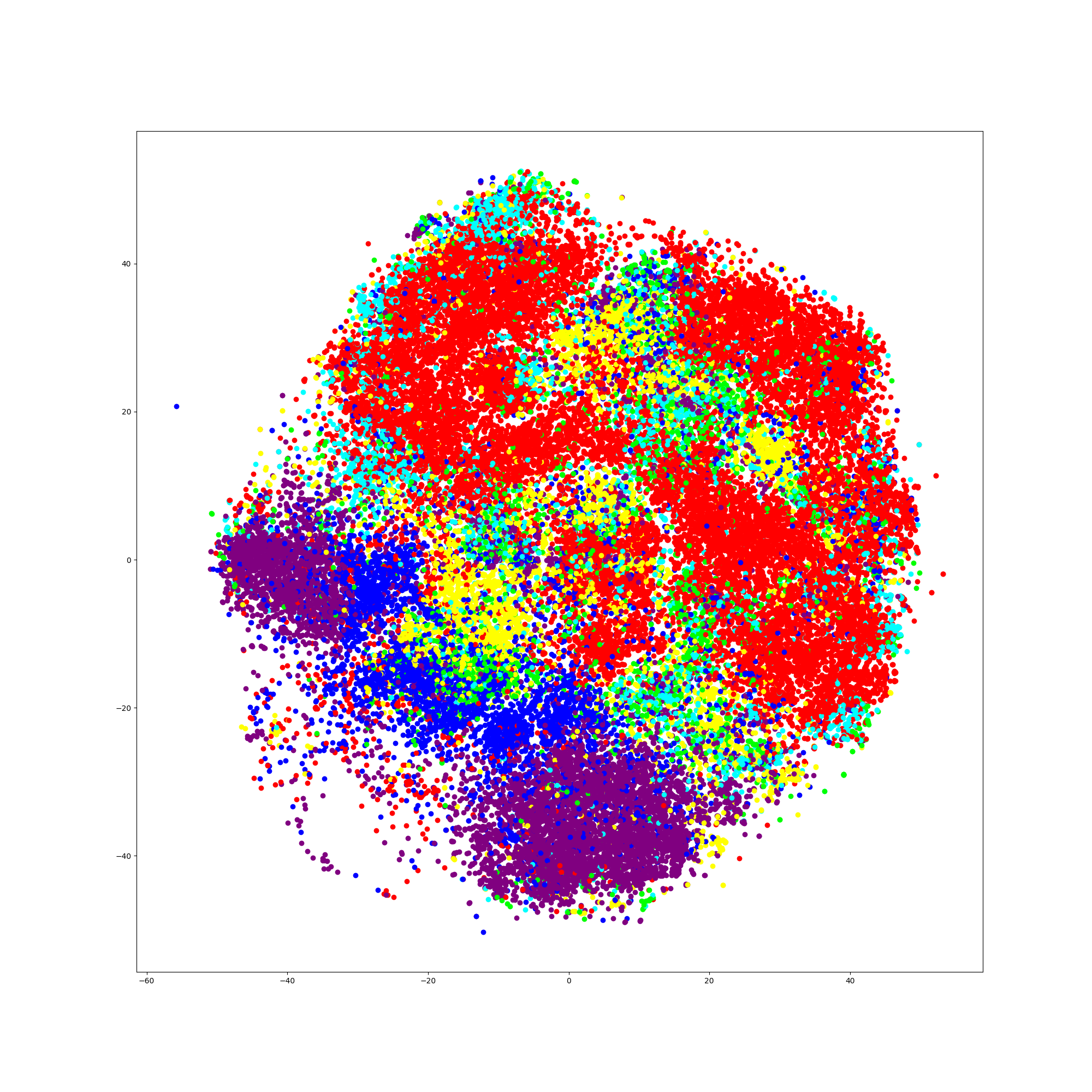}}
\subfigure[BNF from Multilingual DNN]{\label{fig:2c}\includegraphics[width=60mm]{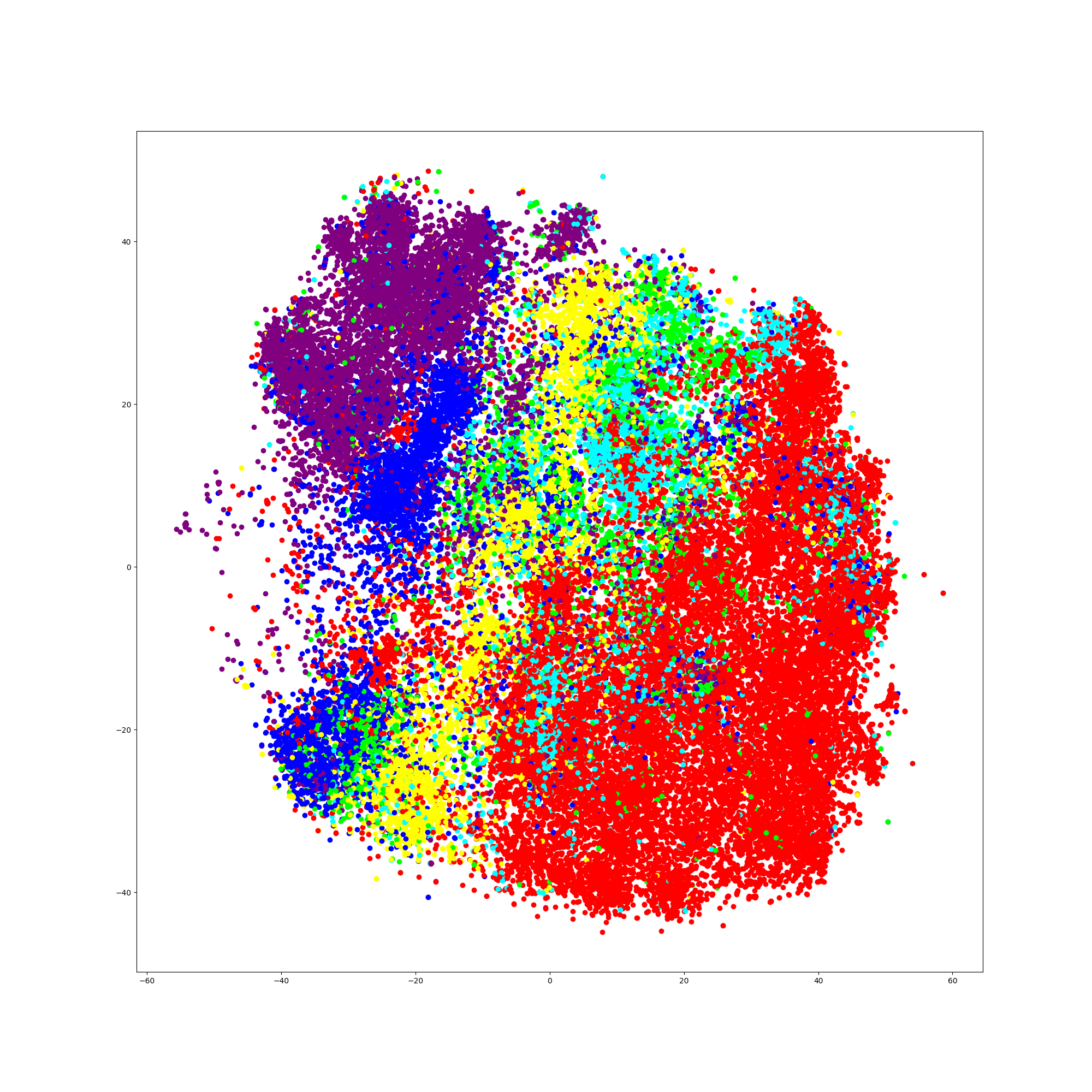}}
\subfigure[Stacked BNF from Multilingual DNN]{\label{fig:2d}\includegraphics[width=60mm]{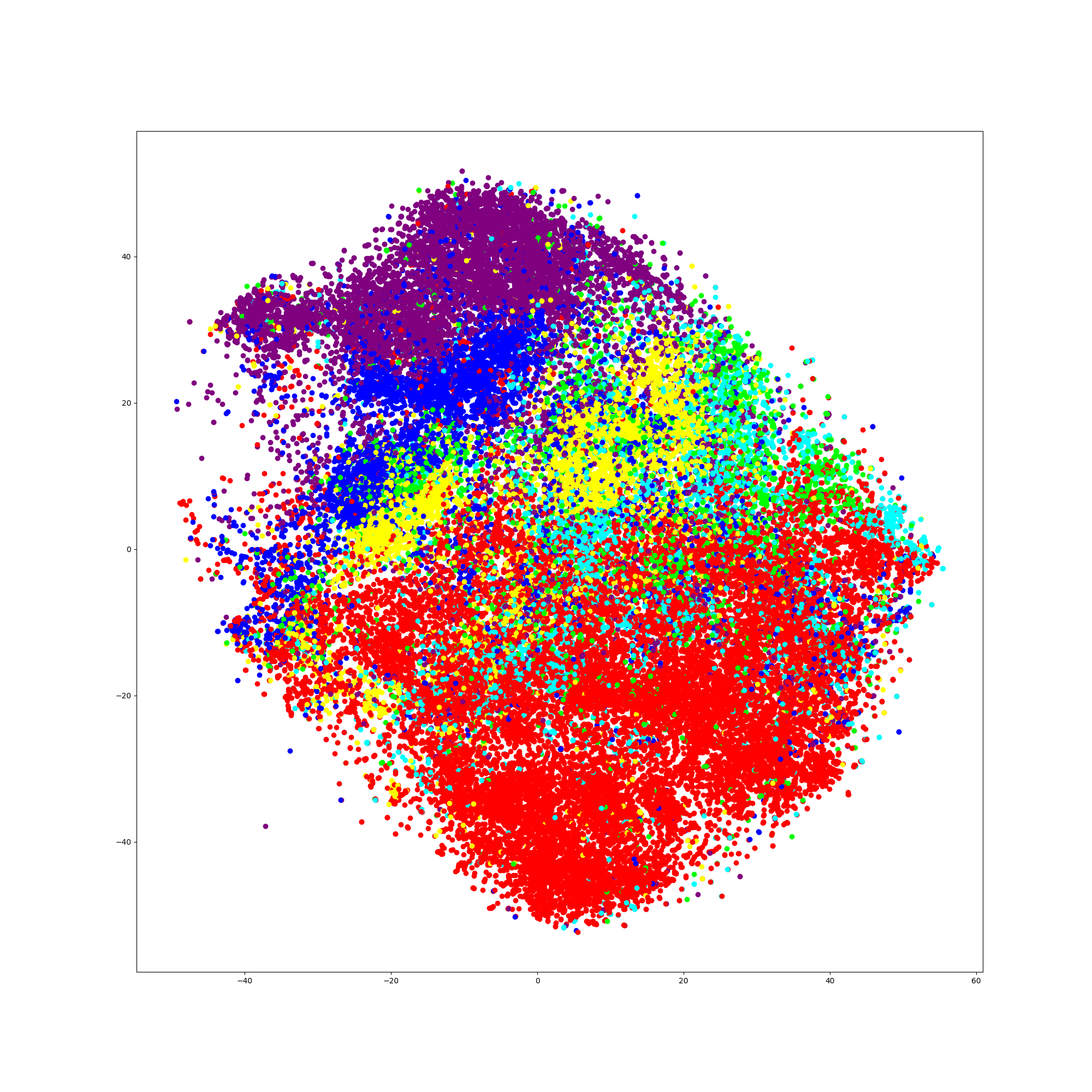}}
\caption{t-sne projection of MFCC and bottleneck features from different models extracted from CALLHOME Spanish. The phoneme groups are represented as: red: vowel; cyan: approximant; blue: fricative; green: vibrant;  yellow: nasal; purple: plosive.
}
\label{fig:tsne-features}
\end{figure}

The phones of the test set are categorized into $6$ phonetic groups, represented with different colours shown in Table \ref{table:phonegroup}.
We visualized the frame-level features in 2 dimensional space using PCA \cite{shlens2014tutorial} follow by t-SNE \cite{maaten2008visualizing} in Figure \ref{fig:tsne-features}. t-SNE maps high-dimension data to lower dimension in non-linear way. t-SNE may have its limitations. in which the mapping may be data sensitive and the data sizes and distances cannot be well illustrated. However, through visualization, it gives us some insight on whether it is comparatively easier to cluster similar data points together unsupervisedly, such that we can understand which features can facilitate segment clustering more. 

 \begin{table}[h] 
 \centering
 \begin{tabular}{|c|c|c|c|c|c|c|}
 \hline
\textbf{Phone group} & vowel & approximant & fricative & vibrant & nasal & plosive \\\hline
\textbf{Colour} & red & cyan & blue & green & yellow & purple \\\hline
 \end{tabular} 
 \caption{Phone groups and their corresponding colour}
 \label{table:phonegroup}
 \end{table}

The features are analyzed based on phone separability. By separability we mean how well data points from same phone groups are projected to same regions, and how well data points from different phone groups are separated. It is expected that if it is easier to observe features group by group in t-SNE and there exist clear boundaries to separate frames from different phone groups, the mapping has higher separability and it is more likely to obtain subwords that can well represent the actual phones in segment clustering.





The visualization is presented in Figure \ref{fig:tsne-features}.
Even though BNF obtained from monolingual DNN has better phone separability than MFCCs, frames from same phone groups are still split into different regions. It may due to the relatively large extend of phonetic difference between Cantonese and Spanish.
BNFs obtained from multilingual DNN achieve better separability than monolingual DNN on vowels, fricatives and plosives. BNFs obtained from Multilingual DNN-BN have slightly more balanced distribution of phone classes, while BNFs obtained from Multilingual DNN-SBN have slightly better separability. 

Overall, compare with conventional feature, BNFs from Multilingual DNN have better generalization ability on phones and provide better representations that group frames of same phone classes together, making it suitable for unsupervised segment clustering.

\subsection{Evaluation of multilingual BNFs on minimal-pair ABX task} 

The multilingual BNFs are evaluated on the task of unsupervised subword modelling in the ZeroSpeech 2017 Challenge Track 1 \cite{dunbar2017zero}. The challenge covers three languages for development and evaluation, i.e. English, French and Mandarin. 
The minimal-pair ABX discriminability metric is used to measure the quality of learned feature representations. It involves three stimuli A, B and X. A and B are a pair of triphone segments with minimal segmental difference, e.g., ``beg'' vs.``bag'', , ``api''vs.``ati'', and X is a speech segment that contains either A or B. In the within-speaker condition, A, B and X are spoken by same speaker. In the across-speaker condition, A and B are spoken by the same speaker, while X is from a different speaker.

BNFs are extracted with the multilingual DNN-BN and the multilingual DNN-SBN as described in Section \ref{sec:BNF_visual}, with the same training data and procedures. The performance of multilingual BNFs is compared with the best submitted systems in the ZeroSpeech 2017 Challenge, and in the relevant works of Heck et al. \cite{heck2017feature} and Chen et al. \cite{chen2017multilingual}. In \cite{heck2017feature}, the Dirichlet process Gaussian mixture model (DPGMM) was applied to cluster speech feature vectors into subword classes. The input features were processed by speaker adaptation in a multi-stage clustering framework.
In \cite{chen2017multilingual}, DPGMMs were used to cluster the unlabelled speech into subword units. The input MFCC features were processed by vocal tract length normalization (VTLN). A multilingual DNN with bottleneck layer was trained on the subword units to extract feature representations.
Our method is different from the above studies in that out-of-domain transcribed data is employed to facilitate supervised DNN training.


The similarity between a pair of speech segments is computed as the average frame-level cosine distance averaged along the optimal frame alignment obtained by dynamic time warping. The ABX discriminability is given by the error rate, which is the mean 
over all possible minimal difference triphone pairs in the test set.  
The within-speaker and across-speaker minimal-pair ABX discriminability of different systems are compared in Table \ref{table:abx_across} and \ref{table:abx_within} respectively. The baseline system is based on standard MFCC+$\Delta$+$\Delta\Delta$ features. In most cases, BNFs from multilingual DNNs could achieve performance comparable to or better than the best system in the Challenge. With similar system structures, the proposed features give better performance than those reported in \cite{chen2017multilingual}. This indicates that, despite he mismatch between training data and evaluation data, increasing the variety of out-of-domain training data is beneficial to improving the quality of BNFs. Even though French is a language not involved in the training of multilingual DNNs, the learned features are capable of representing this language well.

It is also noted that BNFs extracted from multilingual DNN-SBN achieve lower ABX discriminability across speakers, while BNFs extracted from multilingual DNN-BN achieve lower ABX discriminability within speakers. The reasons for such incoherent performance need further investigation.

\begin{table}[]
\resizebox{14.5cm}{!}{%
\begin{tabular}{|l|c|c|c|c|c|c|c|c|c|c|}
\hline
                     & \multicolumn{3}{c|}{English}      & \multicolumn{3}{c|}{French}          & \multicolumn{3}{c|}{Mandarin}     & \multicolumn{1}{l|}{\multirow{2}{*}{Average}} \\ \cline{1-10}
Duration             & 1s           & 10s          & 120s         & 1s            & 10s           & 120s          & 1s           & 10s          & 120s         & \multicolumn{1}{l|}{}                                  \\ \hline
Heck et al.          & 10.1         & 8.7          & 8.5          & 13.6          & \textbf{11.7} & \textbf{11.3} & 8.8          & 7.4          & 7.3          & 9.71                                                   \\ \hline
Chen et al.          & 12.7         & 11           & 10.8         & 17            & 14.5          & 14.1          & 11.9         & 10.3         & 10.1         & 12.49                                                  \\ \hline
Multilingual DNN-BN  & 9.3          & 8.3          & 8.8          & 13.4          & 12.2          & 12.0          & 8.8          & 7.6          & 7.5          & 9.76                                                   \\ \hline
Multilingual DNN-SBN & \textbf{8.2} & \textbf{7.3} & \textbf{7.2} & \textbf{12.7} & 11.7          & 11.5          & \textbf{8.5} & \textbf{7.3} & \textbf{7.2} & \textbf{9.06}                                          \\ \hline
Baseline       & 23.4         & 23.4         & 23.4         & 25.2          & 25.5          & 25.2          & 21.3         & 21.3         & 21.3         & 23.33                                                  \\ \hline
\end{tabular}
}
\caption{Results of minimal-pair ABX discriminability across speakers}
\label{table:abx_across}
\end{table}

\begin{table}[]
\resizebox{14.5cm}{!}{%
\begin{tabular}{|l|c|c|c|c|c|c|c|c|c|c|}
\hline
                     & \multicolumn{3}{c|}{English}               & \multicolumn{3}{c|}{French}                & \multicolumn{3}{c|}{Mandarin}              & \multirow{2}{*}{Average} \\ \cline{1-10}
Duration             & 1s           & 10s          & 120s         & 1s           & 10s          & 120s         & 1s           & 10s          & 120s         &                          \\ \hline
Heck et al.          & 6.9          & 6.2          & 6            & 9.7          & 8.7          & 8.4          & 8.8          & 7.9          & 7.8          & 7.82                     \\ \hline
Chen et al.          & 8.5          & 7.3          & 7.2          & 11.2         & 9.4          & 9.4          & 10.5         & 8.7          & 8.5          & 8.97                     \\ \hline
Multilingual DNN-BN  & 6.6          & 5.8          & 8.6          & \textbf{8.9} & \textbf{8.2} & \textbf{8.0} & \textbf{8.6} & \textbf{7.3} & \textbf{7.2} & \textbf{7.68}            \\ \hline
Multilingual DNN-SBN & \textbf{6.2} & \textbf{5.5} & \textbf{5.5} & 9.7          & 8.4          & 8.3          & 9.6          & 8.1          & 7.9          & 7.69                     \\ \hline
Baseline         & 12           & 12.1         & 12.1         & 12.5         & 12.6         & 12.6         & 11.5         & 11.5         & 11.5         & 12.04                    \\ \hline
\end{tabular}
}
\caption{Results of minimal-pair ABX discriminability within speakers}
\label{table:abx_within}
\end{table}



\section{Segment clustering} \label{sec:Clustering}

The multilingual DNNs described in the previous section can be used to decode and segment speech utterances from a new language. As shown in Figures \ref{fig:bnnet} and \ref{fig:sbnnet}, the DNNs use $5$ blocks of softmax output layers that correspond to different training tasks. Given an input utterance, $5$ different sets of phone-level time alignment can be obtained. By contrasting and combining these multilingual phone boundaries, an initial segmentation of the utterance can be derived by simply merging boundaries that are within an interval of $20$ ms. Subsequently segment-level feature representations are obtained by averaging frame-level bottleneck features within the same segment according to the boundary information. Segment-level features can be computed in different ways besides averaging.

The next problem is to automatically ``discover'' a set of segmental units, similar to ``phonemes'' or ``subword units'', by applying clustering algorithms to the initial segments. The major challenge in clustering tasks is that the process is highly data sensitive. Different clustering algorithms are suitable only for specific data structures and problems  \cite{berkhin2006survey}. 


Considering a $40$-dimension BNF obtained from the multilingual DNN, an important process is to understand the structure such that a good clustering method for the problem can be identified. Since speech applications typically involve a large amount of data, computational cost and efficiency need to be considered carefully. 
Traditional clustering algorithms can be divided into different categories, based on approaches of partitioning, hierarchical structure, fuzzy theory, distribution, etc. Neural network based clustering algorithms are also proposed in recent years \cite{xu2015comprehensive}. In this section, a few commonly used clustering algorithms are considered in segment clustering of BNFs. These algorithms include:
\begin{itemize}
    \item $K$-means clustering;
    \item Hierarchical clustering;
    \item Gaussian mixture model;
    \item Density-based spatial clustering of applications with noise (DBSCAN).
\end{itemize}

\subsection{$K$-means clustering}

$K$-means clustering \cite{macqueen1967some} aims to partition observation data into $k$ clusters. As a result, each data sample is assigned to its closest cluster according to a prescribed distance measure. With $n$ observations $\mathbf{x}_1, ... , 
\mathbf{x}_n$, k-means clustering partition the samples into $k$ clusters $S = {S_1, S_2, ... , S_k}$ such that the sum of errors between the samples in the same clusters with cluster means $\mu = \mu_1,\mu_2,..., \mu_k$ are minimized: 

\begin{equation}
{arg\,min}_{s} \sum^k_{i=1} \sum_{\mathbf{x} \in S_i} ||\mathbf{x} - \mathbf{\mu_i}||^2
\end{equation}





$K$-means clustering is suitable for well-separated data classes. However there may not have a clear-cut boundary in BNF, especially when there are phones with very similar phonetic properties. Also the suitable number of clusters $k$ is not known. 
It is difficult for $k$-means clustering to find a good cutting boundary and is hard to analysis the quality of clusters obtained without knowing the uncertainty of a sample in a cluster.

\subsection{Hierarchical clustering}
Agglomerative hierarchical clustering (AHC) \cite{johnson1967hierarchical} clusters the samples with bottom-up approach. It starts by treating each data point as an initial cluster, and subsequently merging pairs of most similar clusters. The clustering result  can be presented in a dendrogram, which allows visualization of inter-cluster similarity at different stages. In AHC, there are two main factors to be considered: a distance metric to determine the similarity between clusters, and a linkage criteria to determine if a pair of clusters should be combined. Bottleneck features are high-dimension continuous-valued data. Euclidean distance is a commonly used metric. Some commonly used linkage criteria are presented in Table \ref{table:linkage}.

 \begin{table}[h] 
 \centering
 \label{table:AHC}
 \begin{tabular}{|c|c|}
 \hline
 
 Linkage & Similarity considered $d(a,b)$\\\hline
Complete-linkage & $max\{d(x_{ai},x_{bj})\}: i \in (i,...,n_a), j \in (j,...,n_b) $ \\\hline
Single-linkage &  $min\{d(x_{ai},x_{bj})\}: i \in (i,...,n_a), j \in (j,...,n_b) $   \\\hline
Centroid-linkage &  $ || \bar{x}_a - \bar{x}_b ||_2$, 
: $\bar{x}_a =  {1 \over N_a} \sum^{N_a}_{i=1}x_{ai}$ \\\hline
Median-linkage & $ || \tilde{x}_a - \tilde{x}_b ||_2$  
: $\tilde{x}_a = {1 \over 2} (\tilde{x}_p + \tilde{x}_q)$
$ {x}_a =  {x}_p \cup {x}_q$\\\hline
Ward's method  & $\sqrt{2n_an_b \over n_a+n_b}  ||\bar{x}_a - \bar{x}_b||_2$ 
: $\bar{x}_a =  {1 \over N_a} \sum^{N_a}_{i=1}x_{ai}$  \\\hline
    
 \end{tabular}
 \caption{Some of the linkage criteria used in AHC}
 \label{table:linkage}
 \end{table}
 
\begin{figure}[h] 
\centering     
\subfigure[Centroid]{\label{fig:centroid}\includegraphics[width=60mm]{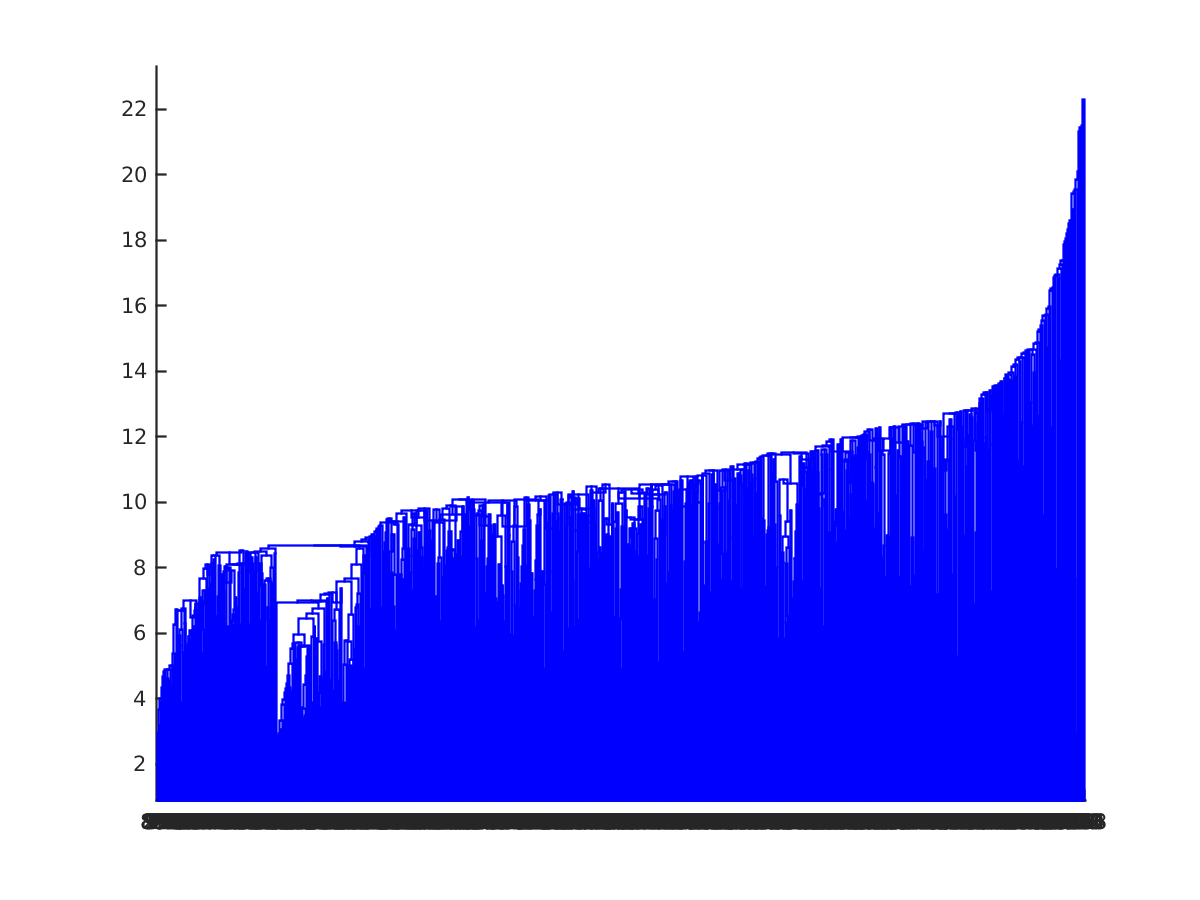}}
\subfigure[Median]{\label{fig:median}\includegraphics[width=60mm]{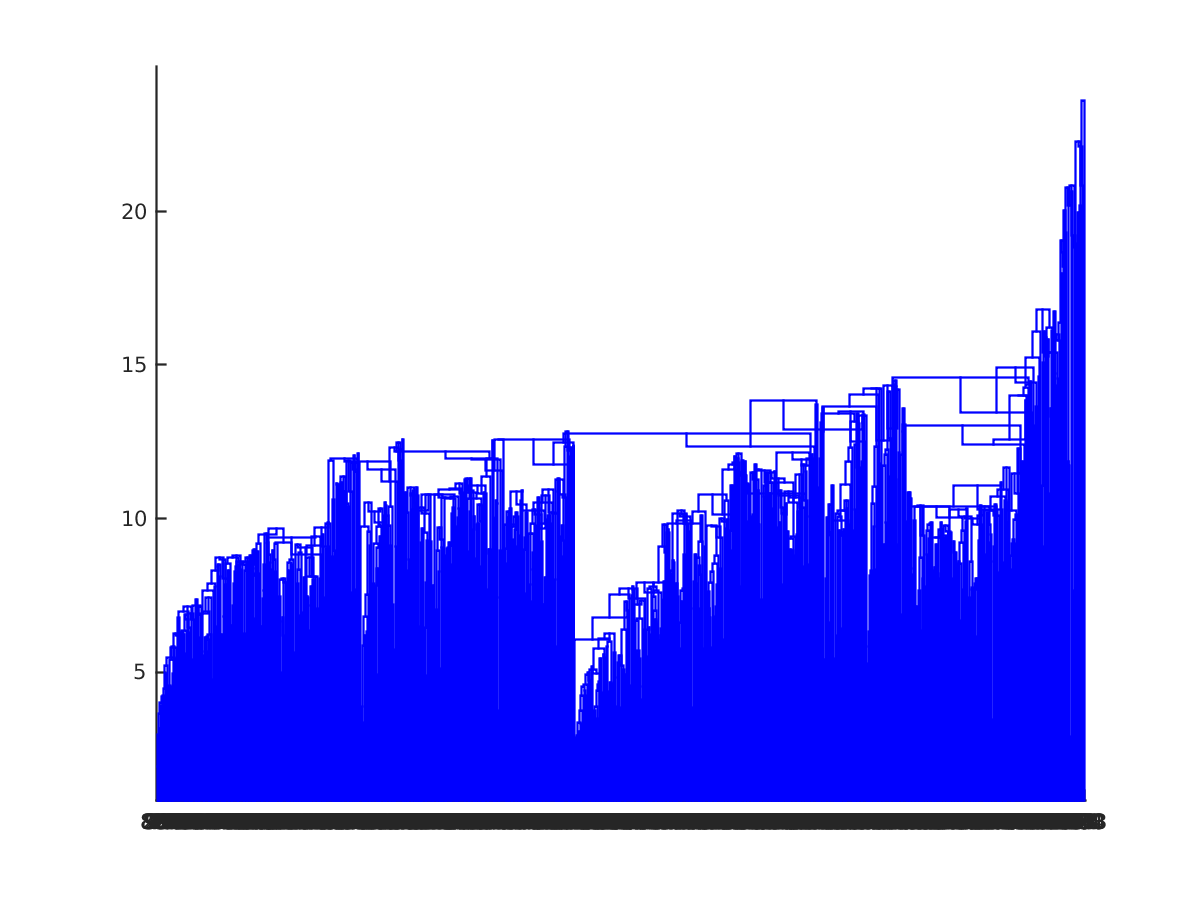}}
\subfigure[Ward]{\label{fig:ward}\includegraphics[width=60mm]{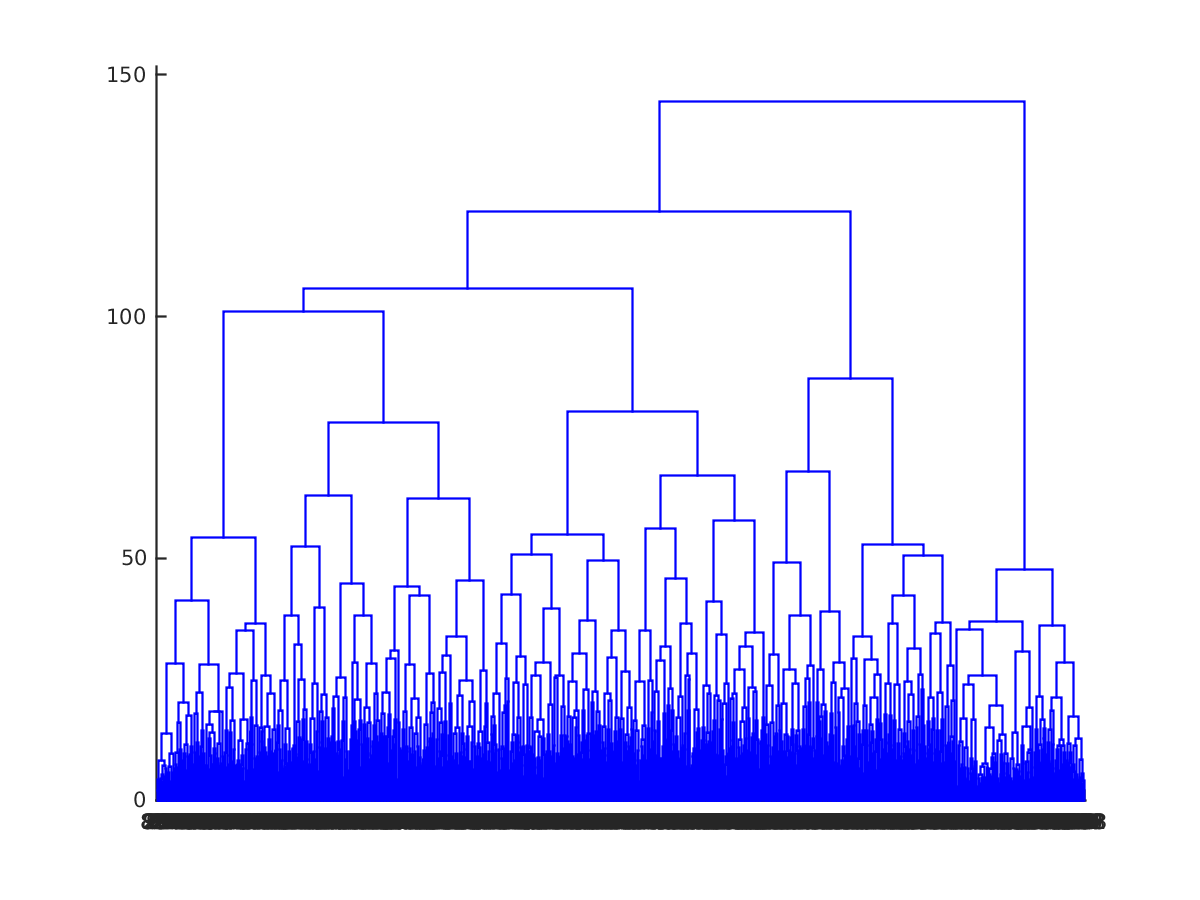}}
\caption{Dendrograms on the clustering result of 1000 segmental BNFs using different linage criteria. They illustrate how the clusters are merged hierarchically with y-axis represents the distances and x-axis represents the BNF clusters.}
\label{dendrogram}
\end{figure}

To understand the properties of segment-level BNF and determine a suitable linkage criteria, let us inspect and compare the dendrograms produced by three different linkage criteria, namely centroid-linkage, median-linkage and Ward's method, as in Figure \ref{dendrogram}. The input data comprises $1000$ frame-level BNFs. Centroid and median linkages are very similar as they merge clusters base on the smallest distance between the clusters. Ward's method attempts to maximize inter-cluster distance and minimize intra-cluster distance.

The dendrograms show that clusters formed by centroid and median linkage tend to be highly unbalanced. Some of the clusters contain a very large number of data points while some have very few. This is undesirable as we expect that phonetic units in a language should be relatively balanced. Ward's method is found to produce better balanced clusters. Therefore, Ward's method will be used for segment clustering.


Although AHC provides very detailed information on the clustering process and distribution in readable form, it requires very large computation time of $O(n^3)$ and memory size of $O(n^2)$ for $n$ samples. This makes AHC impossible to apply on large dataset. Therefore, a hybrid use of two clustering methods is suggested: AHC can be first applied to a subset of manageable data to determine a decent number of clusters, such that other clustering methods with higher efficiency but less determining information can be used, such as K-means clustering.



\subsection{Gaussian mixture model}

Gaussian mixture model is a soft clustering technique, where data samples are not forced to have a unique cluster identity. Each data sample is associated with a set of probabilities that indicates its likeliness of belonging to different clusters. GMM is a probabilistic model that assumes the data points are generated from a mixture of Gaussian distributions. 


In GMM based clustering, the number of components of GMM $K$ is fixed for clustering. It uses the EM algorithm to iteratively search for the parameters of the GMM 
that maximize the likelihood of data. The trained model then assigns each sample to the Gaussian it most probably belongs to. 




\subsubsection{Bayesian GMM}

The limitation of GMM is that the number of mixture components is unknown. Instead of assigning data samples into pre-specified numbers of clusters and searching for the optimal number through multiple trials, the number of components can be found by maximizing the likelihood of parameters
of the GMM model using EM algorithm.
However, the computation required for maximizing the likelihood is large, especially when number of data samples $N$ is large. In nonparametric Bayesian approach \cite{marin2005bayesian,gershman2012tutorial}, dirichlet distribution is used to approximate the posterior distribution of GMM parameters. 

%
%



With the upper bound of maximum number of clusters $K$ being defined, 
the parameters are initialized using k-means, then variational inference (which is an extension of EM algorithm) is used to iteratively update the parameters until convergence. The effective number of clusters eventually can be less then K when the weights of unrepresentative clusters become $0$ during inference.



\subsection{Density-based spatial clustering}
In density-based spatial clustering, data points are grouped based on density. Points that lie in high density regions are regarded as clusters, scattered points are regarded as noise.

Density-based spatial clustering of applications with noise (DBSCAN) is a representative non-parametric method of clustering \cite{ester1996density}. Data samples that are closely packed are grouped as clusters. The number of clusters and the size of each cluster are determined by two parameters: radius $r$ and minimum number of points $minPt$.

If a data point has $minPt$ or more neighboring points within radius of $r$, it is called ``core point''. The points that are reachable from a core point within radius of $r$ are included in the same cluster. The searching process continues at these neighbour points, treating them as the centers and searches for their neighbouring points within radius $r$. Searching is iteratively done until all the reachable points are covered.  Reachable points are assigned to clusters that contain their core points. Points that are not reachable are defined as outliers. 
This method is efficient in terms of time and memory complexity.

One limitation of DBSCAN is that the suitable value of $r$ is not known if we are not familiar with the data. The problem is especially harder for high-dimension data. Also, DBSCAN is not able to handle data with uneven densities. An extension of DBSCAN was proposed in \cite{campello2013density}, which is known as the hierarchical DBSCAN (HDBSCAN).

The HDBSCAN operates in a hierarchical manner. In order to deal with data with different regional densities, the data is represented by a graph, in which data points are connected by edges. The weight of an edge is the reachable distance, i.e. the minimum radius such that the point can be assigned to a cluster.  
The radius $r$ increases at each level, from bottom to top, and edges with weight values smaller than $r$ are removed. The clustering result can be represented by a dendorgram. 

The problem of our concern is about clustering frame-level features into segments. Therefore, instead of keeping the outliers outside the clusters, our goal is to assign every speech frame a corresponding subword label. After the HDBCAN is applied, the distances between outliers and core points are computed, and all outliers are assigned to their closest clusters.




\section{Iterative modelling}

With all speech frames of an audio assigned to clusters, we obtain an initial hypothesis of segment sequence. If each segment cluster is treated as a subword unit of speech, the segment sequence could be considered as a pseudo transcription, which could be used for supervised training of acoustic model. An acoustic segment model can then be trained using the pseudo transcription. This leads to a process of iterative modeling, as described below.

\subsection{Training procedure} \label{sec:iter_train}

DNN-HMM acoustic models \cite{dahl2012context} are trained to represent the learned subword units, i.e., segment clusters. The model training is done in an iterative manner with continuous updating of the pseudo transcriptions. The step-by-step procedures are elaborated below:
\begin{enumerate}
\item Train an initial set of DNN-HMM acoustic model with the pseudo transcriptions obtained from segment clustering;
\item Decode the training data with the current acoustic model and obtain updated pseudo transcriptions\label{enum:step2};
\item Train the acoustic model with the updated transcriptions\label{enum:step3};
\item Repeat Step \ref{enum:step2}) and \ref{enum:step3}) until convergence.
\end{enumerate}
The iterative training is carried out with speech data from the target language. In this way, acoustic model and pseudo transcriptions are jointly optimized for the target language. After terminating the training, the final version of pseudo transcriptions, in the form of subword unit sequences, are used for keyword discovery.

\subsection{Models}

Iterative modeling can be applied to all conventional acoustic models, e.g., HMM, GMM, DNN. In this study, the most basic DNN architecture is adopted.
A 6-layer DNN with 1024 nodes per layer is trained. The input features are frame-level BNFs. The subword units being modeled correspond to the segment clusters obtained as in Section \ref{sec:Clustering}.








\section{Experiments}

The proposed method of unsupervised acoustic modeling is evaluated on a dataset that was acquired online. We use this dataset as a representative task of unsupervised speech modeling of low-resource language, despite that the language being spoken in the dataset is actually not low-resource. Our goal is to examine the efficacy of the proposed approach in the context of a real-world application.

\subsection{Dataset}

The experimental dataset is built upon unedited video recordings in the MIT OpenCourseWare \cite{MITopen}. English was used as the primary medium of instruction in these course lectures. The lectures are from $5$ MIT courses, which are named ``Mathematics for Computer Science'' (MATH), ``Principles of Digital Communication II'' (COMM), ``Introduction to Computer Science and Programming in Python'' (PYTH), ``Geometric Folding Algorithms: Linkages, Origami, Polyhedra'' (ALGO) and ``Discrete Stochastic Processes'' (STOC). Presumably the audio part of a lecture should contain primarily the voice of the course instructor (professor). Due to diverse recording environments and hardware conditions, the recorded lecture may also contain students' voice (e.g., asking or responding to questions), and situational sounds from coughing, laughter, chalk-writing, furniture, etc.

\newcolumntype{P}[1]{>{\raggedright\arraybackslash}p{#1}}
 \begin{table}[h]
 \small
 \centering
 \hspace*{-0.3cm}
 \begin{tabular}{|P{4.8cm}|P{1.5cm}|P{1.5cm}|P{2.2cm}|P{2.5cm}|}
 \hline

Course Name & 
No. of Lectures & Lecture duration & Recording environment & Speaker information \\\hline

Mathematics for Computer Science (MATH) &
25	& 60	&	Lecture hall, clip mic &
	Male with French accent \\\hline
	
Principles of Digital Communications I (COMM) & 
24	 &60 & 	Classroom, clip mic &	Male (senior), native, slow pace \\\hline

Discrete Stochastic Processes (STOP) & 
25&	80&	Lecture hall, clip mic &	Male (senior), native, slow pace \\\hline

Geometric Folding Algorithms: Linkages, Origami, Polyhedra (ALGO) & 
21 	&80	&	Lecture hall, distant mic 	& Male, native \\\hline

Introduction to Computer Science and Programming in Python (PYTH) & 
12&	43&	Lecture hall, distant mic & 	female, native, slightly fast pace \\\hline
 \end{tabular}
 \caption{Information about the course lectures being included in the dataset for experiments} 
 \label{table:course_info}
 \end{table}

 The recording conditions for lectures in different courses varied greatly. Some were recorded in classroom, some in lecture hall; There are different types of microphones: clip-on close-talking mic or built-in mic on video recorder. All recordings are in good perceptual quality such that teacher's speech can be heard clearly. Each course consists of $12 - 25$ lectures. The duration of each lecture is in $45-70$ minutes. The course teacher of MATH spoke with French accent. The speaking rate in PYTH is relatively fast and that in COMM is slow.

\subsection{Clustering results}

Experiments are carried out on two scenarios: 1) training data from lectures of the same course with the same speaker; and 2) training data from lectures of various courses (multiple speaker and varying recording environments).

The sklearn library\footnote{https://scikit-learn.org/stable/modules/clustering.html} and \cite{mcinnes2017accelerated} in Python are used to implement segment clustering with different types of features and clustering algorithms. 

\subsubsection{Single-course training}
We experimented on the course COMM. All of the 25 lectures are used in training a course-based ASM. Multilingual DNN-BN as described in Section \ref{multilingual-dnn-sbn} is used to extract BNFs from audio input. The AHC is used to determine the number of clusters, i.e., $55$, and the $k$-means algorithm is applied to cluster the segmental BNFs into subword units, and producing the initial pseudo transcription. The training of DNN-HMM starts with the pseudo transcription, and continue iteratively as described in \ref{sec:iter_train}. ASMs trained on different acoustic features are compared. They include the proposed BNF, conventional MFCC, and filter-bank. The number of iterations is $5$ for each ASM. The degree of convergence is measured by the difference between transcriptions before and after each iteration, which is regarded as subword error rate (SWER) in Table \ref{table:iter_train}. 

 \begin{table}[h]
 \centering
 
 \begin{tabular}{|c|c|c|c|c|c|}
 \hline
 \multicolumn{2}{|c|}{BNF} &  \multicolumn{2}{|c|}{Filter Bank} & \multicolumn{2}{|c|}{MFCC} \\\hline
 iteration  & SWER (\%) & iteration   & SWER (\%) & iteration   & SWER (\%) \\\hline
 iter 1     & 31.43    & iter 1      & 32.59    & iter 1      & 32.66    \\\hline
 iter 2     & 6.26     & iter 2      & 7.36     & iter 2      & 7.12    \\\hline
 iter 3     & 3.75     & iter 3      & 4.64      & iter 3      & 4.54      \\\hline
 iter 4     & 2.97     & iter 4      &  4.27     &   iter 4    & 4.53         \\\hline
 iter 5     & 2.23     & iter 5      &  4.27     &   iter 5     & 4.54     \\\hline        
 \end{tabular}
 \caption{Difference in input and output labels on bottleneck and filter bank feature}
 \label{table:iter_train}
 \end{table}

From the experimental results, it can be seen that the the ASM trained with BNF could achieve lower SWER and its training converges faster than fbank and MFCC. Also the ASM trained with BNF converges at a lower SWER of $2.23\%$, compared with those trained with fbank and MFCC with SWER of $4.2\% - 4.5\%$.

\subsubsection{Multiple-course training}

$11$ Lectures selected the $5$ courses are pooled to train an ASM in the same way as the single-course case. A larger number of clusters, i.e., $100$, is used in order to model more variations. Different clustering algorithms are used to obtain pseudo transcriptions to initialize DNN-HMM training.
The subword mismatch rate of training an ASM with different sets of pseudo transcription are compared, only the first iteration of DNN-HMM training is used for comparison. 

 \begin{table}[h]
 \centering
 
 \begin{tabular}{|c|c|c|}
 \hline
 \textbf{Clustering method} & \textbf{SWER}  \\\hline
 K-means Clustering &  59.10\% \\\hline 
 AHC with ward method & 37.86\% \\\hline 
 Gaussian Mixture Model & 60.51\% \\\hline 
 Bayesian Gaussian Mixture Model & 41.51\% \\\hline     
 HDBSCAN soft clustering & 34.78\% \\\hline     
 \end{tabular}
 \caption{Subword mismatch rate of the first DNN-HMM iterative training using pseudo transcript from different clustering methods}
 \label{table:feature_iter_train}
 \end{table}
 
When analysing the performance of k-means clustering, a control experiment is carried out on training ASM from single-course recordings, with cluster number set to be $100$. The ASM takes in BNFs as input and pseudo transcript generated from the 100 subwords clustered by k-mean, and gives SWER of $36\%-37\%$ on its first iteration. The subword mismatch for the first iteration (59.10\%) in the multi-course case is much higher.

Apparently k-means clustering does not perform well in clustering BNF under complex speaker and channel conditions. Another possible reason is certain clustering algorithms such as k-mean clustering may not be able to well-separate speaker/channel information from phonetic properties in BNF.

Comparing different clustering methods, it is found that HDBCAN gives the lowest SWER. For the clusters generated from GMM and BGMM, there are some clusters that have less than $500$ frames (GMM: $\{290\}$, BGMM: $\{365, 295\}$). The total number of frames is about $600,000$. Merging these minority clusters into large clusters can be considered.

\section{Summary}

In the chapter, the proposed procedures in building an ASM is discussed. It is shown that bottleneck features produced from Multilingual DNN-BN and Multilingual DNN-SBN can both be suitable features for unsupervised subword modelling, compare with conventional acoustic features. However, the training time for Multilingual DNN-SBN is much longer than Multilingual DNN with similar performance. 

Segmental BNF is also suitable for modelling both single speaker and multi-speakers recordings. Among different clustering methods, HDBSCAN soft clustering is more preferred with its lower initial subword input/output mismatch, making convergence of training the ASM easier. 

\subsubsection{Area of improvement}

As a proof of concept, DNN model used in extracting multilingual bottleneck feature and iterative training in this study is the basic neural network architecture. Other advanced architectures such as CNN, BILSTM can be considered in the future. However, larger computation is also needed.

At this moment, theoretical understanding of bottleneck features is still limited. Other clustering algorithms and their relationship with bottleneck feature can also be investigated in the future.
\chapter{Spoken Term Discovery}

Given a set of input utterances, the ASM produces pseudo transcriptions that characterise its content in terms of sequences of subword-like units. For a low-resource language, an interesting and practically useful application is to discover functional spoken terms, e.g., words, from a large collection of speech data in an unsupervised manner. This chapter is focused on automatic spoken term discovery from untranscribed lecture recordings. Part of the work can be found in \cite{sung2018unsupervised}. The proposed method comprises the following procedures:
\begin{enumerate}
    \item obtain a bag of keyword candidates by identifying repeated sequences of subword units in the pseudo transcriptions;
    \item cluster the keyword candidates to discover keywords.
\end{enumerate}
 
\begin{figure}[h]
\hspace*{-2cm}
\begin{center}
\includegraphics[width=16cm]{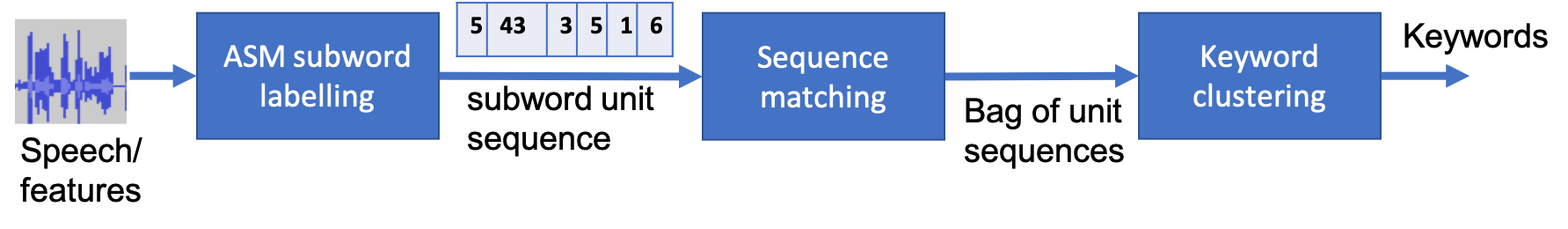}
\end{center}
\caption{Automatic discovery of keywords from pseudo transcription}
\label{fig:system2}
\end{figure}

\section{Generating keyword candidates} \label{ssec:bag_of_seq}
Keyword candidates are hypothesized by detecting repetitively occurred sequences of subword units in the pseudo transcriptions. The underlying assumption is that if a subword sequence appears multiple times in the transcriptions, it likely corresponds to a keyword or key-phrase. It is expected that same word may correspond to different subword sequences, due to pronunciation variation, speaker and environment changes. A robust algorithm of candidate sequence identification is needed to cope with these variabilities.

\subsection{Sequence alignment}

Identifying similar or closely related sequence segments is an important problem in the area of bio-informatics. In \cite{smith1981identification}, an algorithm of inexact matching between a pair of short symbol sequences was described.  The algorithm finds an alignment between two symbol sequences that gives the best-matching sub-sequence.


In traditional symbol matching algorithms, all symbols are weighted equally. In this study, subword-level acoustic information is used to determine the importance of each symbol. The symbol weight is assigned to reflect the likelihood that the symbol represents a speech unit.

A voice activity detection (VAD) algorithm \cite{tan2010low} is first applied to locate speech regions in an utterance. The likelihood for a subword unit to represent a phoneme is estimated as the average number of frames with voice activity in each class, with 1 being the highest score and 0 the lowest. 
Expectedly, hypothesized subwords in non-speech regions would have low likelihood values.


Similarity of two sequences is measured based on local weighted alignment. The pseudo-code of this algorithm is given as in Algorithm \ref{LocalAlign}. 
Given $2$ subword sequences $A = a_1a_2...a_n$ and $B = b_1b_2...b_m$. A score matrix with size of $n \times m$ is initiated to store the similarity scores of sub-sequence match \{$a_1...a_i, b_1...b_j$\} at each location $i, j$. The scores are calculated based on the accumulative scores of matched and mismatched subwords between $A$ and $B$. The local maximum scores are used to identify termination points of the sub-sequences, and each sub-sequence is obtained by tracing back from the maximum score to its minimum. 

\begin{algorithm}[h]
  \caption{Modified local sequence alignment}\label{LocalAlign}
  \begin{algorithmic}[1]
    \Procedure{LocalAlign}{$A = a_1 a_2 ... a_n, B = b_1 b_2 ... b_m$}
      \State $s(a_i,b_j)= \begin{cases}
    +w_{a_i}w_{b_j}&\text{, if $a_i,b_j$ match}\\
    -w_{a_i}w_{b_j}&\text{, if $a_i,b_j$ mismatch}
    \end{cases}$  \Comment{Similarity score between sequence elements $a_i$ and $b_j$}
    
        \State Compute an $(n+1) \times (m+1)$ matrix $\mathbf{P}$, where the element $p_{i, j}$ is, 
        \Statex $p_{i, j} = \begin{cases}
		0&\text{, $i$ or $j = 0$}\\
		\max 
		\begin{cases}
		p_{i-1, j-1} + s(a_i, b_j)\\
		p_{i-1, j}\\
		p_{i, j-1}\\
		0
		\end{cases}&\text{, elsewhere}
		\end{cases}$  
      \State Traceback from $p_{n,j^*}$ ending with an element of $\mathbf{P}$ equal to $0$, where $p_{n,j^*}$ are local maxima of $\{p_{n,j} | 0 \leq j \leq m \}$, to obtain common sub-sequence pairs in $A$ and $B$.

      \State Store all obtained sub-sequences with reasonable length ($\geq 4$ units) into the ``bags of sequences''.

     \EndProcedure
  \end{algorithmic}
\end{algorithm}

Traditional local alignment traces back from global maximum score and returns only one sub-sequence. Whilst the method of using local maximum scores is able to identify multiple matching sub-sequences, regardless of their different order of appearance. For example, given the two English sentences: ``\textbf{Today} \textbf{we went} hiking'' and ``\textbf{We went} to a park \textbf{today}'', searching for the global maximum score would result in``we went'', and the proposed method could discover both ``we went'' and ``today''.
Based on this algorithm, the ``bag of sequences''  containing only the matching sub-sequences is generated. 

\section{Keyword clustering}
With the ``bag of candidate sequences'' obtained, the next step is to generate a dictionary from this bag. This is a process of sequence pattern clustering. Each of the resulted clusters is expected to represent a keyword in the speech database.
Unlike segment clustering in Chapter 3, the number of keyword clusters is not fixed. It depends greatly on the content of the database. It is also allowed to have some of the candidate sequences not assigned to any cluster, i.e., outliers.


\subsection{Distance metric for subword sequence matching}

Distance metric for symbol string comparison is a critical element in a sequence clustering algorithm. It quantifies the similarity between two strings, which in our case are sequences of subword units.

Levenshtein distance \cite{Vreda1999Lev} measures the difference between two string sequences by counting the minimum number of character changes required to transform one string to another. The types of changes could be insertions, deletions and substitutions, similar to the way of calculating the error rate of an ASR system. 

In terms of Levenshtein distance, the dissimilarity between a pair of string sequences $x$ and $y$ is measured as,

\begin{equation}
lev_{x,y}(|x|,|y|)=
  \begin{cases}
  \max(i,j) & \text{if $\min(i,j)=0$,} \\[1ex]
  \begin{aligned}[b]
  \min\bigl(lev_{x,y}&(i-1,j)+1, \\
            lev_{x,y}&(i,j-1)+1, \\
            lev_{x,y}&(i-1,j-1)+1_{(x_i\ne y_j)}
      \bigr)
  \end{aligned} & \text{otherwise.}
\end{cases} \label{eqt:lev_dis}
\end{equation}
where $i$ and $j$ are the subword indexes of sequences $x$ and $y$ respectively, $lev_{x,y}(i,j)$ is the distance between the first $i$ subwords of sequence $x$ (i.e. $x_1x_2...x_i$) and first $j$ subwords of sequence $y$ (i.e. $y_1y_2...y_j$) . $1_{(x_i\ne y_j)}$ is a function, which takes the value of $0$ if $x_i = y_j$, and $1$ otherwise. 

When using the Levenshtein distance to search for keywords from pseudo transcriptions, a high degree of robustness is required to anticipate that the same keyword could correspond to many variants of subword sequences in natural speech. For a short word, deviation of $2-3$ symbols could be very significant, while for a long word we should tolerate a larger number of symbol differences. To cope with this issue, the normalized Levenshtein distance $||L(x,y)||$ is defined as,
\begin{equation}
||L(x,y)|| =  \dfrac{L(x,y)} {\sqrt{|x|^2 + |y|^2}} \label{eqt:norm_lev_dis}.
\end{equation}

Furthermore, subword weights are applied in computing the Levenshtein distance. Similar to the modified local alignment in Section \ref{ssec:bag_of_seq}, the weight of subword is computed as the proportion of frames with speech activity. The weighted distance $L_w(x,y)$ is similar with $L(x,y)$ in Equation (\ref{eqt:lev_dis}), except $1_{(x_i\ne y_j)}$ is replaced by $w$, where insertion and deletion cost for subword $x_i$ in $y$ is $w_{x_i}$, and the substitution cost from ${x_i}$ to $y_j$ is $w_{x_i}w_{y_j}$.
Normalized weighted Levenshtein distance is denoted as $||L_w(x,y)||$, similar to Equation (\ref{eqt:norm_lev_dis}).

\subsection{Leader clustering}
\label{subsec:seq_clus}
The ``bag of subword sequences" created as in Section \ref{ssec:bag_of_seq} contains a large number of subword sequences of different lengths. These sequences are clustered into groups using the leader clustering algorithm  \cite{hartigan1975clustering}, as depicted in  Algorithm \ref{leader}.
 
 Let $T$ be the radius of each cluster. To prevent clusters from overlapping significantly, the minimum distance between a pair of cluster centroids is set to be $a*T$, where $a > 1$.
\begin{figure}[h]
\begin{center}
\includegraphics[width=8cm]{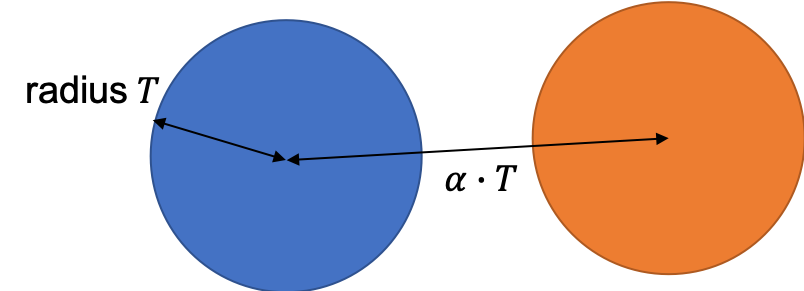}
\end{center}
\caption{Illustration of the meaning of parameters $T$ and $a$.}
\label{fig:leader}
\end{figure}
Leader clustering is sensitive to initialization of centroids.
To avoid poor initialization (e.g. assigning outliers as centroids), the centroid is updated with the most representative sequence, i.e., the sequence having the least total distance with all intra-cluster members, as shown in Line \ref{algo:update_centroid} of Algorithm \ref{leader}. 
The clustering process iterates until the number of clusters does not change further.

\begin{algorithm}
  \caption{Leader clustering}\label{leader}
  \begin{algorithmic}[1]
    \Procedure{Leader}{bag of sequences $S$}
      \State Initial a point $i$ to $\mathrm{centroid}$
      \For{each point $p$ in $S$ }
        \If{$||L(i,p)|| > a*T$ for all $i$ in $\mathrm{centroid}$, $a>1$:}
        \State Add $p$ to $\mathrm{centroid}$
        \EndIf
      \EndFor
        \State 	Assign each point $p$ in $S$ to its closest cluster $i$ with $||L(i,p)|| < T $.
        \For{each group}
        \State Update $\mathrm{centroid}$ with the representative of the cluster (measured by smallest total distance with same group members).\label{algo:update_centroid}
        \EndFor
      \State Repeat steps 2-11 until the number of clusters converge.
    \EndProcedure
  \end{algorithmic}
\end{algorithm}

After clustering, sequences in the same cluster that are found to be overlapping in time are removed. We then obtain the finalized ``keywords'' of the recordings.

\section{Analysis of metric and algorithm }
Determining suitable parameters for the clustering algorithm is not a straightforward and tractable task. In this section, the proposed weighted metric and parameters used in leader clustering are evaluated.

\subsection{Weighted distance metric}
The clustering results obtained using the unweighted and the weighted distance metrics are compared. In this experiment, all lecture recordings of the course MATH are used for training of the ASM model as well as the generation of subword units. Discovery of keywords is performed with the pseudo transcription of only a single lecture, namely ``Number Theory I''.

For leader clustering, we set $a = 1.6$, $T = 0.325$. As a result, the use of weighted distance metric leads to $36,255$ sub-sequences and $292$ keyword clusters, and the unweighted distance metric produces $30,699$ sub-sequences and $264$ clusters. 

A few keywords with different sequence lengths are selected for comparison between the two distance metrics. Table \ref{table:VAD} shows the cluster size (number of discovered words in a cluster) and the purity (the proportion of target keyword included in a cluster).



\begin{table}[h]

\begin{tabular}{|p{1.5cm}|l|c|c|c|c|}
\hline
\multirow{2}{1.5cm}{Subword length} & \multirow{2}{*}{Keywords}    & \multicolumn{2}{c|}{Original}                 & \multicolumn{2}{c|}{Weighted} \\ \cline{3-6} 
                               &                              & Count               & Purity                  & Count         & Purity        \\ \hline
\multirow{2}{*}{18.32}         & greatest common divisor      & \multirow{2}{*}{8}  & \multirow{2}{*}{100\%}  & 4             & 100\%         \\ \cline{2-2} \cline{5-6} 
                               & greatest common divi-        &                     &                         & 6             & 100\%         \\ \hline
15.17                          & linear combination           & 19                  & 100\%                   & 20            & 100\%         \\ \hline
13.00                          & a number theory              & 5                   & 100\%                   & 5             & 100\%         \\ \hline
10.88                          & m divides/and divides/divide & 30                  & 96.7\%                  & 36            & 97.2\%        \\ \hline
10.66                          & x plus y                     & 8                   & 100\%                   & 9             & 100\%         \\ \hline
9.50                           & *chalk-writing sounds*       & 30                  & 100\%                   & 53            & 100\%         \\ \hline
8.50                           & plus one /this one           & 8                   & 75\%                    & 21            & 90.5\%        \\ \hline
8.14                           & 3 gallon                     & 10                  & 30\%                    & 12            & 67\%          \\ \hline
\multirow{2}{*}{7.81}          & equal to                     & \multirow{2}{*}{31} & \multirow{2}{*}{77.4\%} & 25            & 72\%          \\ \cline{2-2} \cline{5-6} 
                               & equal                        &                     &                         &               &               \\ \hline
7.00                           & prime/t prime/b prime        & 18                  & 89\%                    & 20            & 70\%          \\ \hline

\end{tabular}
\caption{Clusters containing the target keywords using unweighted and weighted metric in alignment and clustering.}
\label{table:VAD}
\end{table}

For clusters with relatively long sequences ($> 8.5$ symbols), the tendency is that more keywords could be found by using weighted metric, with high purity maintained. 
For partially similar keywords, e.g., ``greatest common divisor'' and ``greatest common divi-'', clustering with unweighted metric tends to merge them in the same cluster, while using weighted metric seems to treat them as being different.

For clusters of shorter sequences ($< 8$ symbols), the purity decreases more significantly when using weighted metric than unweighted. A possible reason is that the radius measured in weighted metric is more loosen than unweighted metric since the respective sequence distance is usually smaller. 
In general, using weighted metric provides more favourable result, i.e., identifying more keywords with good cluster purity, especially for clusters of long sequences.


\subsection{Effect of clustering parameters}

The effect of the radius $T$ and the margin factor $a$ is investigated in an experiment on the same lecture as in the previous section. The performance of keyword clustering is evaluated with the following parameter values: 

\begin{itemize}
\item radius $T = \{ 0.2 , 0.25, 0.3, 0.35, 0.4 \}$
\item margin factor $a = \{ 1.2 , 1.4, 1.6 \}$
\end{itemize}

\begin{figure}[h]
\begin{center}
\includegraphics[width=12cm]{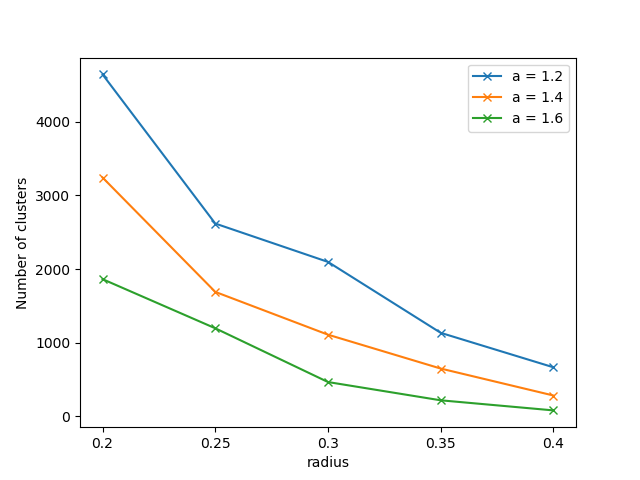}
\end{center}
\caption{Number of discovered keywords corresponding to different parameters $T$ and $a$.}
\label{fig:leader_param}
\end{figure}

A total of $36,255$ sub-sequences are extracted using weighted local sequence alignment. With different clustering parameters, the number of clusters generated varies greatly from $82$ to $4,644$, as shown in Figure \ref{fig:leader_param}. 
It is noted that larger values of $T$ lead to significantly smaller number of clusters. Increasing the value of $a$ reduces the number of clusters, but the effect is not as significant as increasing the radius.

A few clusters that correspond to target keywords or key-phrases are analyzed. They are ``greatest common divisor'', ``linear combination'', ``number theory'', ``algorithm'' and ``divides'', arranged in the descending order of sequence length. The clusters information are presented in Table \ref{table:leader_para}.

\begin{table}[h]
\centering
\resizebox{14cm}{!}{%
\begin{tabular}{|l|l|P{1.0cm}|P{1.0cm}|P{0.8cm}|P{1.3cm}|p{0.7cm}|p{0.9cm}|p{0.7cm}|p{0.9cm}|p{0.7cm}|p{1.2cm}|}\hline
\multicolumn{2}{|c|}{\textbf{parameters}} & \multicolumn{10}{|c|}{\textbf{keywords/key-phrases(number of clusters, average purity)}}                                                                                                                                                \\\hline
radius                      & a         & \multicolumn{2}{|P{2.9cm}|}{greatest common divisor} & \multicolumn{2}{|P{2.1cm}|}{linear combination} & \multicolumn{2}{|P{1.6cm}|}{number theory} & \multicolumn{2}{|P{1.6cm}|}{algorithm} & 
\multicolumn{2}{|P{1.9cm}|}{divides}                   \\\hline
\multirow{3}{*}{0.2}        & 1.2 & 8    & 100\% & 6   & 100\% & 2             & 100\% & 2         & 59.8\% & 5       & 98.3\%  \\\cline{2-12}
                            & 1.4 & 6    & 100\% & 5   & 100\% & 2             & 100\% & 2         & 100\%  & 3       & 97.7\%  \\\cline{2-12}
                            & 1.6 & 5    & 100\% & 4   & 100\% & 2             & 100\% & 2         & 100\%  & 4       & 100\%   \\\hline
\multirow{3}{*}{0.3}        & 1.2 & 4    & 100\% & 3   & 100\% & 1             & 60\%  & 1         & 100\%  & 2       & 93\%    \\\cline{2-12}
                            & 1.4 & 4    & 100\% & 1   & 100\% & 2             & 100\% & 1         & 100\%  & 2       & 97.58\% \\\cline{2-12}
                            & 1.6 & 3    & 100\% & 1   & 100\% & 1             & 100\% & 1         & 67\%   & 2       & 91.35\% \\\hline
\multirow{3}{*}{0.4}        & 1.2 & 3    & 87\%  & 2   & 100\% & 1             & 100\% & \multicolumn{2}{|c|}{not found} & 2       & 81.14\% \\\cline{2-12}
                            & 1.4 & 2    & 100\% & 1   & 100\% & 1             & 100\% & 1         & 100\%  & 2       & 72\%    \\\cline{2-12} 
                            & 1.6 & 1    & 100\% & 1   & 100\% & \multicolumn{2}{|c|}{not found}   & \multicolumn{2}{|c|}{not found} & 1 & 75\%   \\\hline

\end{tabular}
}
\caption{The effect of number of clusters and their purity representing specific keywords.}
\label{table:leader_para}
\end{table}


When a small radius $T$ is used, long keywords or key-phrases tend to be split into shorter constituents. 
For example, when radius $T = 0.2$, $a = 1.6$, the cluster representing ``greatest common divisor'' is split into multiple clusters that represent ``greatest common divisor'', ``-test common divisor'', ``greatest common di-'' and ``the greatest com-''. with $T = 0.4$, all these clusters merge into one.
The same observation is noticed also for clusters of shorter sequences, ``the algorithm'' and ``this algorithm'' are assigned in different clusters when radius $T = 0.2$, but they are assigned to the same cluster when $T \geq 0.3$. The purity of clusters increases when $T$ is decreased.
Reducing the margin factor $a$ tends to encourage to splitting of clusters, though the effect is not as significant as reducing $T$. 

While using a large radius and large margin help discover long phrases precisely, it may lead to the missing of relatively short phrases (e.g., ``number theory'' and ``algorithm'' with $T = 0.4$, $a =1.6$). It also causes undesirable grouping of phrases with similar sound sequences, e.g., ``the five'' is grouped with ``divide''.

\section{Evaluation on ZeroSpeech 2017 spoken term discovery task}

The proposed spoken term discovery system is evaluated with spoken term discovery task in ZeroSpeech 2017 Challenge Track 2 \cite{dunbar2017zero}. Similar as Track 1, the challenge provides three languages for evaluation, namely English, French and Mandarin.

\begin{figure}[h]
\hspace*{-2cm}
\begin{center}
\includegraphics[width=10cm]{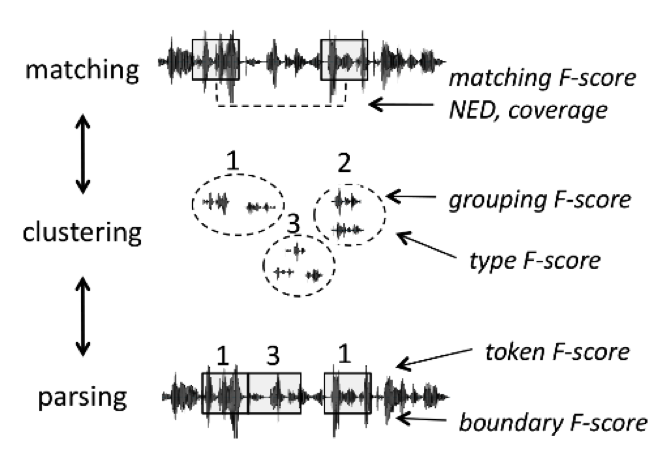}
\end{center}
\caption{Evaluation of term discovery in Zerospeech Challenge 2017.}
\label{fig:zerospeech}
\end{figure}

In the challenge, the spoken term discovery evaluation process is presented as a 3-stages work described as in Figure \ref{fig:zerospeech}: 1) Matching repeating sequence; 2) Clustering the matching sequences; 3) Token representation of the audio with boundary information. Evaluation is done on each of the stage.


Evaluation is done on the discovered clusters with boundary information and cluster members provided by the performers. The result is compared with the actual lexicon and word transcription. $2-3$ metrics are used in each stage. Most of the measures are defined in terms of precision, recall and f-score. Precision measures the probability that the discovered element belongs to the actual set (e.g. lexicon, word in the transcription). Recall measures the probability that the actual element is in the discovered set. F-score is the harmonic mean between precision and recall. The evaluation in each of the stage are as follow:

\begin{itemize}
    \item Matching quality
        \begin{itemize}
        \item The accuracy of the spoken terms discovered are evaluated using two metrics: NED and Coverage. NED measures the normalized edit distance of the word cluster and the word from actual transcription. Coverage is corpus that contains the matching pairs discovered.
        \end{itemize}
    \item Clustering Quality
        \begin{itemize}
        \item The performance of the system is discovered in terms of lexicon discovery. It uses group precision, recall and f-score to evaluate how well the word clusters match on the sequence of phonemes, and type precision, recall and f-score to evaluate how well the cluster match the actual lexicons.
        \end{itemize}
    \item Parsing Quality
        \begin{itemize}
        \item The metrics evaluate the performance in terms of word segmentation. It measures the word clusters that are aligned with the actual word transcription. It uses token precision, recall and f-score to evaluate the number of words that are correctly segmented. Boundary precision, recall and F-score is used to evaluate the number of actual word boundaries found.
        \end{itemize}
\end{itemize}

To evaluate our proposed spoken term discovery system, we experimented it on the Mandarin data set of $2.5$ hours long. Subword units are discovered by clustering on bottleneck features extracted from multilingual DNN-BN presented in Chapter 3. For the sequence matching process, we tried two different minimum matching length, which is the minimum matching sequence that will be considered for clustering. The experimented matching length are $length = 2$ and $length = 4$. For the word clustering process, we want a straighter matching, with less duplicate word clusters for same keywords, therefore we set radius $T = 0.2$ and margin factor $a = 2.0$. Weighted normalized levenstein distance is used.

There were not many participants in the spoken term detection track, only $2$ systems are presented in the challenge. We evaluated our system with the baseline and the $2$ participated systems.
The baseline is a spoken term system proposed by Jansen et al. \cite{jansen2011efficient}. It performs spoken term discovery by DTW matching with random projections, follow by connected-component graph clustering.
One of the performer is Kamper et al. \cite{kamper2017embedded}, which uses k-means to discover repeating acoustic patterns, jointly optimized with exhaustive segmentation. The word segments for clustering are represented in fix-dimension. Another is Garcıa-Granada et al., which uses supervised Hungarian ASR to decode the speech, followed by sequence matching on the decoded transcription. Autoencoder is used to learn the word representation for word filtering using DTW.





\begin{table}[h]
\resizebox{14.5cm}{!}{%
\begin{tabular}{l|c|c|c|c|c|c|c|c|c|}
\cline{2-10}
                                            & \multicolumn{3}{c|}{Grouping} & \multicolumn{3}{c|}{Token}  & \multicolumn{3}{c|}{Type}   \\ \cline{2-10} 
                                            & Precision  & Recall  & Fscore & Precision & Recall & Fscore & Precision & Recall & Fscore \\ \hline
\multicolumn{1}{|l|}{Baseline}              & 30.2       & 96.7    & 44.7   & 4         & 0.1    & 0.1    & 4.5       & 0.1    & 0.2    \\ \hline
\multicolumn{1}{|l|}{Kamper et al.}         & 2.9        & 10.6    & 4.6    & 2.5       & 3.4    & 2.9    & 2.5       & 4.1    & 3.1    \\ \hline
\multicolumn{1}{|l|}{García-Granada et al.} & 2.4        & 45      & 4.6    & 3.6       & 2      & 2.6    & 4.5       & 2.9    & 3.5    \\ \hline
\multicolumn{1}{|l|}{$T=0.2, a=2.0, length=2$}      & 0	& 45.1&	0.1&	2.3 & 2	& 2.2	& 3.4	& 2.9	& 3.1 \\ \hline
\multicolumn{1}{|l|}{$T=0.2, a=2.0, length=4$}    & 0	& 0	& 0	& 1.5	& 0.2	& 0.4	& 1.7	& 0.3	&  0.6	\\ \hline
\end{tabular}
}

\bigskip\bigskip
\resizebox{13cm}{!}{%
\begin{tabular}{l|c|c|c|c|c|c|c|ll}
\cline{2-8}
                                            & \multicolumn{3}{c|}{Boundary} & \multicolumn{4}{c|}{NLP}            &  &  \\ \cline{2-8}
                                            & Precision  & Recall  & Fscore & NED  & Coverage & n-words & n-pairs &  &  \\ \cline{1-8}
\multicolumn{1}{|l|}{baseline}              & 37.5       & 0.9     & 1.8    & 30.7 & 2.9      & 156     & 160     &  &  \\ \cline{1-8}
\multicolumn{1}{|l|}{Kamper et al.}         & 36.2       & 46.7    & 40.8   & 88.1 & 117.7    & 2967    & 356585  &  &  \\ \cline{1-8}
\multicolumn{1}{|l|}{García-Granada et al.} & 22.8       & 18.6    & 20.5   & 80.2 & 43.4     & 2887    & 17845   &  &  \\ \cline{1-8}

\multicolumn{1}{|l|}{$T=0.2, a=2.0, length=2$}                  & 	28.4& 	27.9& 	28.1& 	94.7& 	76.6&	 2003& 	74874 \\ \cline{1-8}
\multicolumn{1}{|l|}{$T=0.2, a=2.0, length=4$}              & 28.5	& 6.2	& 10.2	& 94.8	& 26	& 80	& 9173 \\ \cline{1-8}

\end{tabular}
}
\caption{Zerospeech Challange 2017 Track 2 metrics for the spoken term discovery systems on Mandarin dataset.}
\label{table:zerospeech_std}
\end{table}

The results of the systems are presented in percentage in Table \ref{table:zerospeech_std}. It is observed that our proposed system, when having a small minimum matching length, can achieve reasonable performance compared with other systems in terms of token, type, boundary and NLP quality. The word coverage rate $76.6$ is quite high.
In terms of grouping, our system has reasonable recall rate, but the precision rate is very low, which means although the discovered word clusters maintain a good consistency of subword sequences, the subword sequences in the cluster may not have exact match with actual phonemes. It is noted that the evaluation is done only with the cluster boundaries and members information available. The discovered subwords are not measured with actual lexicons directly.
However, it is also understandable since our system perform inexact matching. It does not perform any segmentation or further process matching sequences nor word clusters. This suggests more work toward word level segmentation can be consider in improving our work.

Although using a high minimum matching length of $4$ gives very low quality in terms of grouping, token, type and boundary, it is interesting to note that even only few words of $80$ are discovered, the coverage rate is higher than the baseline. Possible reason is that since a straight matching criteria is used, the resulted discovered word clusters contain less non-representing clusters.

From this evaluation it seems that using a shorter minimum matching length is better, however, the actual parameters used for the algorithm is language and data dependent. For example, words in Mardarin usually have a shorter formation of phonemes, while words in English can have various number of phonemes. More work on determining the suitable set of parameters for different languages is needed.

\section{Summary}
In this chapter, local alignment follow by leader clustering is proposed for keyword discovery on the pseudo transcriptions generated from ASM. 
The proposed algorithms are shown to be effective in discovering repeating keywords that is compatible with other available spoken term discovery systems. 

While comparing with the algorithm and metric used, it is observed that subword weighting based on acoustic properties can improve sequence alignment and keyword clustering. The selection of clustering parameters also effects the resulting clusters. In this set of video lectures, the radius of the weighted normalized Levenshtein distance $T$ is preferred to be around $0.3$, and the margin factor between clusters $a$ is around $1.4$.

\subsubsection{Area of improvement}

Weighting subwords based on voice activity detection improves the clustering performance in general. However, there are much more speech characteristics that can be considered. 
Other weighted factors such as 1) weighting based on similarity of the subwords, giving lower penalty when acoustic similar subwords are swapped. 2) Weighting based on posterior probabilities of the subwords from the ASM, or even weighting each subword unit based on the confusion scores of the ASM decoding result.

The relationship of algorithm parameters and the experimented languages requires further investigation. Improvement can be done on the resulting matching sequences and clusters, such as use of word segmentation.

\chapter{Results Analysis and Potential Applications}


In this chapter, we analysis the proposed system with real world data. The system combines the ASM in Chapter 3 and keyword discovery in Chapter 4. Online lectures are experimented. Quality and coverage of the discovered keywords are evaluated in a softer manner.
The relation between discovered keywords and lecture topics is investigated in further detail. Tools for visualizing and analyzing text documents are used. They include word embedding and TF-IDF, which reflect the goodness of discovered linguistic information.

\section{Results of keyword discovery}

The performance of the proposed system is evaluated by comparing and relating the discovered keywords with word-level transcriptions provided at the MIT OpenCourseWare website. The same set of course recordings mentioned in Table \ref{table:course_info} in Section $3.4$ is used. The ASM that generates the pseudo transcriptions for keyword discovery is trained on single course only.

The performance is analyzed in terms of the quality of clusters and the clusters' coverage on lecture content. For the quality of discovered clusters, the goal is to have similar subword sequences included in the same cluster, similarity is measured by normalized Levenshtein distance. These subword sequences are expected to represent the topic-related keywords or key-phrases frequently spoken in the lecture. On the other hand, it is desired that a significant portion of content-related words could be discovered by clustering of subword sequences. 

\subsection{Quality of clusters}
\label{ssec: cluster_analysis}

The clustering results on two selected lectures in the course MATH are examined. The parameter values used for leader clustering are $a = 1.6$, $T = 0.35$.  For Lecture 4 (``Number Theory'', $80$ min. long), the system generates $95$ keyword clusters from $34,313$ candidate subword sequences. For Lecture 8 (``Graph Theory II: Minimum Spanning Trees'', $83$ min. long), there are $119$ clusters from $25,899$ candidate subword sequences.

Tables \ref{table:lec_4} and \ref{table:lec_8} list the words corresponding to the $10$ longest sequence clusters in the two lectures respectively. It is observed that most of these clusters ($\sim90\%$) correspond to words that are related to the lecture topic.

\begin{table}[h]
\centering
\small
\begin{tabular}{|l|l|l|l|}
\hline
Cluster \# & Corresponding words & Cluster size & Purity                     \\\hline
58         & divide any result                       & 2   & 100\%                       \\\hline
63         & the zero steps                        & 2   & 100\%                       \\\hline
70         & Multiple of                          & 2  & 100\%                       \\\hline
56         & linear combination                      & 17   & 100\%                       \\\hline
40         & the/a number theory                             & 5    & 100\%                       \\\hline
70         & a and b & 14       & 100\% \\\hline
28         & use the lemma again                                           & 2     & 100\%                       \\\hline
43         & Greatest common, The greatest common                     & 18     & 100\%                       \\\hline
77 & *chalk-writing sounds*  & 8   & 100\%                       \\\hline
76  & gallon jug & 15   & 100\%                       \\\hline
\end{tabular}
\caption{Discovered clusters from lecture ``Number Theory I''}
\label{table:lec_4}
\end{table}


\begin{table}[h]
\centering
\small
\begin{tabular}{|l|P{7cm}|l|l|}

\hline
Cluster \# & Corresponding words                               & Cluster size & Purity     \\\hline
70         & b equal to                              & 2     & 100\%      \\\hline
66         & this particular edge, this particular err             & 4   & 75\%       \\\hline
57         & connected subgraph, the subgraph         & 7     & 100\%      \\\hline
89         & A subgraph,The smaller part                       & 4      & 50\%, 50\% \\\hline
87         & still connected, both connected                     & 4       & 100\%      \\\hline
83         & Double star is                                    &  2       & 100\%      \\\hline
80  & So we know that  & 2   & 100\%      \\\hline
116  & Vertices, vertices that, -ices have       &  7     & 85.7\%    \\\hline
45   & the spanning tree, a spanning tree, spanning tree & 25  & 100\%      \\\hline
109        & vertices  &  4  & 100\%      \\\hline 
\end{tabular}
\caption{Discovered clusters from lecture ``Graph Theory II: Minimum Spanning Trees''}
\label{table:lec_8}
\end{table}

It is noted that clusters with sequence length of $12$ subword units or more generally have high purity. Each of these clusters provides a valid representation of a specific word or phrase. As the sequence length decreases, the cluster's purity tends to decrease. Sequences containing less than $5$ subword units typically correspond to parts of different words that have similar pronunciations, e.g., ``so'', ``(al)so'', and ``so(lve)''; ``in'' and ``in(teger)''.

It is noted that the same word may be represented by more than one clusters. For example, clusters \#116 and \#109 of Lecture 8 (Table \ref{table:lec_8}) both correspond to ``vertices''. Also, some discovered words may be constituted by shorter sequences of other clusters.

Some of the clusters represent non-speech sounds, e.g., cluster \#77 in Table \ref{table:lec_4} corresponds to ``chalk-writing sounds'', which is very common in live recordings of lectures. There exists several clusters with ``chalk-writing sounds'' of various length.


While listening to the segments in a same cluster, different variations of same word can be observed. For example, different tones or different stressed syllables, different paces of speaking a same word, and
same word with different recording channels (very clear and distant). 
Even there is influence of background noise to a spoken term such as chalk writing sound, the model is still able to assign it in the same group along with the clean speech of the same keyword.


Besides speaker and channel variations, phoneme variation is also observed in same clusters, for example, ``prime'' and ``find'' , ``m divides'', ``n divides'' and ``and divides'', which are not exactly the same but very similar in pronunciation.



\subsection{Coverage of discovered words}

For the intended task of keyword discovery, it is desired that a significant portion of the content-related words could be covered by the unsupervisedly generated clusters. In this section, we analyzed the automatically discovered keyword clusters with respect to the frequently occurred words and phrases in the ground-truth transcriptions (available at the MIT OpenCourseWare website\footnote{\fontfamily{pcr}\selectfont ocw.mit.edu/index.htm}). Word-level trigrams, bigrams and unigrams are computed from the transcription for each lecture session or all lectures in a course, with the function words ``is'', ``a'', ``the'', etc, being discarded.

For a specific lecture session, the most frequent $N$-grams are examined one by one, to determine whether the corresponding word(s) can be matched with any of the discovered word clusters. There are cases that a cluster may partially match a trigram or bigram. If the unmatched part is a function word, e.g. ``linear combination'' versus ``linear combination of'', it is regarded as a case of match. If the unmatched part is a content word, e.g., ``divisor'' versus ``common divisor'', it is regarded as mismatch.

\subsubsection{Coverage in lecture base}


\begin{table}[]
\resizebox{14cm}{!}{%
\begin{tabular}{|c|c|c|c|c|c|c|}
\cline{1-3}\cline{5-7} 
\textbf{Trigram} & \textbf{Count} & \textbf{Match?} &  & \textbf{Unigram} & \textbf{Count} & \textbf{Match?} \\\cline{1-3}\cline{5-7} 
greatest common divisor & 44    & partly &  & jug          & 80    & \checkmark  \\
the greatest common     & 39    & \checkmark &  & times        & 74    & \checkmark  \\
a and b                 & 31    & \checkmark  &  & zero         & 63    & \checkmark  \\
is equal to             & 26    & \checkmark  &  & equal/equals & 63    & \checkmark  \\
x plus y                & 22    & \checkmark  &  & y            & 60    & \checkmark  \\
linear combination of   & 20    & \checkmark  &  & divides      & 59    & \checkmark  \\
common divisor of       & 18    & partly &  & gallon       & 56    & \checkmark  \\
a linear combination    & 16    & \checkmark  &  & m            & 54    & \checkmark  \\
the b jug               & 12    & \checkmark  &  & right        & 54    &        \\
three gallon jug        & 11    & partly &  & x            & 53    & \checkmark  \\\cline{1-3}
\textbf{Bigram} & \textbf{Count} & \textbf{Match?} &  & gallons      & 50    & \checkmark  \\\cline{1-3}
common divisor          & 44    & partly &  & plus         & 48    & \checkmark  \\
greatest common         & 44    & \checkmark  &  & divisor      & 45    & \checkmark  \\
the greatest            & 39    & \checkmark  &  & one          & 45    & \checkmark  \\
gallon jug              & 39    & \checkmark  &  & three        & 44    &        \\
equal to                & 39    & \checkmark  &  & greatest     & 44    & \checkmark  \\
linear combination      & 38    & \checkmark  &  & common       & 44    & \checkmark  \\
a and                   & 38    & \checkmark  &  & linear       & 43    & \checkmark  \\
and b                   & 32    & \checkmark  &  & minus        & 43    & \checkmark  \\
is equal                & 26    & \checkmark  &  & combination  & 38    & \checkmark  \\
to prove                & 25    &        &  & number       & 37    & \checkmark  \\
plus y                  & 22    & \checkmark  &  & prime        & 37    &        \\
x plus                  & 22    & \checkmark  &  & prove        & 30    &        \\
m divides               & 21    & \checkmark  &  & theorem      & 27    &        \\
combination of          & 20    & \checkmark  &  & five         & 26    & \checkmark  \\
s prime                 & 20    &        &  & jugs         & 24    & \checkmark  \\
times a                 & 19    &        &  & example      & 23    &        \\
divisor of              & 19    & \checkmark  &  & algorithm    & 23    & \checkmark  \\
the remainder           & 17    & \checkmark  &  & state        & 22    &        \\
that m                  & 17    &        &  & remainder    & 20    & \checkmark  \\\cline{1-3}\cline{5-7}
\end{tabular}
}
\caption{Matching results for lecture ``Number Theory I''}
\label{table:ngram_CS_Maths}
\end{table}


\begin{table}[]
\resizebox{14cm}{!}{%
\begin{tabular}{|c|c|c|c|c|c|c|}
\cline{1-3}\cline{5-7} 
\textbf{Trigram}  & \textbf{Count} & \textbf{Match ?}  & & \textbf{Unigram} & \textbf{Count} & \textbf{Match ?} \\\cline{1-3}\cline{5-7} 
a coordinate object   & 14    & \checkmark  &   &   object    & 118   & \checkmark  \\
can interact with     & 13    & \checkmark  &   &  coordinate & 70    & \checkmark  \\
a fraction object     & 9     & partly &   &  class      & 60    & \checkmark  \\
an object of          & 7     & \checkmark  &   &  type       & 47    & \checkmark  \\
of the class          & 7     & \checkmark  &   &  method     & 46    & \checkmark  \\
of type coordinate    & 7     & partly &   &  data       & 45    & \checkmark  \\
the exact same        & 6     &        &   &  objects    & 44    & \checkmark  \\
you can create        & 6     &        &   &  right      & 38    &        \\
is equal to           & 6     &        &   &  i          & 35    &        \\
going to define       & 6     & \checkmark  &   &  list       & 32    &        \\\cline{1-3}	
\textbf{Bigram}  & \textbf{Count} & \textbf{Match ?} & &  python               & 30    & \checkmark  \\\cline{1-3}
an object             & 25    & \checkmark  &   &  x         & 29    &       \\      
the class             & 22    & \checkmark  &   & create     & 28    &        \\
coordinate object     & 20    & partly &   &   self     & 28    &        \\
interact with         & 19    & \checkmark  &   & print      & 27    &        \\
a list                & 17    &        &   & one        & 25    & \checkmark  \\
a coordinate          & 17    & \checkmark  &   & c          & 24    & \checkmark  \\
 the object            & 14    & \checkmark &   & dot        & 24    &        \\
fraction object       & 14    & partly &   & attributes & 22    & \checkmark  \\
object that           & 14    & \checkmark  &   & fraction   & 21    & \checkmark  \\
the x                 & 13    &        &   & underscore & 20    &        \\
this method           & 13    & \checkmark  &   & value      & 20    & \checkmark  \\
can interact          & 13    & \checkmark  &   & interact   & 19    & \checkmark  \\
create a              & 13    &        &   & y          & 18    &        \\
  a fraction            & 12    & \checkmark&   & define     & 18    &        \\
data attributes       & 12    & \checkmark  &   & init       & 17    &        \\
of type               & 11    & \checkmark  &   & add        & 17    &        \\
for example           & 11    & \checkmark  &   & lists      & 16    &        \\
to define             & 10    &        &   & car        & 16    &        \\
the list              & 10    &        &   & code       & 15    &        \\
underscore underscore & 10    &        &   & & &  \\\cline{1-3}\cline{5-7}
\end{tabular}
}
\caption{Matching results for lecture ``Object-Oriented Programming''}
\label{table:ngram_PYTH}
\end{table}


Firstly, the coverage is measured in lecture base, one lecture from two different courses are examined.  For Lecture 4 (``Number Theory I'') of the course MATH and Lecture 8 (``Object-Oriented Programming'') of the course PYTH, we analyzed the top $10$ trigrams, $20$ bigrams and $30$ unigrams and matched them with the $10$ longest sequence clusters and other highly-populated clusters. The details of matching results for both lectures are given as in Table \ref{table:ngram_CS_Maths} and \ref{table:ngram_PYTH}.

The matching rates are found to be 
$73.3\%$ and $51.6\%$ respectively. 
To explain the $20\%$ difference in the coverage range, we analysis the property of the missing words. It is noted that the uncovered unigrams are mostly words with small number of phones, e.g. ``add'', ``car'', ``code'', while the covered words are mostly polysyllabic words, e.g., ``python", ``coordinate''. For the lecture ``Number Theory I'', the matching rate for unigrams is higher, due to more complicated phonetic structure of the words.


\subsubsection{Coverage in course base}
\label{sec:exp_comm}

Same analysis has also been done for the whole course COMM, which contains $24$ lectures of $70$ minutes long. Keyword discovery is done on the whole course. The $100$ most frequent trigrams, bigrams and unigrams are examined by comparing with clusters generated from all lectures. A high matching rate of $85.8\%$ is recorded (Table \ref{table:ngram_Comm_I}).

\begin{table}[]
\vspace*{-0.5cm}
\caption{Matching results for course COMM }
\label{table:ngram_Comm_I}
\resizebox{15cm}{!}{%
\begin{tabular}{|c|c|c|c|c|c|c|c|c|c|c|}
\cline{1-3}\cline{5-7}\cline{9-11}
\textbf{Trigram} & \textbf{Count} & \textbf{Match?} &  & \textbf{Bigram} & \textbf{Count} & \textbf{Match?} &  & \textbf{Unigram} & \textbf{Count} & \textbf{Match?} \\\cline{1-3}\cline{5-7}\cline{9-11}
u of t                    & 247   & \checkmark&  & equal to          & 450   & \checkmark&  & function    & 779   & \checkmark\\
in terms of               & 247   & \checkmark&  & u of              & 286   & partly   &  & one         & 707   & \checkmark\\
is equal to               & 213   & \checkmark&  & set of            & 278   & partly   &  & random      & 671   & \checkmark\\
than or equal             & 123   & \checkmark&  & the probability   & 224   & \checkmark&  & u           & 654   & \checkmark\\
or equal to               & 120   & \checkmark&  & is equal          & 221   & \checkmark&  & sub         & 649   & \checkmark\\
the same thing            & 117   &        &  & number of         & 216   & \checkmark&  & minus       & 642   & \checkmark\\
wind up with              & 109   & \checkmark&  & random variable   & 205   & \checkmark&  & same        & 614   & \checkmark\\
the probability of        & 108   & \checkmark&  & a function        & 180   & \checkmark&  & probability & 588   & \checkmark\\
the fourier transform     & 92    & \checkmark&  & fourier transform & 180   & \checkmark&  & mean        & 584   & \checkmark\\
less than or              & 86    & \checkmark&  & the fourier       & 177   & \checkmark&  & times       & 542   & \checkmark\\
the same as               & 86    & \checkmark&  & the channel       & 166   & \checkmark&  & words       & 542   & \checkmark\\
expected value of         & 85    & \checkmark&  & the noise         & 155   & \checkmark&  & two         & 535   & \checkmark\\
is less than              & 84    & \checkmark&  & the problem       & 149   & \checkmark&  & frequency   & 496   & \checkmark\\
a sequence of             & 84    & \checkmark&  & sequence of       & 148   & \checkmark&  & equal       & 485   & \checkmark\\
equal to the              & 84    & \checkmark&  & of u              & 147   & partly   &  & zero        & 468   & \checkmark\\
the inner product         & 81    &        &  & probability of    & 142   & \checkmark&  & set         & 466   & \checkmark\\
the expected value        & 79    & \checkmark&  & fourier series    & 137   & \checkmark&  & fact        & 464   & \checkmark\\
fourier transform of      & 77    & \checkmark&  & the real          & 124   & \checkmark&  & really      & 463   &        \\
the sum of                & 76    &        &  & or equal          & 123   & \checkmark&  & problem     & 457   & \checkmark\\
the integral of           & 74    & \checkmark&  & inner product     & 123   &        &  & noise       & 442   & \checkmark\\\cline{5-7}
2 to the                  & 72    & \checkmark&     \multicolumn{4}{l}{}              &  & functions   & 434   & \checkmark\\
e to the                  & 70    &        &     \multicolumn{4}{l}{}              &  & \checkmark     & 417   & \checkmark\\
z of t                    & 69    & \checkmark&     \multicolumn{4}{l}{}              &  & real        & 417   & \checkmark\\
h of x                    & 68    & \checkmark&     \multicolumn{4}{l}{}              &  & n           & 410   & \checkmark\\
the number of             & 67    & \checkmark&     \multicolumn{4}{l}{}              &  & terms       & 402   & \checkmark\\
gaussian random variables & 67    & \checkmark&     \multicolumn{4}{l}{}              &  & gaussian    & 386   & \checkmark\\
a function of             & 64    & \checkmark&     \multicolumn{4}{l}{}              &  & point       & 384   & \checkmark\\
the fourier series        & 64    & \checkmark&     \multicolumn{4}{l}{}              &  & plus        & 384   & \checkmark\\
the real part             & 63    & \checkmark&     \multicolumn{4}{l}{}              &  & fourier     & 378   & \checkmark\\
a set of                  & 62    & \checkmark&     \multicolumn{4}{l}{}              &  & different   & 362   & \checkmark\\
inner product of          & 61    &        &     \multicolumn{4}{l}{}              &  & bit         & 348   & \checkmark\\
mean square error         & 60    & \checkmark&     \multicolumn{4}{l}{}              &  & number      & 346   & \checkmark\\
probability of error      & 57    & partly &     \multicolumn{4}{l}{}              &  & variables   & 329   & \checkmark\\
of t and                  & 57    & \checkmark&     \multicolumn{4}{l}{}              &  & start       & 326   & \checkmark\\
degrees of freedom        & 57    & \checkmark&     \multicolumn{4}{l}{}              &  & vector      & 316   & \checkmark\\
the probability density   & 56    & \checkmark&     \multicolumn{4}{l}{}              &  & channel     & 312   & \checkmark\\
a random variable         & 56    & \checkmark&     \multicolumn{4}{l}{}              &  & process     & 309   & \checkmark\\
g of t                    & 51    & partly   &     \multicolumn{4}{l}{}              &  & variable    & 300   & \checkmark\\
equal to this             & 50    & \checkmark&     \multicolumn{4}{l}{}              &  & sequence    & 296   & \checkmark\\
wide sense stationary     & 50    & partly   &     \multicolumn{4}{l}{}              &  & source      & 295   & \checkmark\\\cline{1-3}
\multicolumn{7}{l}{}                                                               &  & complex     & 281   & \checkmark\\
\multicolumn{7}{l}{}                                                               &  & sample      & 275   & \checkmark\\
\multicolumn{7}{l}{}                                                               &  & sense       & 272   & \checkmark\\
\multicolumn{7}{l}{}                                                               &  & energy      & 271   & \checkmark\\
\multicolumn{7}{l}{}                                                               &  & density     & 270   & \checkmark\\
\multicolumn{7}{l}{}                                                               &  & code        & 267   & \checkmark\\
\multicolumn{7}{l}{}                                                               &  & binary      & 266   & \checkmark\\
\multicolumn{7}{l}{}                                                               &  & sum         & 256   &        \\
\multicolumn{7}{l}{}                                                               &  & waveform    & 248   & \checkmark\\
\multicolumn{7}{l}{}                                                               &  & dealing     & 243   &        \\
\multicolumn{7}{l}{}                                                               &  & error       & 243   & \checkmark\\
\multicolumn{7}{l}{}                                                               &  & numbers     & 242   & \checkmark\\\cline{9-11}
\end{tabular}
}
\end{table}







\section{Semantic relation of discovered keywords}





In the field of natural language processing (NLP), word embedding has been commonly adopted to produce vector representations of words or word sequences. It has been shown in numerous applications that the learnt word embedding space is able to reflect the linguistic functions and relations of the words. For example, vector representations of words with similar meanings are close to each other in terms of common-sense distance measure. In this section, the tool of Word2Vec is utilized to evaluate the results of keyword discovery.

\subsection{Word2Vec}

\begin{figure}[h] 
\centering     
\subfigure[CBOW]{\label{fig:cbow}\includegraphics[width=50mm]{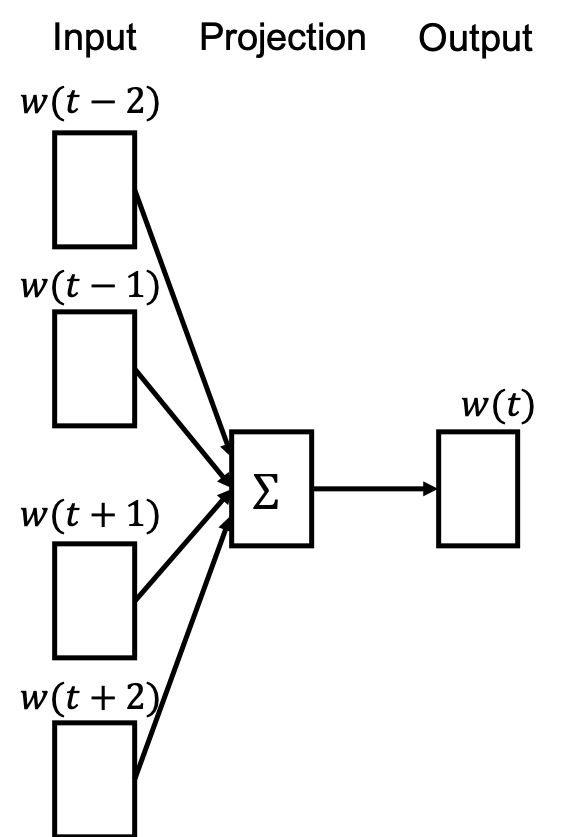}}
\hspace{3cm}
\subfigure[skip-gram]{\label{fig:skipgram}\includegraphics[width=50mm]{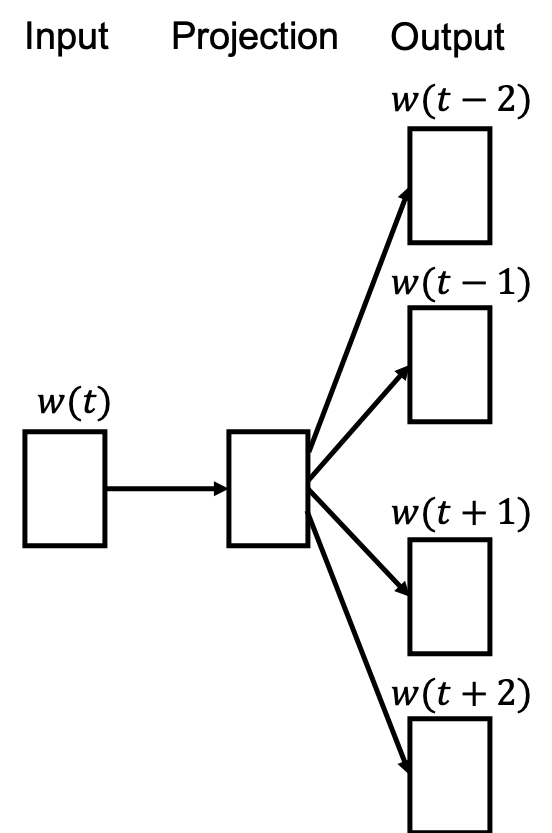} }
\caption{Architecture of Word2Vec models: CBOW and skip-gram.}
\label{fig:wordembed_model}
\end{figure}

Word2Vec \cite{mikolov2013efficient} is developed to train a neural network by which a high-dimension one-hot word vector is converted into a low-dimension word embedding. There are two major approaches, namely continuous bag-of-words model (CBOW) and skip-gram. As shown in Figure \ref{fig:wordembed_model}, the two models share similar simple architecture, with an input layer, a projection layer and and an output layer. The training task of CBOW is to predict a target word (output) from neighbouring words (input). The exact positions of the neighbouring words are not used. The skip-gram model aims to classify a word, given other words in context. It is found that the skip-gram can model infrequent words better than CBOW. After training the model, the embedding space retrieved from the projection layer gives the word vector representations. 

\subsection{Experimental setup}

 \label{sec:exp_word2vec}
 
Experiment is done with the $24$ lectures from course COMM. An ASM is trained with these lecture recordings to obtain the pesudo transcriptions. Keyword clusters are discovered from the pseudo transcriptions by leader clustering with radius $T = 0.3$ and margin factor $a = 1.4$. A total of $9,605$ clusters are found from $1,371,951$ subword sub-sequences.

The skip-gram model with $100$-dimension embedding is used. Illustrated in Figure \ref{fig:skipgram_eg}, the input to the model covers $3$ left-context word clusters (or phrases) and $3$ right-context word clusters, without any overlapping with the target word cluster $w(n)$. 
Sub-sampling is done with selecting $4$ neighbours out of the $6$ in training. $10$ negative samples are considered in batch training of 256 clusters at a time. 

%
%
%

\begin{figure}[h]
\begin{center}
\includegraphics[width=10cm]{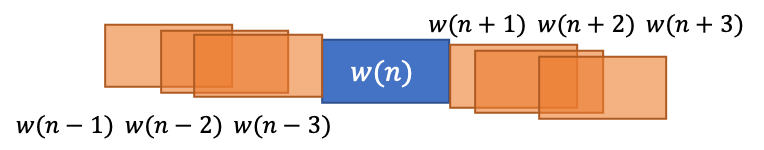}
\end{center}
\caption{How the neighbours of a word are selected for the skipgram.}
\label{fig:skipgram_eg}
\end{figure}

\subsection{Semantic relation of learned keywords}

We analyze a set of selected clusters, which correspond to identifiable semantic content. Each of these clusters is contrasted to its closest neighboring clusters in respect of semantic relation. Table \ref{table:large_word2vec} shows the clusters that contain more than $5$ sequences, and Table \ref{table:short_word2vec} shows those with equal or less than $5$ sequences. Only those recognizable and unique keywords or key-phrases are shown in the tables, By recognizable we mean that the cluster has a high purity value such that it could be clearly related to a specific word or phrase. By unique we mean that if there exist more than one clusters that represent with the same phrase, only one is shown in the tables. The corresponding words of the neighbouring clusters are listed in ascending order of their distances from the selected clusters.

\begin{table}[h]
\centering
\footnotesize
\begin{tabular}{|P{1.3cm}|l|P{9cm}|}
\hline
\textbf{Cluster size} & \multicolumn{1}{|c|}{\textbf{Cluster content}} & \multicolumn{1}{|c|}{\textbf{k nearest neighbour}}\\\hline
62 & probability & probability, probability dense-, proba-, product, -bility, the problem, the probability \\
33 & communication & communication,	-cation ,communica- ,communica- \\
27 & waveform & waveform, a waveform, waveform, waveforms  \\
25 & numbers & something, we can, which is, of that, symbol, sum \\ 
17 & expected value	& expected val-, value, the expected, expected val-, expect	value \\
6 & minus one & x one, each one, minus x one, r one, one, square, times, zero \\\hline

\end{tabular}

\caption{Neighbours of larger clusters (more than $5$)}
\label{table:large_word2vec}
\end{table}

\begin{table}[h]
\centering
\footnotesize
\begin{tabular}{|p{1.3cm}|l|l|}
\hline
\textbf{Cluster size} & \multicolumn{1}{|c|}{\textbf{Cluster content}} & \multicolumn{1}{|c|}{\textbf{k nearest neighbour}}\\\hline
4	& bandwidth	& the bandwidth, input, and to, ahh, as	,bring. where, information \\
4	& gaussian noise & gaussian, in terms, start to, y prime, but, BLT, this \\
3	& probability & study, minimize, entropy, j, -bility, the probability, probability	\\
2	& separation & we study, ahh , people, doesn't, som-, because, sy-, under- \\
2	& combination & understand, one, inherent \\\hline
\end{tabular}

\caption{Neighbours of smaller clusters (equal or less than $5$)}
\label{table:short_word2vec}
\end{table}


It is clear that clusters with larger size are more likely to capture important semantic information and learn meaningful keywords. For example, cluster represents the keyword ``waveform'' has size of $27$ . Its closest neighboring clusters correspond to closely related words or phrases like ``a waveform'', ``waveform'' and ``waveforms''. For some of the recognized keywords, the closest neighboring clusters match parts of the words, e.g., cluster referring to ``probablity'', with size $62$, have closely related clusters that correspond to ``proba-'' and ``-bility'', while for the cluster corresponding to ``communication'', ``communica-'' and ``-cation'' are its closest neighbors. These observations suggest the possibility of merging semantically related keyword clusters. They also provide a basis for understanding the structure of the language.

On the other hand, it is interestingly noted that clusters representing different words/phrases under the same technical topic are closely related in the word embedding space. For example, the cluster ``minus one'' is found to be close to ``x1'', ``r1'', ``square'', ``times'' and ``zero''.

As for clusters of smaller sizes, relationship between the clusters and their neighbours become less relevant. 
For example, in Table \ref{table:short_word2vec}, although cluster referring to ``probability'', with size $3$, has some relevant neighbours ``-bility'' and ``the probability'', they are not its closest neighbours. The closest neighbours ``study'', ``minimize'' and ``entropy'' are less relevant.
For clusters referring to ``separation'' and ``combination'', their neighbours are totally irrelevant.

\subsection{Discussion}

Unlike most applications of Word2Vec, where the words in documents are well defined (linguistically), the automatically discovered word clusters usually do not have 100\% purity. They typically could be mapped to multiple different phrases. 
The embedded relationship may not in fact reflect the true relationship of the actual phrases.

Moreover, the clusters may be established based on acoustic-phonetic similarities rather than semantic relation. This happens especially on short clusters which represent partial words. For example, some clusters referring to ``problem'' and ``proba-(bility)'' are found to be close to each other. 
For clusters of short sequences, the meaning of similarity in embedding space could be different from those of long sequences.

\section{Topic relevance of discovered keywords}


\subsection{TF-IDF} \label{TFIDF}
Term frequency-inverse document frequency (TF-IDF) \cite{ramos2003using} makes use of the occurrence count of terms, which could be language units at any level, to characterize, compare and correlate language documents. TF-IDF has long been applied successfully to text document retrieval and related tasks.

Let $t$ denote a term. The term frequency of $t$ in document $d$ is given by,
\begin{equation}
tf(t,d) = freq_{t,d} 
\end{equation}
where $d \in D$ and $D = \{d_1, ..., d_N\}$ is a collection of documents concerned.

The inverse document frequency measures the importance of $t$ by looking into its occurrence counts across all documents, i.e., 
\begin{equation}
idf(t,D) = log{N \over {| t \in d, d \in D | }}
\end{equation}

TF-IDF combines $tf$ and $idf$ as
\begin{equation}
tfidf(t,d) = tf(t,d)*idf(t,D)
\end{equation}

A term with high TF-IDF score for a document means that it is a key term to the document. We consider the term likely to be a keyword that can be used to identify the document.

\subsection{Experiment on single-course recordings}


In this experiment, the ASM is trained with all lecture recordings of the course COMM. Unsupervised word discovery is performed on the pseudo transcriptions generated with the ASM. The approach of TF-IDF is applied with each of the discovered word clusters regarded as a term and each lecture in the course COMM is treated as a document.

The three word clusters with the highest TF-IDF scores in each lecture are shown as in Table \ref{table:comm_list}. Words or phrases that do not give information specific to the lecture topic are considered as irrelevant, e.g., ``before'', ``this particular'', ``whatever'', ``until here''. Among the $72$ keywords, only $15$ are irrelevant.

\newcommand*{\MyIndent}{\hspace*{0.7cm}}%
\renewcommand{\arraystretch}{0.7}
\begin{table}[]
\centering
\footnotesize
\hspace*{-0.5cm}
\resizebox{15cm}{!}{%
\begin{tabular}{|l|l|}
\hline
\multicolumn{2}{|c|}{\textbf{Principles of Digital Communications I (COMM)}}
\\\hline
1: Introduction  & 13: Random Processes   \\
\MyIndent- you have a sequence of bit coming out & \MyIndent- this is standard PAM  \\
\MyIndent- noise get set & \MyIndent- use of k  \\
\MyIndent- power & \MyIndent- (en)tropy quantiza(tion)  \\

2: Discrete Source Encoding &14: Jointly Gaussian Random Vectors \\
\MyIndent- Kraft inequality  & \MyIndent- QAM system \\
\MyIndent- binary tree & \MyIndent- five pi over \\
\MyIndent- binary digit & \MyIndent- when you do \\

3: Memory-less Sources &15: Linear Functionals \\
\MyIndent- siblings  & \MyIndent- linearly dependent \\
\MyIndent- chance variables & \MyIndent- expected value \\
\MyIndent- represent & \MyIndent- (densi)ty of j \\

4: Entropy and Asymptotic Equipartition Property &16: Review; Introduction to Detection \\
\MyIndent- which comes out of the source  & \MyIndent- one variable covariance function \\
\MyIndent- variable & \MyIndent- zero mean \\
\MyIndent- typical set & \MyIndent- these variables \\

5: Markov Sources &17: Detection for Random Vectors and Processes \\
\MyIndent- abab  & \MyIndent- finite set of e- \\
\MyIndent- typical set & \MyIndent- zero mean \\
\MyIndent- random variable & \MyIndent- whatever \\

6: Quantization &18: Theory of Irrelevance \\
\MyIndent- perpendicular bisector  & \MyIndent- hyphothe- \\
\MyIndent- (en)tropy & \MyIndent- hyphothesis \\
\MyIndent- jumps up to a big value & \MyIndent- binary \\

7: High Rate Quantizers and Waveform Encoding &19: Baseband Detection \\
\MyIndent- you look at this integral here  & \MyIndent- maximum likelihood \\
\MyIndent- they aren't very useful & \MyIndent- vectors \\
\MyIndent- entropy & \MyIndent- this particular \\

8: Measure &20: Introduction of Wireless Communication \\
\MyIndent- which lie between  & \MyIndent- (or)thogonal codes \\
\MyIndent- set is mea(sured) & \MyIndent- think about \\
\MyIndent- ah but anyway & \MyIndent- see this \\

9: Discrete-Time Fourier Transforms &21: Doppler Spread \\
\MyIndent- function of frequency  & \MyIndent- electromagnetic \\
\MyIndent- fourier series & \MyIndent- to figure out \\
\MyIndent- before & \MyIndent- this equation \\

10: Degrees of Freedom &22: Discrete-Time Baseband Models for Wireless Channels\\
\MyIndent- minus w to plus w  & \MyIndent- in terms of the sampling \\
\MyIndent- like the inverse transform & \MyIndent- the real part \\
\MyIndent- ah but anyway & \MyIndent- tau \\

11: Signal Space &23: Detection for Flat Rayleigh Fading and Incoherent Channels \\
\MyIndent- the nomalized form of  & \MyIndent- this max filter \\
\MyIndent- the length & \MyIndent- whenever \\
\MyIndent- the inner product & \MyIndent- until here \\

12: Nyquist Theory &24: Case Study on Code Division Multiple Access \\
\MyIndent- hardly ever talk  & \MyIndent- has four states \\
\MyIndent- zero energy & \MyIndent- want some \\
\MyIndent- timing recovery circuit & \MyIndent- think and take \\\hline

\end{tabular}
}
\caption{List of lecture topics for COMM and their 3 keywords with highest TF-IDF scores}
\label{table:comm_list}
\end{table}


With the TF-IDF scores of all discovered keywords and their corresponding lectures obtained, we make an attempt to visualize the score matrix to analyze the relationship of keywords and lecture content.


Each lecture in COMM is divided into $10$-minutes long audio using a sliding window of $5$-minutes shift, resulting in $15-20$ sessions for each lecture. Each session is treated as one document. 
The TF-IDF score vector of each session is calculated and visualized using t-SNE in Figure \ref{fig:system2}, with the sessions from same lecture labeled in same shape and colour.

\renewcommand{\arraystretch}{1.0}
\begin{figure}[h]
\begin{center}
\includegraphics[width=14cm]{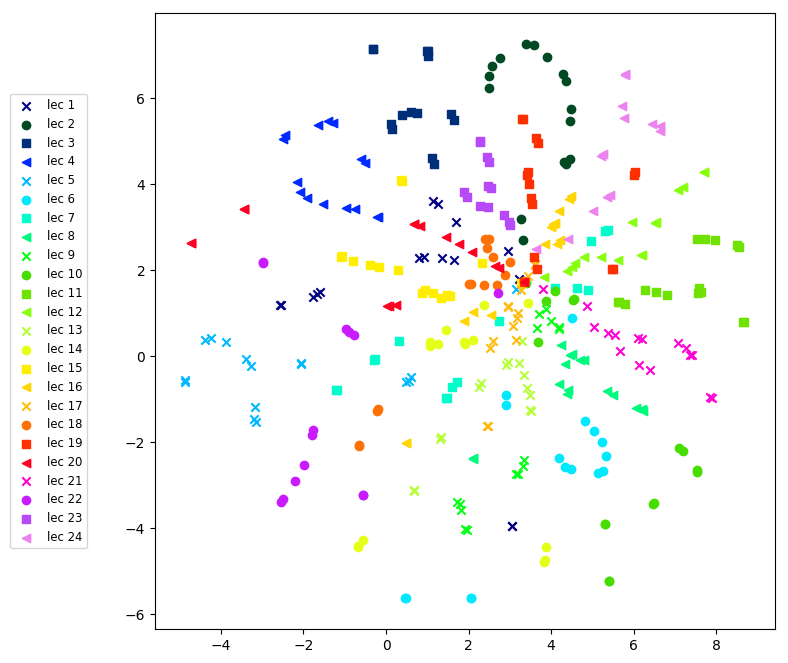}
\end{center}
\caption{t-SNE plot on the TF-IDF score matrix of 24 lectures in COMM}
\label{fig:system2}
\end{figure}

Even though the lectures are from same course, most of the sessions from same lecture are projected to close region.

\subsubsection{Finding most similar sessions/lectures}

An experiment on topic comparison is done. 
Given a session, another session that has the most similar content/topic is found using session TF-IDF scores.

Lectures are cut into $5$-minutes long sessions without overlapping. Each session is treat as a document and TF-IDF scores of all sessions are calculated. The session with TF-IDF score vector that gives the highest cosine similarity with the tested session is selected. 10 fold cross validation is conducted. The average accuracy in finding sessions being in same lectures is 92.73\%, with standard deviation of 5.26\%. This shows that weighting the keywords based on TF-IDF can highlight the distinctive keywords of one lecture from another. 


\subsection{Experiment on multiple-courses recordings}

To evaluate if topic comparison can also be applied to lectures from various courses, with different speakers and recording environments, we select 11 lectures from 5 different courses in Table \ref{table:course_info} in Section $3.4$. The topics of the lectures are listed in Table \ref{table:cross_courses}.

\begin{table}[h]
\centering
\footnotesize
\begin{tabular}{|l|}
\hline

\multicolumn{1}{|c|}{Mathematics for Computer Science (MATH)} \\\hline
Lecture 4: Number Theory I \\
Lecture 8: Graph Theory II: Minimum Spanning Trees \\\hline

\multicolumn{1}{|c|}{Principles of Digital Communications I (COMM)} \\\hline
Lecture 9: Discrete-Time Fourier Transforms \\
Lecture 19: Baseband Detection \\\hline

\multicolumn{1}{|c|}{Discrete Stochastic Processes (STOP)} \\\hline
Lecture 8: Markov Eigenvalues and Eigenvectors \\
Lecture 22: Random Walks and Thresholds \\\hline

\multicolumn{1}{|c|}{Geometric Folding Algorithms: Linkages, Origami, Polyhedra (ALGO)} \\\hline
Lecture 4: Efficient Origami Design \\
Lecture 10: Kempe's Universality Theorem \\
Lecture 13: Locked Linkages \\\hline

\multicolumn{1}{|c|}{Introduction to Computer Science and Programming in Python (PYTH)} \\\hline
Lecture 6: Recursion and Dictionaries \\
Lecture 12: Searching and Sorting \\\hline
\end{tabular}
\caption{Selected courses for the 5 courses}
\label{table:cross_courses}
\end{table}

%
%
The same parameter setting of $T=0.3$ and $a=1.4$ is used. There are a total of $12,200$ keyword clusters discovered, with each lecture having around $5,640$ clusters on average. Around $2,000-3,000$ keywords are shared among different courses. 

Similar with the single-course experiment, each lecture is split into $5$-minute sessions. The TF-IDF scores are calculated in session-based. TF-IDF score vectors of the sessions are projected using t-SNE in Figure \ref{fig:system2}. It is observed that lectures from the same course are in the same region, expect for the course COMM, with lecture 19 being at the same region as STOP.

\begin{figure}[h]
\begin{center}
\includegraphics[width=14cm]{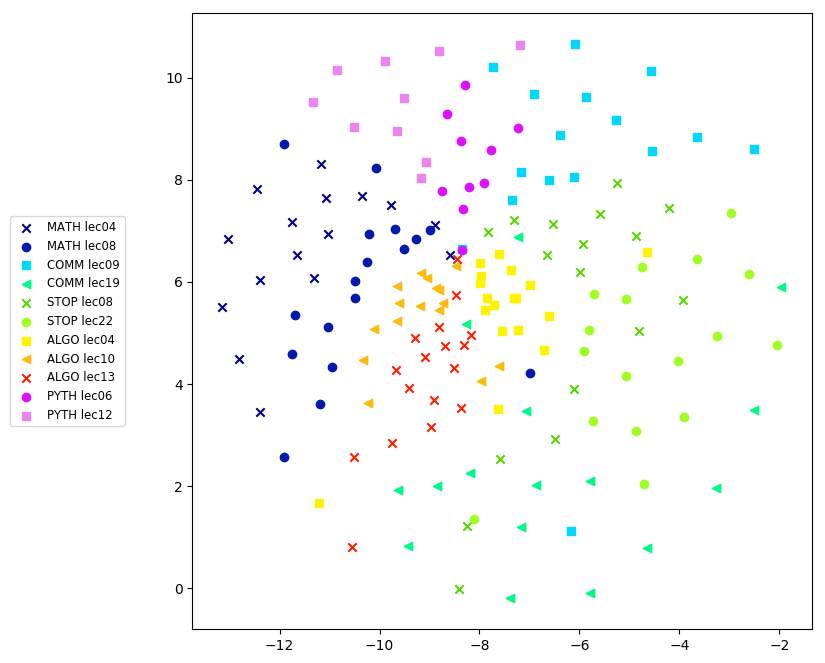}
\end{center}
\caption{t-SNE plot on the TF-IDF score matrix of 11 lectures from 5 courses}
\label{fig:system2}
\end{figure}

\subsubsection{Finding most similar lectures}

Lecture retrieval based on TF-IDF score is also performed. Given a lecture $D$, the most similar lecture among the rest of the 10 lectures are obtained by finding the lecture with TF-IDF score vector that gives the highest cosine similarity with $D$. It is found that most of the lectures retrieved are from the same course, expect for COMM lecture 19 ``Baseband Detection'', its most similar lecture is ``Random Walks and Thresholds'' from STOP lecture 22.

The two lectures' topics do not look similar at first sight, but when we analysis their content, we found that the content is very similar. When analysing the $60$ most frequent unigrams excluding function words,  it is found that $20\%$ of the words are overlapping, such as ``probability'' , ``likelihood'', ``variable'', ``function'' and ``minus''. There are also words with very similar meaning, such as ``product'' in COMM and ``times'' in STOP, both lectures have mathematics symbols for calculation but with different notations, such as ``k'', ``t'', ``n0'' in COMM, ``e'', ``f'',``h0'' in STOP.

It is shown that even though the two lectures are from different courses, both lectures use very similar methods in solving their problems. It would be preferable if the relationship of words with similar meaning, e.g. mathematics notation, can be learnt to benefit the topic comparison process. Combining the use of Word2Vec can be considered in the future.

\subsection{Potential applications}

After analysing the keywords discovered and the result on lecture comparison, the proposed model shows its potential in applications regarding topic comparison and modelling. In this section, several potential real world applications are discussed.

\subsubsection{Scenario 1}
There are platforms that provide online courses with different languages available, such as Coursera\footnote{Coursera is an online learning platform that provides free courses taught by various universities and organizations. Link: https://www.coursera.org}. When the user is watching a lecture and come across some terms that he wants to know more. Service can be provide for him to search for  other fundamental courses/lectures to learn more about the specific terms.

\begin{figure}[h]
\begin{center}
\includegraphics[width=11cm]{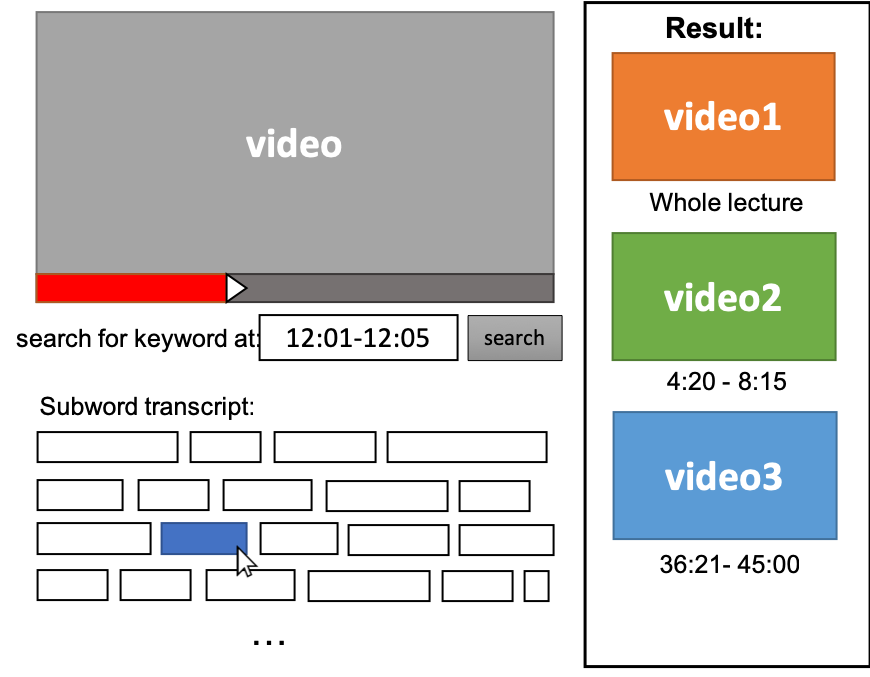}
\end{center}
\caption{Illustration of the lecture recommendation based on segments system.}
\label{fig:lec_recommand}
\end{figure}

One application for the proposed system is to allow user to search for similar courses/lectures by clicking and identifying segments he wants to look into. The frontend design is illustrate in Figure \ref{fig:lec_recommand}, the keywords discovered are presented as segment transcription, which may or may not be displayed to the user. Very similar to interactive transcript, the current segment that is playing will be located and highlighted. If the segments are displayed, the user can click on the segment that he is interested to look into. If not, he can specify the time interval of the words in the search bar. The search engine can look for additional lectures that are related to the segments specified.

At the back end, the system preprocess all lectures in the database by discovering their spoken terms and compute the TF-IDF scores. The information is stored in the database. The TF-IDF score of selected session is calculated upon request and is compared with the TF-IDF scores of lectures from all courses. The most similar lectures are retrieved and displayed in the result bar. The comparison process can also be done by splitting the lectures into sessions, only sessions with highest similarities are displayed.

\subsubsection{Scenario 2}

Another possible application for our system is video suggestion in multimedia sharing platforms such as Youtube\footnote{https://www.youtube.com}.

Traditional video recommendation system uses collaborative filtering  approach that suggests related videos based on their popularity and correlation with the current video \cite{chen2010collaborative}. Limitation of this approach is that similarity between videos are measured based on users' activity instead of the content. Newly uploaded videos or less popular videos may be underweighted. Same situation may happen for videos in low resource languages, which may not have transcription or video tags. 

\begin{figure}[h]
\begin{center}
\includegraphics[width=10cm]{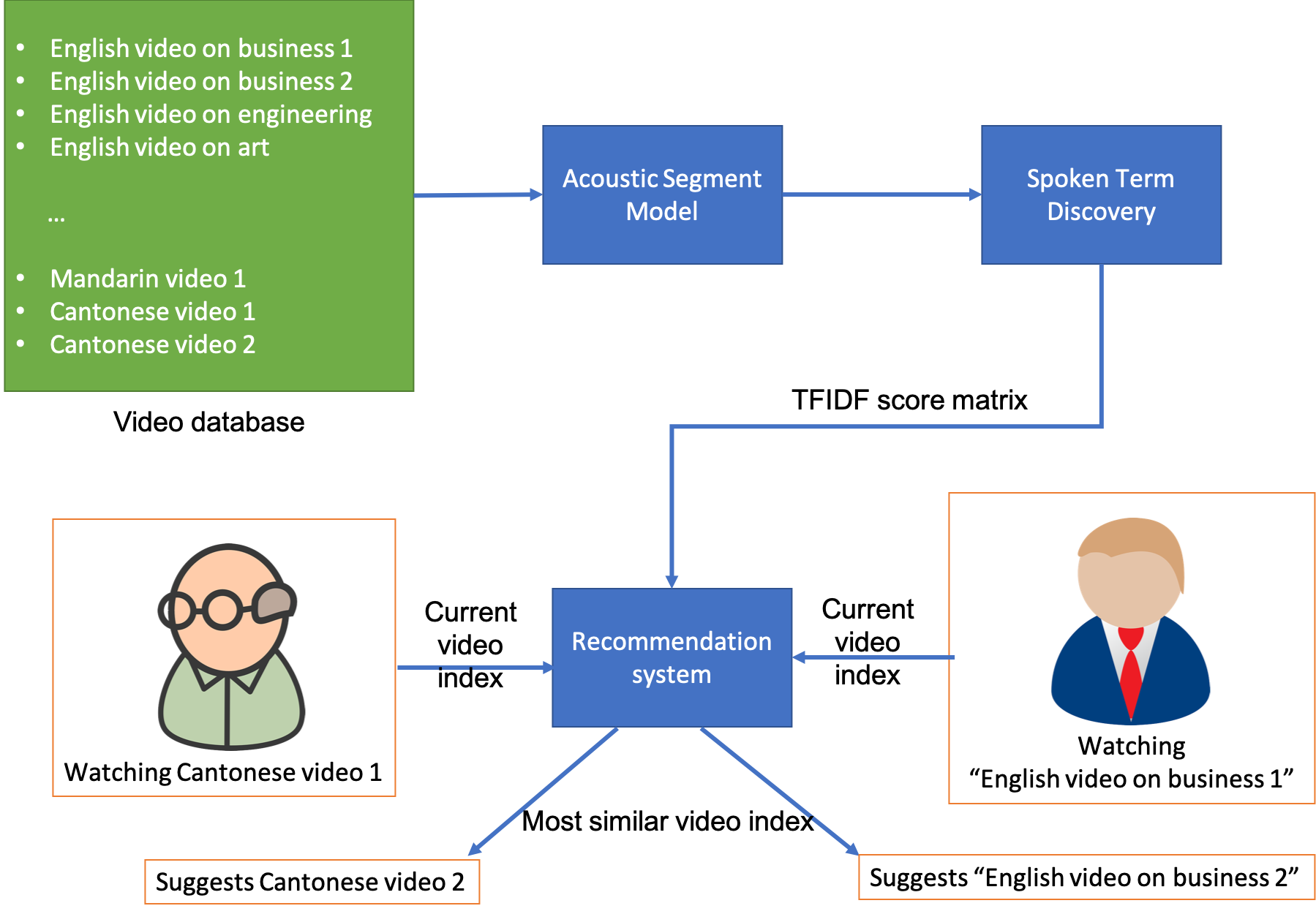}
\end{center}
\caption{Illustration of the video recommendation system.}
\label{fig:skipgram_eg}
\end{figure}

The proposed system can be applied to any languages without the need of pre-labelling the videos with languages or tags. Unsupervised spoken term discovery can be done on all videos in the database, follow by calculating the TF-IDF scores of each video based on the discovered keywords. The TF-IDF score matrix is stored. When a user is watching a video, next suggested videos can be retrieved by searching for the videos whose TF-IDF scores give the highest cosine similarities with the current one.

The system can also facilitate elders who only speak their own dialects in finding videos they can understand. 
Since ASM training and spoken term discovery process are language independent and they can unsupervisely discover keywords that are acoustically and linguistically similar, the system is able to suggest videos with the same dialects.

\section{Summary}

In this chapter, the proposed model is experimented with online lectures to evaluate the quality and coverage of keywords discovered. The semantic relationship and topic relevance of the discovered keywords are analyzed.

The proposed model is shown to be effective in discovering keywords that can mostly align with frequent words in actual transcription. The discovered keywords are also observed to be context related.

Word2Vec is used to understand the relationship among the keywords. It is found that Word2Vec is able to group similar clusters together. Potential improvement to the spoken term discovery result is to merge different clusters corresponding to same phrases, and linking between multiple short clusters and long clusters.

TF-IDF is used to analysis the relationship of a lecture topic and its discovered keywords. It is shown that topic comparison can be done by computing the cosine similarities of the TF-IDF score vectors of the lectures. Potential applications on video sharing platform are discussed as well.

\subsubsection{Area of improvement}

The proposed system needs improvement in its ability in identifying keywords of relatively short length. In fact, candidate sequences that are shorter than 4 subword units are not included when generating the candidate sequences for clustering (see Algorithm \label{LocalAlign} in Section \ref{ssec:bag_of_seq}). 

Computation time for clustering process is very long, it may take days to cluster the whole course with $24$ lectures. Other more robust clustering methods besides leader clustering can be considered, such as Burkhard-Keller Trees \cite{baeza1998fast}, KNN-based clustering using k-dimensional tree \cite{otair2013approximate}.

Potential improvement for the application of topic comparison is to combine the use of TF-IDF and Word2Vec, so that topics with similar content but different terminologies can also be discovered easier.
\chapter{Conclusion and Future Work}
\section{Conclusion}


Unsupervised acoustic modelling is a critical technology in dealing with zero resource languages. 
Our study has shown that it is possible to develop unsupervised model to represent a language without requiring any linguistic information and transcriptions.

In acoustic segment modelling, feature extraction is an important part in unit learning and discovery. Our study shows that introducing bottleneck layer to DNN is effective method in extracting language independent features that suit any language. We also found that bottleneck features can be learnt from training multilingual DNN using resource rich languages. The bottleneck features learnt suits the application of subword unit discovery compared to conventional acoustic features by segment clustering.
Among different types of clustering methods, density-based clustering can better cluster the bottleneck features proposed.

By treating untranscribed freestyle lecture as an example of zero resource recordings, an acoustic segment model is built on top of a few courses. With reference to bioinformatics symbol searching problems, the subword unit sequences generated from the ASM can be used for spoken pattern discovery. It is shown that spoken terms discovered are aligned with real transcription of the lectures, providing evidence that ASM trained using bottleneck feature extracted from multilingual DNN can effectively model untranscribed recordings.

After experimenting on a set of multi-speakers, multi-topics courses, it is shown that the proposed model is able to perform effectively under both single-speaker-single-channel and multiple-speakers-multiple-channels scenarios. The discovered spoken terms can also be further used for lectures comparison, analysing similarity between lectures based on discovered words. The relationship between each spoken term can also be understood using Word2Vec.  

\section{Contribution}

This thesis has made the following contributions towards the developing in unsupervised acoustic modelling and spoke term discovery:

\begin{itemize}
    \item We proposed using bottleneck feature obtained from multilingual DNN for ASM training. The features are language independent and completely data driven, and can effectively represent the training recordings. The clustering effectiveness of several clustering algorithms are also investigated.
    \item Tools that are originated from text and document modelling are used in speech application. TF-IDF and Word2Vec are used in understanding relationship between discovered spoken terms and topics of recordings.
    \item The proposed system can be applied to real word applications such as videos, lectures recommendation in multilingual multimedia platform.
\end{itemize}

\section{Limitations}

The evaluation presented in this work requires a more comprehensive analysis. In this thesis, the proposed system is only evaluated based on observation of its performance on real data. Although metrics from ZeroSpeech Challange are used, the metrics can only evaluate part of the tasks separately and it is difficult to evaluate the whole discovery process.
Currently, there only exist evaluations on certain tasks, such as feature representations, subword modelling, segmentation. However, there are little work regrading the development and evaluation of the complete spoken term discovery system. There are also lack of evaluation on the applications of this technology. The system can only be concluded as functioning with reasonable performance, but a comprehensive measurement is hard to conduct.


Also, the proposed system is evaluated on only very few languages. Evaluation and extension towards language-independent system still require more work to be done.

\section{Suggestion for future work}

There are a few directions that can be conducted for future work.

\subsubsection{Better clustering algorithms}
Although it is shown that the clustering algorithms used in segment clustering in Chapter 3 and keywords discovering in Chapter 4 are effective, there may exist other algorithms that can better cluster the data to yield better system performance.

One possible direction is to look for clustering algorithm that can cluster large amount of data more efficient, so the system can process larger set of recordings. In terms of clustering bottleneck features, it is observed that hierarchical clustering (HDBSCAN, AHC) generally have better clustering performance than other clustering algorithms. Algorithms that cluster in hierarchical based but are relatively less computation expensive can be investigated.

Since bottleneck features is being introduced recent years and there are still a lot work to be done in understanding its proprieties. More research work in analyzing the data complexity and property of bottleneck features is needed, and to determine suitable clustering algorithms accordingly.

While clustering subword sequences into keyword clusters, other string clustering algorithms and different metrics can be considered. Moreover, alternative linguistic proprieties can be considered in weighting the subword sequences, as summarized in Chapter 4.

\subsubsection{Experiment on low resource language}

In this thesis, experiments are mostly done on opencourse lectures, without a carefully collected and documented recordings as LDC/published corpora. Although the recordings used are more daily life related, it is hard to conduct a very detail and structured analysis due to lack of man power in labelling the lectures. Tasks specified corpora such as Switchboard, broadcast news for topic classification and keyword search can be experimented for more solid analysis.

Moreover, the ability of the purposed system in processing zero resource language is not fully utilized. Experiment is suggested to conduct on low resource languages to examine the ability in process languages other than English. The system and tools discussed in Chapter 5 also be used to discover linguistics knowledge and provide better understanding to unfamiliar languages.


\pagestyle{plain}

\bibliography{references}

\end{document}